\pdfoutput=1
\documentclass[%
 reprint,
 amsmath,amssymb,
 aps,
prb,
]{revtex4-1}

\usepackage{multirow}
\usepackage{graphicx}
\usepackage{dcolumn}
\usepackage{bm}
\usepackage{hhline}
\usepackage{subfigure}
\usepackage{xcolor}
\usepackage{comment}
\usepackage{soul}
\usepackage{stmaryrd}
\usepackage{array,booktabs}
\usepackage{hyperref}

\newcommand{\NA}{\;\;--}
\DeclareFontFamily{U}{mathb}{}
\DeclareFontShape{U}{mathb}{m}{n}{
  <-5.5> mathb5
  <5.5-6.5> mathb6
  <6.5-7.5> mathb7
  <7.5-8.5> mathb8
  <8.5-9.5> mathb9
  <9.5-11.5> mathb10
  <11.5-> mathb12
}{}
\DeclareSymbolFont{mathb}{U}{mathb}{m}{n}
\DeclareMathSymbol{\ulsh}{3}{mathb}{"E8}
\DeclareMathSymbol{\ursh}{3}{mathb}{"E9}
\DeclareMathSymbol{\dlsh}{3}{mathb}{"EA}
\DeclareMathSymbol{\drsh}{3}{mathb}{"EB}

\allowdisplaybreaks

\begin{document}


\title{{Symmetric} U(1) and $\mathbb{Z}_2$ spin liquids on the pyrochlore lattice}

\author{Chunxiao Liu}
\affiliation{Department of Physics, University of California, Santa
Barbara, CA 93106-9530, USA}
\author{G\'abor B. Hal\'asz}
\affiliation{Materials Science and Technology Division, Oak Ridge
National Laboratory, Oak Ridge, TN 37831, USA} \affiliation{Quantum
Science Center, Oak Ridge, TN 37831, USA} \affiliation{Kavli
Institute for Theoretical Physics, University of California, Santa
Barbara, CA 93106-4030, USA}
\author{Leon Balents}
\affiliation{Kavli Institute for Theoretical Physics, University of
  California, Santa Barbara, CA 93106-4030, USA}
\affiliation{Canadian Institute for Advanced Research, 661
University Ave., Toronto, ON M5G 1M1 Canada}





\date{\today}

\begin{abstract}
The geometrically frustrated 3D pyrochlore lattice has been long
predicted to host a quantum spin liquid, an intrinsic long-range
entangled state with fractionalized excitations. To date, most
proposals for pyrochlore materials have focused on
quantum spin ice, a U(1) quantum spin liquid whose only low energy
excitations are emergent photons of Maxwell type. In this work, we
explore the possibility of finding pyrochlore quantum spin liquids
whose low energy theories go beyond this standard one. We give a
complete classification of symmetric $\mathrm{U(1)}$ and
$\mathbb{Z}_2$ spin liquids on the pyrochlore lattice within the
projective symmetry group framework for fermionic spinons. We find 18
$\mathrm{U(1)}$ spin liquids and 28 $\mathbb{Z}_2$ spin liquids that
preserve pyrochlore space group symmetry while, upon further
imposing time reversal symmetry, the numbers of classes become 16
and 48, respectively. For each class, the most general
symmetry-allowed spinon mean-field Hamiltonian is given.
Interestingly, we find that several U(1) spin liquid classes possess
an unusual gapless multi-nodal-line structure (``nodal star'') in
the spinon bands, which is protected by the projective actions of
the three-fold rotation and screw symmetries of the pyrochlore space
group. Through a simple model, we study the effect of gauge
fluctuations on such a nodal star spin liquid and propose that the
leading terms in the low temperature specific heat have the scaling
form $C/T\sim \sqrt{T}+\sqrt{T}/\ln T$, in contrast to the form $C/T
\sim T^2$ of the standard U(1) pyrochlore spin liquid with gapped
{spinons}.
\end{abstract}

\pacs{Valid PACS appear here}
\maketitle


\section{Introduction}

Quantum spin liquids (QSLs) are zero temperature phases of quantum
magnets in which localized spins evade magnetic long-range order due
to strong quantum fluctuations and form liquid-like states
\cite{Savary_2016}. Such states are fundamentally characterized by
intrinsic long-range entanglement and support non-local
excitations carrying fractionalized quantum numbers. These non-local
fractionalized excitations interact with each other via an emergent
gauge field. Therefore, QSLs are naturally described in terms of
gauge theories.

At a coarse level, different QSLs can be classified by their
underlying low energy effective theories.
Depending on whether a mass exists, the fractionalized spinon
excitations (the matter fields) may or may not bear relevance to the
low energy description of QSLs. The gauge field may also be gapless
as is a U(1) gauge field of photons, or gapped as is a
$\mathbb{Z}_2$ gauge field which is of topological nature. All these
types have been extensively studied in two dimensions (2D). Perhaps
the most well-known $\mathbb{Z}_2$ QSL is the Kitaev model on the 2D
honeycomb lattice: this model supports either gapped or gapless
Majorana fermions \cite{KITAEV20062}. On the other hand, a compact
U(1) gauge theory is confined in 2D \cite{POLYAKOV1977429}, and a
QSL corresponding to the deconfined phase can appear only in the
presence of gapless matter fields. So far, the two most studied
examples are spinon Fermi surface and Dirac U(1) QSLs. On the
experimental side, promising Kitaev materials include a family of
honeycomb iridates \cite{PhysRevB.82.064412} and $\alpha$-RuCl$_3$
\cite{2016NatMa..15..733B}, and new proposals are still in progress
\cite{doi:10.1146/annurev-conmatphys-033117-053934}. A spinon Fermi
surface U(1) QSL has been speculated to emerge in the 2D layered
triangular materials YbMgGaO$_4$ and NaYbSe$_2$
\cite{paddison2017continuous,shen2016evidence,dai2020spinon}, while
a U(1) Dirac QSL may be relevant for the 2D layered kagome material
Herbertsmithite \cite{PhysRevLett.98.117205, Yan1173}. The recent
material NaYbO$_2$ may also realize a Dirac U(1) spin liquid
\cite{PhysRevB.100.144432,bordelon2019field}. These spin liquid
candidates provide a natural ground for the experimental realization
of exotic quantum phenomena such as quantum electrodynamics in three
dimensional spacetime (QED3) and 2D topological order.

Moving to the three-dimensional (3D) world, arguably the most
studied examples are QSLs on the pyrochlore lattice. Consisting of
corner-sharing tetrahedra, the geometrically frustrated pyrochlore
lattice has been proposed to host a QSL phase since the birth of the
concept of QSLs \cite{PhysRev.102.1008}. An important
theoretical advance occurred in 2004: through a rigorous mapping,
Ref.~\cite{hermele2004pyrochlore} showed that the pyrochlore
Heisenberg model with local Ising anisotropy has a U(1) QSL phase,
commonly known as quantum spin ice, which is described by the
Maxwell theory at low energies. Since then, the properties of such
pyrochlore QSLs have been extensively studied
\cite{hermele2004pyrochlore,
PhysRevLett.108.037202,PhysRevResearch.2.013334,PhysRevLett.124.127203},
and numerous experiments have reported liquid-like behaviors in
rare-earth pyrochlore materials
\cite{2017NatCo...8..892S,PhysRevLett.115.097202,2019NatPh..15.1052G,gaudet2019quantum,2021arXiv210109049L}.

Given the interest in quantum spin ice that realizes a prototypical
low energy theory, it is compelling to ask whether there are other
spin liquids whose low energy theory is of a new prototype. In a
narrower sense, this amounts to asking if gauge fields can interact
with novel forms of gapless matter fields. This has indeed been
considered in other works. For example,
Refs.~\cite{xu2012spin,mishmash13theory} considered a class of QSLs
with symmetry protected quadratic spinon band touchings for the
triangular spin liquid candidates Ba$_3$NiSb$_2$O$_9$,
$\kappa$-BEDT, and DMIT. In three dimensions,
Refs.~\cite{huang2017interplay, PhysRevLett.101.197202} studied
possible symmetric spin liquids on the hyperkagome lattice and found
that certain classes possess gapless nodal lines of spinons along
high symmetry paths in the Brillouin zone. A similar conclusion was
drawn in Ref.~\cite{burnell2009monopole} for a QSL on the pyrochlore
lattice. However, in these two 3D examples it is not clear whether
such gapless nodal structures are stable against perturbations.
Nodal lines of excitations also appear in other spin liquids, which
are either robust against gauge fluctuations
\cite{mandal2009exactly, PhysRevLett.117.017204} or symmetry
protected \cite{zhangyahuiposter}.

Another more systematic, yet formal, way of classifying QSLs is
based on their symmetry properties. Due to the absence of magnetic
long-range order, a QSL state usually preserves the full symmetry of
the lattice and it may also preserve time reversal symmetry.
Crucially, due to the emergent gauge structure, the fractionalized
excitations of the system carry a \emph{projective} representation
of the symmetry group \--- a representation of the group extension
of the original symmetry group by the gauge structure. The
classification of symmetric spin liquids can therefore be achieved
by classifying all the distinct projective representations for a
given lattice symmetry and a given gauge group type. In his seminal
work \cite{wen2002quantum}, Wen coined this procedure the
classification of projective symmetry groups (PSGs). The PSG
approach has led to many fruitful results in our understanding of
symmetric spin liquids. For example, we now know that there are at
most 20 different $\mathbb{Z}_2$ QSL classes on the kagome
\cite{PhysRevB.83.224413} and triangular \cite{PhysRevB.93.165113}
lattices with distinct projective symmetries for fermionic spinons,
while the analogous numbers for the 3D hyperkagome
\cite{PhysRevLett.101.197202} and hyperhoneycomb
\cite{PhysRevB.97.195141} lattices are 3 and 160, respectively.
The idea of the PSG has also been applied to the pyrochlore lattice,
in the hope of identifying experimental spin liquid candidates
within the classification
\cite{bergman2006ordering,burnell2009monopole,gang2016magnetic,2018arXiv180601276E,liu2019competing}.

In this work, we apply the PSG method for Abrikosov fermions to give
a complete classification of symmetric QSLs on the pyrochlore
lattice with either $\mathbb{Z}_2$ or U(1) gauge type. For each
gauge type, we first consider only space group symmetry, and later
add time reversal symmetry. In general, we allow spin-orbit coupling
in the underlying spin system and do not require SU(2) spin rotation
symmetry. By following the general PSG principle to solve the
gauge-symmetry consistency equations, we find that there can be at
most 18 and 28 symmetric quantum spin liquids preserving the
pyrochlore PSG for the U(1) and $\mathbb{Z}_2$ gauge types,
respectively. When time-reversal symmetry is imposed, the number of
possible symmetric spin liquids is reduced to 16 for the U(1) type
and is increased to 48 for the $\mathbb{Z}_2$ type. For each class,
the most general symmetry-allowed spinon mean-field Hamiltonian is
given. Importantly, we find that a large family of spinon
Hamiltonians possesses gapless nodal lines along the four equivalent
(111) directions of the Brillouin zone. We call this unusual nodal
structure a ``nodal star'' and show that it is stable at the
mean-field level as it is protected by the projective three-fold
rotation and screw symmetries of the system. We then go beyond the
mean-field level and consider a full-fledged low energy theory of
the spinon nodal star coupled to a U(1) gauge field. Specifically,
we obtain thermodynamic properties of the system by computing the
photon contribution to the free energy. {We find
that the two most dominant low temperature contributions to the
specific heat are $C \sim T^{3/2}$ from the bare spinons and $C \sim
T^{3/2} / \ln T$ from the photon-spinon interactions.} This scaling
of the low temperature specific heat may serve as a clear evidence
for the experimental discovery of a nodal star U(1) QSL.

The rest of this paper is organized as follows. In
Sec.~\ref{sec:level1}, we discuss the symmetry properties of the
pyrochlore lattice, explain the main idea of the PSG, and apply the
PSG procedure to the classification of pyrochlore QSLs with or
without time reversal symmetry. In Sec.~\ref{secMF}, we construct
mean-field Hamiltonians for the fermionic spinons and analyze their
symmetry properties. We prove that several mean-field Hamiltonians
obtained from the U(1) PSG possess symmetry protected nodal lines.
In Sec.~\ref{sec:exp}, we construct a continuum model for the spinon
nodal lines coupled to a U(1) gauge field, and consider the
thermodynamic properties of the system. Finally, we summarize and
discuss our results in Sec.~\ref{sec:discussion}.

\section{\label{sec:level1}Projective symmetry group}

Upon imposing symmetries, a spin liquid phase may split (or
``fractionalize'') into several distinct phases that all preserve
the symmetry action. The phases are distinguished by the distinct
quantum numbers carried by the fractionalized excitations under the
symmetry action.  These spin liquids phases are called symmetric
spin liquids. The classification of symmetric spin liquids can be
viewed as a generalized symmetry analysis which explores the
symmetry action on fractionalized excitations (in our case, the
fermionic spinons). The purpose of this section is to describe the
procedure for this classification.

\subsection{Lattice and time reversal symmetries}

In this subsection, we establish the convention and notation for
this work and give a brief introduction to the symmetry properties
of the pyrochlore lattice. A more detailed analysis of the
pyrochlore space group can be found in Ref.~\cite{liu2019competing}.

The pyrochlore lattice consists of four FCC-type sublattices which
we label by $\mu =0,1,2,3$. The lattice vectors $\bm{e}_1$,
$\bm{e}_2$, and $\bm{e}_3$ are defined as
\begin{subequations}
\begin{align}
\bm{e}_1 & =  \frac{a}{2}(\hat{\bm{y}}+\hat{\bm{z}}),\\
\bm{e}_2 & =  \frac{a}{2}(\hat{\bm{z}}+\hat{\bm{x}}),\\
\bm{e}_3 & =  \frac{a}{2}(\hat{\bm{x}}+\hat{\bm{y}}),
\end{align}
\end{subequations}
where $a$ is the cubic lattice constant, the Cartesian coordinate
has its basis $\hat{\bm{x}}$, $\hat{\bm{y}}$, and $\hat{\bm{z}}$
aligned with the cubic system of the pyrochlore lattice, and its
origin sits on a $\mu=0$ site. We define the following
sublattice-dependent coordinate:
\begin{align}
(r_1,r_2,r_3)_\mu &\equiv \bm{r}_\mu \equiv r_1
\bm{e}_1+r_2\bm{e}_2+r_3\bm{e}_3 + \frac{1}{2} \bm{e}_\mu,
\end{align}
where it is implicitly understood that $\bm{e}_0=0$.

The space group of the pyrochlore lattice group is
$Fd\overline{3}m$. It is generated by the following five symmetry
operations:
\begin{subequations}
\begin{align}
T_1 \colon &\bm{r}_\mu\rightarrow (r_1+1,r_2,r_3)_\mu,\\
T_2 \colon &\bm{r}_\mu\rightarrow (r_1,r_2+1,r_3)_\mu, \\
T_3 \colon &\bm{r}_\mu\rightarrow (r_1,r_2,r_3+1)_\mu, \\
\overline{C}_6\colon&\bm{r}_\mu\rightarrow (-r_3-\delta_{\mu,3},-r_1-\delta_{\mu,1},-r_2-\delta_{\mu,2})_{\overline{C}_6(\mu)}, \\
S\colon   &\bm{r}_\mu \rightarrow \nonumber\\
&(-r_1-\delta_{\mu,1},-r_2-\delta_{\mu,2},r_1+r_2+r_3+1-\delta_{\mu,0})_{S(\mu)},\nonumber\\
\end{align}
\end{subequations}
where $T_1$, $T_2$, and $T_3$ are translations along the lattice
vectors $\bm{e}_1$, $\bm{e}_2$, and $\bm{e}_3$, respectively,
$\overline{C}_6$ is a sixfold rotoinversion around the $[111]$ axis,
and $S$ is a nonsymmorphic screw operation which is the composition
of a twofold rotation around $\bm{e}_3$ and a translation by
$\bm{e}_3/2$. In the above equations, we defined the symmetry action
on the sublattice indices: $\overline{C}_6(\mu)=0,2,3,1$ and
$S(\mu)=3,1,2,0$ for $\mu=0,1,2,3$. By definition, the rotoinversion
can be written as the composition of an inversion (with respect to
the origin) and a threefold rotation around the $[111]$ axis:
$\overline{C}_6=I\circ {C}_3$, with $I=\overline{C}_6^3$ and
${C}_3=\overline{C}_6^4$. The point group of the pyrochlore lattice
is the cubic group $O_\text{h}$. It is generated by $\overline{C}_6$
and $S'$, where $S'$ is a twofold rotation around $\bm{e}_3$.

In addition to the pyrochlore space group symmetries, time-reversal
operation $\mathcal{T}$ is an internal symmetry that commutes with
all space-group operations and satisfies $\mathcal{T}^2 = -1$ when
acting on a half-integer spin state. The pyrochlore symmetry group
is then completely characterized by the following group relations:
\begin{subequations}\label{SGtr}
\begin{align}
T_iT_{i+1} T_i^{-1} T_{i+1}^{-1} &= 1,\quad i=1,2,3, \label{4a}\\
\overline{C}^6_6 &= 1,  \\
S^2 T^{-1}_3 &= 1, \\
\overline{C}_6 T_i \overline{C}^{-1}_6 T_{i+1} &=1,\quad i=1,2,3, \label{eq:4d}\\
ST_iS^{-1} T^{-1}_3T_i&= 1, \quad i=1,2,\\
ST_3 S^{-1} T_3^{-1} &=1,\\
(\overline{C}_6S)^4&=1, \\
(\overline{C}_6^3S)^2 &=1, \label{4h}\\
\mathcal{T}^2 &= -1, \label{4i}\\
\mathcal{T}\mathcal{O}\mathcal{T}^{-1}\mathcal{O}^{-1}&=1,\quad
\mathcal{O}\in\{T_1,T_2,T_3,\overline{C}_6, S\}, \label{4j}
\end{align}
\end{subequations}
where it is implicitly understood that $i+3 \equiv i$.

\subsection{Projective symmetry group}

In this subsection, we describe the basic idea of the PSG and
classify all symmetric $\mathbb{Z}_2$ and U(1) spin liquids on the
pyrochlore lattice. To start with, one expresses spins in terms of
the Abrikosov partons
\begin{equation}\label{parton}
\hat{\bm{S}}_{\bm{r}_\mu} = \frac{1}{2} f^\dag_{\bm{r}_\mu}
\bm{\sigma} f^{\vphantom\dagger}_{\bm{r}_\mu},
\end{equation}
where $\bm{\sigma}=(\sigma^1,\sigma^2,\sigma^3)$ are the Pauli
matrices. Formally, partons are mathematical representations of the
fractionalized excitations in a spin liquid phase, and the Abrikosov
partons $f_{\bm{r}_\mu}=
\left(\begin{array}{c}f_{\bm{r}_\mu,\uparrow}\\f_{\bm{r}_\mu,\downarrow}\end{array}\right)$
are introduced here to describe the fermionic spinon excitations
that are of our interest. The transformation in Eq.~\eqref{parton},
however, is not a faithful one. The partons live in an enlarged
Hilbert space at each site $\bm{r}_\mu$, while the original Hilbert
space is recovered under the constraint
\begin{equation}
\sum\limits_{\sigma=\uparrow,\downarrow}
f^\dag_{\bm{r}_\mu,\sigma}f^{\vphantom\dagger}_{\bm{r}_\mu,\sigma} =
1,\quad \forall \, \bm{r}_\mu.
\end{equation}
As a consequence, the parton description contains redundant
information: a local U(1) gauge transformation
$f_{\bm{r}_\mu}\rightarrow e^{i \theta(\bm{r}_\mu)} f_{\bm{r}_\mu}$
leaves $\hat{\bm{S}}_{\bm{r}_\mu}$ invariant. In fact, such a gauge
redundancy can be enlarged to SU(2), which can be seen from the
identity
\begin{equation}\label{Ssss}
\hat{\bm{S}}_{\bm{r}_\mu} = \frac{1}{4}
\mathrm{Tr}(\Psi^\dag_{\bm{r}_\mu} \bm{\sigma} \Psi_{\bm{r}_\mu}),
\end{equation}
with $\Psi_{\bm{r}_\mu} = \left(\begin{array}{cc}
f_{\bm{r}_\mu,\uparrow} & f^\dag_{\bm{r}_\mu,\downarrow}\\
f_{\bm{r}_\mu,\downarrow} & -
f^\dag_{r_\mu,\uparrow}\end{array}\right)$, and from the fact that
any site-dependent SU(2) gauge transformation,
\begin{equation}
G \colon \Psi_{\bm{r}_\mu} \rightarrow \Psi_{\bm{r}_\mu}
W(\bm{r}_\mu),\quad W(\bm{r}_\mu)\in \text{SU(2)}, \label{op-G}
\end{equation}
leaves the spins $\hat{\bm{S}}_{\bm{r}_\mu}$ invariant. The
enlargement of the parton Hilbert space and the gauge redundancy
must be properly treated to validate the parton description.

The PSG method is a way to resolve this redundancy in the parton
description of a spin liquid with full lattice symmetries (a
so-called symmetric spin liquid). The crucial step is to realize
that physical symmetries act \emph{projectively} on the parton
operators, and that seemingly different parton Hamiltonians describe
the same physics if they are related by gauge transformations.
Conversely, if two parton Hamiltonians cannot be related by gauge
transformations, they must carry different projective
representations of the physical symmetry \--- this suggests that the
classification of projective symmetry will lead to a full
classification of symmetric spin liquids. We now formulate this
statement in a more concrete way. Consider a spin-orbit coupled spin
system on a pyrochlore lattice. Under a space group operation
$\mathcal{O}$ the spins transform as $\mathcal{O}\colon
\hat{\bm{S}}_{\bm{r}_\mu}\rightarrow U_{\mathcal{O}}
\hat{\bm{S}}_{\mathcal{O}(\bm{r}_\mu)} U^\dag_{\mathcal{O}}$, where
$U_{\mathcal{O}}$ is the SU(2) rotation matrix associated with the
operation $\mathcal{O}$. (When $\mathcal{O}$ is a pure translation,
the SU(2) matrix $U_{\mathcal{O}}$ is simply the identity matrix.)
According to Eq.~\eqref{Ssss}, we na\"ively expect that the partons
transform as
\begin{equation}
\mathcal{O}\colon \Psi_{\bm{r}_\mu} \rightarrow U^\dag_{\mathcal{O}}
\Psi_{\mathcal{O} (\bm{r}_\mu)}\label{op-O}
\end{equation}
with nontrivial rotations
\begin{equation}\label{su2_under_o}
U_{\overline{C}_6} = U_{C_3}=e^{-\frac{i}{2}
\frac{2\pi}{3}\frac{(1,1,1)}{\sqrt{3}}\cdot \bm{\sigma}},\quad U_S =
e^{-\frac{i}{2}\pi \frac{(1,1,0)}{\sqrt{2}}\cdot \bm{\sigma}}.
\end{equation}
However, due to the SU(2) gauge redundancy, any operation
$\mathcal{O}$ can be accompanied by a site-dependent SU(2) gauge
transformation of the form in Eq.~\eqref{op-G}. The partons thus
transform projectively as
\begin{equation}\label{b_under_o}
\widetilde{\mathcal{O}} = G_{\mathcal{O}} \circ \mathcal{O} \colon
\Psi_{\bm{r}_\mu} \rightarrow  U^\dag_{\mathcal{O}}
\Psi_{O(\bm{r}_\mu)}W_{\mathcal{O}}[\mathcal{O}(\bm{r}_\mu)],
\end{equation}
where the symbol ``$\circ$'' indicates that the projective operation
$\widetilde{\mathcal{O}}$ is the composition of the physical
symmetry operation $\mathcal{O}$ and the gauge transformation
$G_{\mathcal{O}}$.

The projective symmetry can be extended to include internal
symmetries, and here we consider time reversal operation
$\mathcal{T}$ as an example. The spins transform under $\mathcal{T}$
as $ \hat{\bm{S}}_{\bm{r}_\mu}\xrightarrow{\mathcal{T}}
\mathcal{K}^\dag U_{\mathcal{T}} \hat{\bm{S}}_{\bm{r}_\mu}
U^\dag_{\mathcal{T}} \mathcal{K}$, where $U_{\mathcal{T}} = i
\sigma^2$, and $\mathcal{K}=\mathcal{K}^\dag=\mathcal{K}^{-1}$
applies complex conjugation to everything on its right. Using the
special property of the SU(2) algebra, one can design the projective
action of ${\mathcal{T}}$ on $\Psi$ to be unitary (see
App.~\ref{app:u1_TR} for detailed derivation):
\begin{equation}\label{timereversaldef3}
\widetilde{\mathcal{T}}=G_{\mathcal{T}}\circ \mathcal{T}\colon
\Psi_{\bm{r}_\mu}\rightarrow
U_{\mathcal{T}}\Psi_{\bm{r}_\mu}{W}_{\mathcal{T}}(\bm{r}_\mu),
\end{equation}
Note that this does not modify the anti-unitary nature of time
reversal symmetry.

For a symmetric spin liquid, the projective operations
$\widetilde{\mathcal{O}}$ and $\widetilde{\mathcal{T}}$ generate the
symmetry group of the parton Hamiltonian, commonly known as the
projective symmetry group (PSG). The classification of symmetric
spin liquids amounts to the classification of PSGs. To achieve this,
one needs to find all the gauge-inequivalent solutions for the gauge
transformations $G_{\mathcal{O}}$ and $G_{\mathcal{T}}$ that are
consistent with the symmetry group of the system. Any group relation
of Eq.~\eqref{SGtr} can be written in the general form of
\begin{equation}
\mathcal{O}_1 \circ \mathcal{O}_2 \circ \dots = 1, \label{grprlt0}
\end{equation}
which translates into the gauge-enriched group relation
\begin{equation}\label{grprlt1}
\widetilde{\mathcal{O}}_1 \circ \widetilde{\mathcal{O}}_2 \circ
\dots = (G_{\mathcal{O}_1}\circ \mathcal{O}_1)\circ
(G_{\mathcal{O}_2}\circ \mathcal{O}_2)\circ\dots = \mathcal{G},
\end{equation}
where $\mathcal{G}$ is a pure gauge transformation and corresponds
to the identity operation for the spins. We say that $\mathcal{G}$
is an element of the invariant gauge group (IGG), the group of all
pure gauge transformations that leave the parton Hamiltonian
invariant. The IGG transformation on each site is a subgroup of
SU(2), typically $\mathbb{Z}_2$ or U(1). In most cases, there exists
a gauge choice (the canonical gauge \cite{wen2002quantum}) in which
the IGG transformation can be made ``global'' of the form
$\mathcal{G} = e^{i \sigma^3 \chi}$ with a constant $\chi$. In this
paper, we will be classifying both $\mathbb{Z}_2$ and U(1) spin
liquids, therefore we consider both IGG $= \mathbb{Z}_2$ and U(1),
for which $\chi = \{0,\pi\}$ and $\chi \in [0,2\pi)$, respectively.

Making use of the general conjugation rule
\begin{equation}\label{OGO}
\mathcal{O}_i \circ G_{\mathcal{O}_j} \circ \mathcal{O}^{-1}_i
\colon \Psi_{\bm{r}_\mu} \rightarrow
\Psi_{\bm{r}_\mu}W_{\mathcal{O}_j}[\mathcal{O}^{-1}_i(\bm{r}_\mu)],
\end{equation}
which follows directly from Eqs.~\eqref{op-O} and \eqref{b_under_o},
Eq.~\eqref{grprlt1} can be rewritten as
\begin{equation}\label{pgg}
\begin{aligned}
G_{\mathcal{O}_1} &\circ (\mathcal{O}_1 \circ G_{\mathcal{O}_2}
\circ \mathcal{O}^{-1}_1) \\
&\circ (\mathcal{O}_1 \circ \mathcal{O}_2 \circ G_{\mathcal{O}_3}
\circ \mathcal{O}^{-1}_2 \circ \mathcal{O}^{-1}_1) \circ \dots =
\mathcal{G},
\end{aligned}
\end{equation}
which then becomes an SU(2) equation:
\begin{eqnarray}\label{phaseequationO}
W_{\mathcal{O}_1}(\bm{r}_\mu)
W_{\mathcal{O}_2}[\mathcal{O}_1^{-1}(\bm{r}_\mu)]
W_{\mathcal{O}_3}\{\mathcal{O}_2^{-1}[\mathcal{O}_1^{-1}(\bm{r}_\mu)]\}\dots = \mathcal{G}. \nonumber\\
\end{eqnarray}
The PSG classification is obtained by listing all group relations
and finding all solutions to the corresponding SU(2) equation
\eqref{phaseequationO}. We emphasize that solutions must be
discriminated by the principle of gauge equivalence, rather than
resemblance. Indeed, by means of a general gauge transformation $G$
as in Eq.~\eqref{op-G}, the gauge-enriched group relations in
Eq.~\eqref{grprlt1} can be rewritten as
\begin{equation}
(G\circ G_{\mathcal{O}_1} \circ \mathcal{O}_1 \circ G^{-1}) \circ
(G\circ G_{\mathcal{O}_2}\circ \mathcal{O}_2\circ G^{-1})\circ\dots
= \mathcal{G},
\end{equation}
which transforms $W_{\mathcal{O}_i} (\bm{r}_\mu)$ according to
\begin{equation}
W_{\mathcal{O}_i}(\bm{r}_\mu)\rightarrow
W(\bm{r}_\mu)W_{\mathcal{O}_i}(\bm{r}_\mu)W^{-1}[\mathcal{O}_i^{-1}(\bm{r}_\mu)].
\end{equation}
This indicates that two seemingly distinct solutions to the PSG
equations can in fact be equivalent.

\subsection{Classification result}

\begin{figure}[t]
\centering
    \centering
    \subfigure[]{\label{fig:loop1}\includegraphics[height=0.245\textwidth]{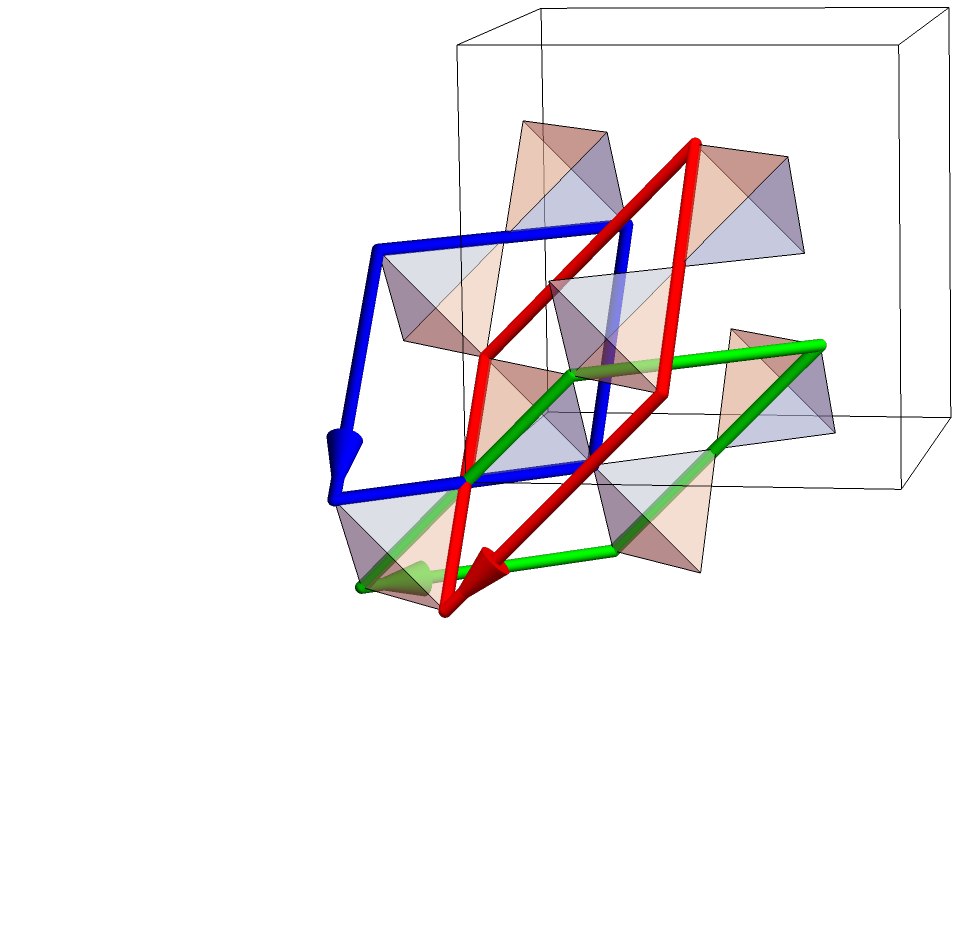}}
    \hspace{-7mm}%
    \subfigure[]{\label{fig:loop2}\includegraphics[height=0.245\textwidth]{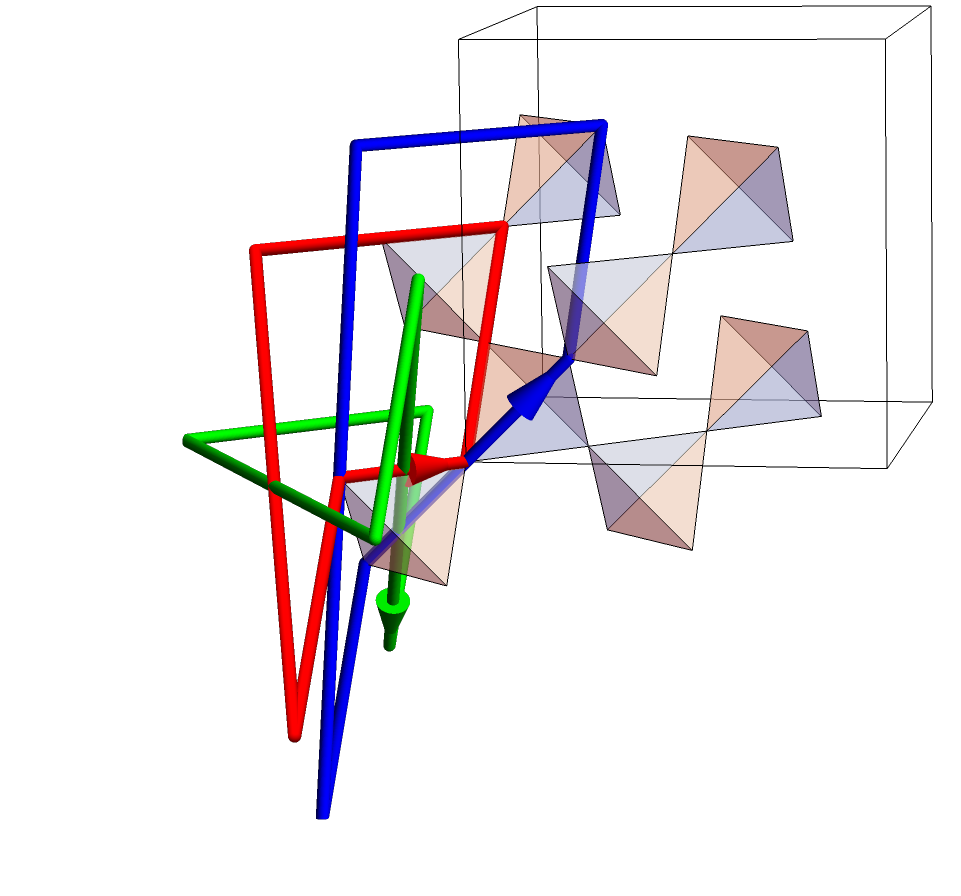}}
\caption{Examples of loops for (a) $T_iT_{i+1}T_i^{-1}T_{i+1}^{-1}=1$ and (b) $ST_1S^{-1}T_3^{-1}T_1 = 1$.}\label{loop}
\end{figure}

We first solve the PSG equations obtained from space group
symmetries. The PSG equations for U(1) and $\mathbb{Z}_2$ gauge
groups are solved in App.~\ref{app:u1_SG} and App.~\ref{app:z2_SG},
respectively. The results are presented in Table
\ref{table:u1_psg_table} for the U(1) gauge group and in Table
\ref{table:z2_psg_table} for the $\mathbb{Z}_2$ gauge group. We find
18 gauge-inequivalent PSG solutions for the U(1) gauge type and 28
gauge-inequivalent PSG solutions for the $\mathbb{Z}_2$ gauge type.
As a result, there can be at most 18 U(1) and 28 $\mathbb{Z}_2$
symmetric spin liquids, ignoring possible time reversal symmetry.
Both the U(1) and the $\mathbb{Z}_2$ solutions have the following
form:
\begin{subequations}\label{www}
\begin{align}
W_{T_i}(\bm{r}_\mu) &= e^{i\sigma^3\phi_{T_i}(\bm{r}_\mu)},\quad i=1,2,3,\\
W_{\overline{C}_6}(\bm{r}_\mu) &= W_{\overline{C}_6,\mu}e^{i \sigma^3\phi_{\overline{C}_6}(\bm{r}_\mu)},\label{eq:wc6}\\
W_S(\bm{r}_\mu) &= W_{S,\mu}e^{i \sigma^3\phi_S
(\bm{r}_\mu)},\label{eq:ws}
\end{align}
\end{subequations}
with
\begin{subequations}
\begin{align}
\phi_{T_1}(\bm{r}_\mu) &= 0,\\
\phi_{T_2}(\bm{r}_\mu) &= -\chi_1 r_1,\\
\phi_{T_3}(\bm{r}_\mu) &= \chi_1(r_1-r_2),\\
\phi_{\overline{C}_6}(\bm{r}_\mu) &= -\chi_1r_1(r_2-r_3)  \notag\\
&\quad-[{2\chi_{ST_1}+2 \chi_1}+(\delta_{\mu,2}-\delta_{\mu,3}) \chi_1]r_1\notag\\
&\quad+\delta_{\mu,2}\chi_1r_3,\\
\phi_S(\bm{r}_\mu) &=
\chi_1\left[\frac{(r_1+1)r_1}{2} - \frac{(r_2+1)r_2}{2}-r_1r_2\right]\notag\\
&\quad+
[(\delta_{\mu,1}-\delta_{\mu,2})\chi_1+(2\chi_1 - \chi_{ST_1})]r_1\notag\\
&\quad+
[(2\delta_{\mu,1}-\delta_{\mu,2})\chi_1 + 3\chi_{ST_1}]r_2\notag\\
&\quad+ [(\delta_{\mu,1}-\delta_{\mu,2}) + 2]\chi_1r_3.
\end{align}
\end{subequations}
The parameters $\chi_1$ and $\chi_{ST_1}$ are elements of the IGG
defined on the right-hand sides of the PSG equations obtained from
Eq.~\eqref{SGtr}. Concretely:
\begin{itemize}
\item The parameter $\chi_1$ is associated with $T_iT_{i+1}T_i^{-1}T_{i+1}^{-1}=1$, and physically quantifies the Aharonov-Bohm (AB) phase a spinon accumulates under such a sequence of translations (see Fig.~\ref{fig:loop1} for examples of the path). The AB phase is a gauge invariant quantity. In U(1) PSG, such a phase is allowed to take on values $0$, $\pi$ and $\pi/2$. They give rise to zero-flux, $\pi$-flux and $\frac{\pi}{2}$-flux spin liquids, respectively. The $\frac{\pi}{2}$-flux spin liquid explicitly breaks time reversal symmetry since a $\frac{\pi}{2}$ flux changes sign under time reversal operation.  In $\mathbb{Z}_2$ PSG, only 0- and $\pi$-flux spin liquids are found.
\item The parameter $\chi_{ST_1}$ is associated with  $ST_1S^{-1}T_3^{-1}T_1 = 1$, and physically quantifies the AB phase a spinon accumulates under the sequence of operations $ST_1S^{-1}T_3^{-1}T_1$  (see Fig.~\ref{fig:loop2} for examples of the path). Such a phase is allowed to take on values  $0$ and $\pi$ in both U(1) and $\mathbb{Z}_2$ PSGs.
\end{itemize}
$W_{\overline{C}_6,\mu}$ and $W_{S,\mu}$ in Eqs.~\eqref{eq:wc6} and
\eqref{eq:ws} are the SU(2) matrices at the origin, $\bm{r}_\mu=0$,
which depend on additional discrete parameters as given in Tables
\ref{table:u1_psg_table} and \ref{table:z2_psg_table} for the two
gauge types. The parameters $\chi_{\overline{C}_6S}$,
$\chi_{\overline{C}_6}$ and $\chi_{S\overline{C}_6}$ are elements of
the IGG associated with $(\overline{C}_6S)^4=1$, $I^2=1$ and
$(IS)^3=1$, respectively, and have the same AB phase interpretation
as explained above. We note that two additional parameters
$w_{\overline{C}_6}$ and $w_S$ appear in the U(1) PSG
classification: they are $\mathbb{Z}_2$-valued ($w=0,1$) and
determine whether or not the SU(2) matrices $W_{\overline{C}_6,\mu}$ and
$W_{S,\mu}$ belong to the IGG. It is necessary to
introduce these two parameters in the U(1) case a priori in order to
simplify the SU(2) PSG equations to U(1) phase equations for
$\phi_{\mathcal{O}}(\bm{r}_\mu)$. This is not required in the
$\mathbb{Z}_2$ case, since the phases
$\phi_{\mathcal{O}}(\bm{r}_\mu)$ are $\mathbb{Z}_2$-valued and
commute with $W_{\overline{C}_6,\mu}$ and $W_{S,\mu}$. The SU(2)
equations for $W_{\overline{C}_6,\mu}$ and $W_{S,\mu}$, however, do
rely on an additional discrete parameter $j$ in some classes (see
Table \ref{table:z2_psg_table}).

\begin{table*}
\begin{ruledtabular}
\caption{The 18 U(1) PSG classes of pyrochlore space group symmetry only and 16 U(1) PSG classes of pyrochlore space group symmetry combined with time reversal symmetry. The tabulation is such that the latter are embedded in the former; and that daggered square brackets ``$\left[\,\,\,\right]^\dag$'' are used to enclose data that are specific to the former. The parameters $\chi_1$, $\chi_{ST_1}$, and $\chi_{\overline{C}_6S}$ are discrete elements of the IGG and $w_{\overline{C}_6}$ and $w_S$ are $\mathbb{Z}_2$-valued parameters introduced in the canonical gauge. Together they label the U(1) PSG classes. The gauge fixing of the SU(2) matrices $W_{\overline{C}_6,\mu}$ and $W_{S,\mu}$ defined in Eq.~\eqref{www} is given.
\label{table:u1_psg_table}}
\begin{tabular}{ccc|cc|cc|cc}
\multirow{3}{*}{$\chi_1$} & \multirow{3}{*}{$\chi_{ST_1}$} & \multirow{3}{*}{$\chi_{\overline{C}_6S}$} &
\multirow{3}{*}{$w_{\overline{C}_6}$} & \multirow{3}{*}{$w_S$} & \multirow{3}{*}{$(W_{\overline{C}_6,0},W_{\overline{C}_6,1},W_{\overline{C}_6,2},W_{\overline{C}_6,3})$} & \multirow{3}{*}{$(W_{S,0},W_{S,1},W_{S,2},W_{S,3})$} & \multicolumn{2}{c}{Number}\\
&&&&&&&\multicolumn{2}{c}{of classes}\\
\cline{8-9}
&&&&&&& $\left[\#\right]^\dag$ & $\#$\\
\hline
$0$ or $\pi$ & $0$ 
& $0$ or $\pi$ &
0 & 0
& $(1,1,1,e^{i\sigma^3(\chi_1-\chi_{ST_1})})$ & $(1,e^{i\sigma^3(\chi_{ST_1}+\chi_{\overline{C}_6S})},1, e^{i \sigma^3\chi_1})$ & $\left[4\right]^\dag$ & 4 \\
$0$ or $\pi$ & $0$                      & $0$ or $\pi$                      & 0 & 1
& $(1,1,1,e^{i\sigma^3(\chi_1-\chi_{ST_1})})$ & $(i\sigma^1,i\sigma^1e^{i\sigma^3(\chi_{ST_1}+\chi_{\overline{C}_6S})},i\sigma^1, i\sigma^1e^{i \sigma^3\chi_1})$ & $\left[4\right]^\dag$ & 4 \\
$0$ or $\pi$ & $0$                      & $0$ or $\pi$                     & 1 & 0
& $(i\sigma^1,i\sigma^1,i\sigma^1,i\sigma^1e^{i\sigma^3(\chi_1-\chi_{ST_1})})$ & $(1,e^{i\sigma^3(\chi_{ST_1}+\chi_{\overline{C}_6S})},1, e^{i \sigma^3\chi_1})$ & $\left[4\right]^\dag$ & 4 \\
$0$ or $\pi$ & $0$                      & $0$ or $\pi$                      & 1 & 1
& $(i\sigma^1,i\sigma^1,i\sigma^1,i\sigma^1e^{i\sigma^3(\chi_1-\chi_{ST_1})})$ & $(i\sigma^1,i\sigma^1e^{i\sigma^3(\chi_{ST_1}+\chi_{\overline{C}_6S})},i\sigma^1, i\sigma^1e^{i \sigma^3\chi_1})$ & $\left[4\right]^\dag$ & 4 \\
{$\left[\frac{\pi}{2}\right]^\dag$} & {$\left[\pi\right]^\dag$}           & {$\left[0\text{ or }\pi\right]^\dag$}                 & {$\left[0\right]^\dag$} & {$\left[1\right]^\dag$}
& {$\left[(1,1,1,e^{i\sigma^3(\chi_1-\chi_{ST_1})})\right]^\dag$} & {$\left[(i\sigma^1,i\sigma^1e^{i\sigma^3(\chi_{ST_1}+\chi_{\overline{C}_6S})},i\sigma^1, i\sigma^1e^{i \sigma^3\chi_1})\right]^\dag$} & $\left[2\right]^\dag$ & 0 \\
\end{tabular}
\end{ruledtabular}
\end{table*}

\renewcommand{\arraystretch}{1.1}
\begin{table*}
\centering
\caption{The 28 $\mathbb{Z}_2$ PSG classes of pyrochlore space group symmetry only and 48 $\mathbb{Z}_2$ PSG classes of pyrochlore space group symmetry combined with time reversal symmetry. The tabulation is such that if a class of the latter is derived from a class of the former then they are written in the same line; and that the double-daggered double-square brackets ``$\left\llbracket\,\,\,\right\rrbracket^\ddag$'' are used to enclose data that are specific to the latter.  The parameters $\chi_1$, $\chi_{ST_1}$, $\chi_{\overline{C}_6S}$, $\chi_{S\overline{C}_6}$, and $\chi_{\overline{C}_6}$ are discrete elements of the IGG and $j$ is an extra parameter introduced during solving the PSG equations. Together they label the $\mathbb{Z}_2$ PSG classes. {Note that the value of the $\mathbb{Z}_2$ parameter $\chi_{ST_1}$ can be either $\chi_1$ or $\pi-\chi_1$, and we denote the latter case by ``$\neg\chi_1$''.} The gauge fixing of the SU(2) matrices $W_{\overline{C}_6,\mu}$ and $W_{S,\mu}$ defined in Eq.~\eqref{www}, and the time reversal SU(2) matrix defined in Eq.~\eqref{t_gf_z2} for $\mu=0$ are given ($k=1,2,3$ corresponds to different gauge fixing). The time reversal parameters $\chi_{\mathcal{T}\overline{C}_6}$ and $\chi_{\mathcal{T}S}$ are related to $\eta_\mu$ in Eq.~\eqref{t_gf_z2} by $(\eta_0,\eta_1,\eta_2,\eta_3) = (1,\eta_{\mathcal{T}\overline{C}_6}\eta_{\mathcal{T}S}, \eta_{\mathcal{T}S}, \eta_{\mathcal{T}S})$. We defined the shorthand notation $W^{\pi\pi0}_{S,2} = e^{i \chi_{\overline{C}_6S}}i\sigma^{k+1}e^{i\frac{\pi}{6}\sigma^{k-1}}$.}
\label{table:z2_psg_table}
\begin{ruledtabular}
\begin{tabular}{cc|cccc|c|cc|cc}
\multirow{3}{*}{$\chi_1$}&\multirow{3}{*}{$\chi_{ST_1}$}&\multirow{3}{*}{$\chi_{\overline{C}_6S}$}&\multirow{3}{*}{$\chi_{S\overline{C}_6}$}&\multirow{3}{*}{$\chi_{\overline{C}_6}$} &\multirow{3}{*}{ $j$} &\multirow{3}{*}{ $\left\llbracket \begin{array}{{>{$}p{0.6cm}<{$}>{\raggedleft$}p{0.5cm}<{$}>{\raggedleft$}p{0.6cm}<{$}}}\chi_{\mathcal{T}\overline{C}_6}& \chi_{\mathcal{T}S}& W_{\mathcal{T},0}\end{array}\right\rrbracket^\ddag$} & \multirow{3}{*}{$(W_{\overline{C}_6,0},W_{\overline{C}_6,1},W_{\overline{C}_6,2},W_{\overline{C}_6,3})$} &  \multirow{3}{*}{$(W_{S,0},W_{S,1},W_{S,2},W_{S,3})$} &\multicolumn{2}{c}{Number}\\
&&&&&&&&& \multicolumn{2}{c}{of classes} \\
\cline{10-11}
&&&&&&&&& $\#$ & $\left\llbracket\#\right\rrbracket^\ddag$\\
\hline
$0$ or $\pi$&$\chi_1$&$0$ or $\pi$ & $0$&$0$ & \NA      & $\left\llbracket\begin{array}{{l>{\centering$}p{0.6cm}<{$}l}}{0\,}& {0\,}&{i\sigma^{k\hphantom{+1}}}\end{array}\right\rrbracket^\ddag$ & $(1,1,1,1)$ & $(1,1,e^{i\chi_{\overline{C}_6S}},1)$ & 4& $\left\llbracket 4\right\rrbracket^\ddag$\\
\multirow{2}{*}{$0$ or $\pi$}&\multirow{2}{*}{$\chi_1$}&\multirow{2}{*}{$0$ or $\pi$} & \multirow{2}{*}{$\pi$}&\multirow{2}{*}{$\pi$} & \multirow{2}{*}{\NA}   & \multirow{2}{*}{$\left\llbracket \begin{array}{{l>{\centering$}p{0.6cm}<{$}l}}{0\,}&{0\,}&{i\sigma^k}\\{\pi}&{0\,}&{i \sigma^{k-1}}\end{array}\right\rrbracket^\ddag$}  & \multirow{2}{*}{$(-i\sigma^k,1,1,i\sigma^k)$} & \multirow{2}{*}{$(1,1,e^{i\chi_{\overline{C}_6S}},1)$} & \multirow{2}{*}{4} & \multirow{2}{*}{$\left\llbracket 8\right\rrbracket^\ddag$} \\
&&&&& &   & & & & \\
$0$ or $\pi$&$\neg\chi_1$&$0$&$\pi$&$0$ & $1$& $\left\llbracket\begin{array}{{l>{\centering$}p{0.6cm}<{$}l}}{0\,}& {\pi}& {i\sigma^{k-1}}\end{array} \right\rrbracket^\ddag$ & $(-e^{i\chi_{\overline{C}_6S}}e^{i\frac{\pi j}{3}\sigma^{k-1}},1,1,1)$ & $(1,i\sigma^k,i\sigma^k e^{i\frac{\pi j}{3} \sigma^{k-1}},1)$ & 2 & $\left\llbracket 2\right\rrbracket^\ddag$ \\
\multirow{2}{*}{$0$ or $\pi$}&\multirow{2}{*}{$\neg\chi_1$}&\multirow{2}{*}{$0$}&\multirow{2}{*}{$\pi$}&\multirow{2}{*}{$0$} & \multirow{2}{*}{$3$} & \multirow{2}{*}{$\left\llbracket\begin{array}{{l>{\centering$}p{0.6cm}<{$}l}}{0}&{0}&{i\sigma^k}\\{0\,}&{\pi}&{i \sigma^{k-1}}\end{array}\right\rrbracket^\ddag$} & \multirow{2}{*}{$(-e^{i \chi_{\overline{C}_6S}}e^{i\frac{\pi j}{3}\sigma^{k-1}},1,1,1)$} & \multirow{2}{*}{$(1,i\sigma^k,i\sigma^k e^{i\frac{\pi j}{3} \sigma^{k-1}},1)$} & \multirow{2}{*}{2}&  \multirow{2}{*}{$\left\llbracket 4\right\rrbracket^\ddag$} \\
&&&&&& & & & \\
\multirow{2}{*}{$0$ or $\pi$} &\multirow{2}{*}{$\neg\chi_1$}&\multirow{2}{*}{$\pi$}&\multirow{2}{*}{$\pi$}&\multirow{2}{*}{$0$} & \multirow{2}{*}{$0$} & \multirow{2}{*}{$\left\llbracket\begin{array}{{l>{\centering$}p{0.6cm}<{$}l}}{0}&{0}&{i\sigma^k}\\{0\,}&{\pi}&{i \sigma^{k-1}}\end{array}\right\rrbracket^\ddag$} & \multirow{2}{*}{$(-e^{i \chi_{\overline{C}_6S}}e^{i\frac{\pi j}{3}\sigma^{k-1}},1,1,1)$} & \multirow{2}{*}{$(1,i\sigma^k,i\sigma^k e^{i\frac{\pi j}{3} \sigma^{k-1}},1)$} & \multirow{2}{*}{2}&  \multirow{2}{*}{$\left\llbracket 4\right\rrbracket^\ddag$} \\
&&&&&& & &  & & \\
$0$ or $\pi$&$\neg\chi_1$&$\pi$&$\pi$&$0$ & $2$ & $\left\llbracket\begin{array}{{l>{\centering$}p{0.6cm}<{$}l}}{0\,}&{\pi}& {i\sigma^{k-1}}\end{array}\right\rrbracket^\ddag$ & $(-e^{i\chi_{\overline{C}_6S}}e^{i\frac{\pi j}{3}\sigma^{k-1}},1,1,1)$ & $(1,i\sigma^k,i\sigma^k e^{i\frac{\pi j}{3} \sigma^{k-1}},1)$ & 2& $\left\llbracket 2\right\rrbracket^\ddag$ \\
\multirow{2}{*}{$0$ or $\pi$}&\multirow{2}{*}{$\neg\chi_1$}&\multirow{2}{*}{$0$ or $\pi$}&\multirow{2}{*}{$0$}&\multirow{2}{*}{$\pi$} & \multirow{2}{*}{\NA} & \multirow{2}{*}{$\left\llbracket\begin{array}{{l>{\centering$}p{0.6cm}<{$}l}}{0}&{0}&{i\sigma^k}\\{\pi}&{\pi}&{i \sigma^{k-1}}\end{array}\right\rrbracket^\ddag$} & \multirow{2}{*}{$(i\sigma^k,1,1,i\sigma^k)$} & \multirow{2}{*}{$(1,i\sigma^k,e^{i \chi_{\overline{C}_6S}} i \sigma^k,1)$} &  \multirow{2}{*}{4} &  \multirow{2}{*}{$\left\llbracket 8\right\rrbracket^\ddag$} \\
 &&&&& & &  & & \\
$0$ or $\pi$&$\neg\chi_1$&$0$ or $\pi$&$\pi$&$\pi$ & $0$ & $\left\llbracket\begin{array}{{l>{\centering$}p{0.6cm}<{$}l}}{0\,}&{\pi}&{i\sigma^{k-1}}\end{array}\right\rrbracket^\ddag$ & $(e^{i\frac{\pi}{6}\sigma^{k-1}},1,1,i\sigma^{k-1})$ & $(1,i\sigma^k,W^{\pi\pi0}_{S,2},1)$ & 4 &  $\left\llbracket4\right\rrbracket^\ddag$\\
\multirow{3}{*}{$0$ or $\pi$}&\multirow{3}{*}{$\neg\chi_1$}&\multirow{3}{*}{$0$ or $\pi$}&\multirow{3}{*}{$\pi$}&\multirow{3}{*}{$\pi$} & \multirow{3}{*}{$1$}  &
\multirow{3}{*}{$\left\llbracket\begin{array}{{l>{\centering$}p{0.6cm}<{$}l}}{0}&{\pi}&{i\sigma^{k-1}}\\{\pi}&{0}&{i \sigma^{k}}\\{\pi}&{\pi}&{i \sigma^{k+1}}\end{array}\right\rrbracket^\ddag$}
 & \multirow{3}{*}{$(-i\sigma^{k-1},1,1,i\sigma^{k-1})$} & \multirow{3}{*}{$(1,i\sigma^k,e^{i \chi_{\overline{C}_6S}} i\sigma^k ,1)$} & \multirow{3}{*}{4}& \multirow{3}{*}{$\left\llbracket 12\right\rrbracket^\ddag$}\\
 &&&&&  & &  &  &  \\
 &&&&&  & &  &  &  \\
\end{tabular}
\end{ruledtabular}
\end{table*}

In the second step, we include the time reversal operation
$\mathcal{T}$ in the symmetry group. In an appropriate gauge, the
solution for time reversal operation can be chosen as
\begin{equation}\label{t_gf_u1}
{W}_{\mathcal{T}}(\bm{r}_\mu) = i \sigma^1
\end{equation}
for U(1) gauge type and
\begin{equation}\label{t_gf_z2}
{W}_{\mathcal{T}}(\bm{r}_\mu) = i \eta_\mu \sigma^k
\end{equation}
for $\mathbb{Z}_2$ gauge type, where $k=1,2,3$ depends on the PSG
class and $\eta_\mu = \pm $ is a sublattice-dependent sign factor.
Applying this gauge choice and the space group PSG solutions in
Eq.~\eqref{www} to the time reversal PSG equations associated with
Eqs.~\eqref{4i} and \eqref{4j},
we obtain the PSG solutions for time reversal invariant symmetric
spin liquids. We find that there are 16 classes for the U(1) gauge
type and 48 classes for the $\mathbb{Z}_2$ gauge type.  We list
these classes again in Tables \ref{table:u1_psg_table} and
\ref{table:z2_psg_table} for U(1) and $\mathbb{Z}_2$ gauge groups,
respectively {and explicitly mark the data that are specific to the time reversal symmetric classes}.

It is worth pointing out that including time reversal symmetry to
the PSG has opposite effects on the U(1) and the $\mathbb{Z}_2$
classes. In U(1) PSG, the presence of time reversal symmetry forbids
the two $\frac{\pi}{2}$-flux PSG classes {(last line of
Table \ref{table:u1_psg_table})}, thereby reducing the total number
of PSG classes from 18 to 16. On the
other hand, in $\mathbb{Z}_2$ PSG, adding time reversal symmetry
introduces two additional discrete parameters
$\chi_{\mathcal{T}\overline{C}_6}$ and $\chi_{\mathcal{T}S}$ in the
labeling of the time reversal invariant classes {(see columns 7 and 8 of  Table \ref{table:z2_psg_table})}.
These two parameters are the IGG elements associated with
Eq.~\eqref{4j} for $\mathcal{O} = \overline{C}_6$ and $S$,
respectively, and characterize the phases that a spinon acquires in
completing the corresponding spacetime processes. We find that
{20 classes} obtained from the pure space group PSG
are further ``fractionalized'' as a result of these additional
parameters, thereby increasing the total number of PSG classes from
28 to 48.
In U(1) PSG, $\chi_{\overline{C}_6\mathcal{T}}$ and
$\chi_{S\mathcal{T}}$ do not increase the number of classes since
they are fully determined by the $\mathbb{Z}_2$ parameters
$w_{\overline{C}_6}$ and $w_S$ that are already introduced for the
pure space group PSG. The phenomenon described here in fact also
happens in the classification of other projective lattice
symmetries: it is generally true that adding time reversal symmetry
will increase the number of $\mathbb{Z}_2$ PSG classes and reduce
the number of U(1) PSG classes.

\section{Analysis of the mean-field ans\"atze\label{secMF}}

The PSG classification in the last section provides the symmetry
constraints on constructing Hamiltonians that describe
fractionalized spinons in symmetric U(1) and $\mathbb{Z}_2$ spin
liquids. Since the gauge fields are deconfined in a spin liquid
phase, a good description for the spinons is already achieved at the
mean-field level. In this section, we present a complete list of
parton mean-field Hamiltonians for symmetric spin liquids on the
pyrochlore lattice. 
These Hamiltonians can either be
analyzed in their own right, or serve as a first step towards a more
realistic description of spin liquids upon adding spinon
interactions or fluctuating gauge fields. 

\subsection{Construction of the mean-field ans\"atze}

We are now in the position to construct the mean-field ansatz for
each PSG class. The most general mean-field ansatz for fermionic
spinons can be written as
\begin{equation}
\begin{aligned}
H = \sum\limits_{\alpha={0,x,y,z}} H^\alpha,\qquad H^\alpha = \sum\limits_{\bm{r}_\mu,\bm{r}'_\nu} H^\alpha_{\bm{r}_\mu,\bm{r}'_\nu},\\
H^\alpha_{\bm{r}_\mu,\bm{r}'_\nu} = \mathrm{Tr}\left[\sigma^\alpha
\Psi_{\bm{r}_\mu} u^{(\alpha)}_{\bm{r}_\mu,\bm{r}'_\nu}
\Psi^\dag_{\bm{r}'_\nu}\right],
\end{aligned}
\end{equation}
where $u^{(\alpha)}_{\bm{r}_\mu,\bm{r}'_\nu}$ with $\alpha =
0,x,y,z$ contain all the sixteen real parameters for the bond
$\bm{r}_\mu\rightarrow \bm{r}'_\nu$,
\begin{equation}\label{bondsss}
\begin{aligned}
u^{(0)}_{\bm{r}_\mu,\bm{r}'_\nu}
&= ia^0_{\bm{r}_\mu,\bm{r}'_\nu}1  - (b^0_{\bm{r}_\mu,\bm{r}'_\nu}\sigma^1+ c^0_{\bm{r}_\mu,\bm{r}'_\nu} \sigma^2+d^0_{\bm{r}_\mu,\bm{r}'_\nu} \sigma^3),\\
u^{(x)}_{\bm{r}_\mu,\bm{r}'_\nu}
&= a^x_{\bm{r}_\mu,\bm{r}'_\nu}1 + i (b^x_{\bm{r}_\mu,\bm{r}'_\nu}\sigma^1+ c^x_{\bm{r}_\mu,\bm{r}'_\nu} \sigma^2+d^x_{\bm{r}_\mu,\bm{r}'_\nu} \sigma^3),\\
u^{(y)}_{\bm{r}_\mu,\bm{r}'_\nu}
&= a^y_{\bm{r}_\mu,\bm{r}'_\nu}1 + i (b^y_{\bm{r}_\mu,\bm{r}'_\nu}\sigma^1+ c^y_{\bm{r}_\mu,\bm{r}'_\nu} \sigma^2+d^y_{\bm{r}_\mu,\bm{r}'_\nu} \sigma^3),\\
u^{(z)}_{\bm{r}_\mu,\bm{r}'_\nu} &= a^z_{\bm{r}_\mu,\bm{r}'_\nu}1 +
i (b^z_{\bm{r}_\mu,\bm{r}'_\nu}\sigma^1+
c^z_{\bm{r}_\mu,\bm{r}'_\nu} \sigma^2+d^z_{\bm{r}_\mu,\bm{r}'_\nu}
\sigma^3).
\end{aligned}
\end{equation}
Note that $1$ denotes the $2 \times 2$ identity matrix.

The bond parameters are subject to constraints provided by the PSG.
The PSG operators $\widetilde{\mathcal{O}}$ and
$\widetilde{\mathcal{T}}$ are the symmetry operators of the
Hamiltonian $H$, meaning $\widetilde{\mathcal{O}} \colon
H\rightarrow H$ and $\widetilde{\mathcal{T}} \colon H\rightarrow H$.
Since the spinons transform under $\widetilde{\mathcal{O}}$ and
$\widetilde{\mathcal{T}}$ according to Eqs.~\eqref{b_under_o} and
\eqref{timereversaldef3}, we have the following rules:
\begin{itemize}
\item For a general translation $\bm{t}=t_1 \bm{e}_1+t_2\bm{e}_2+t_3\bm{e}_3$, we have
\begin{equation}\label{uuu0}
u^{(\alpha)}_{\bm{r}_\mu,\bm{r}'_\nu}
 = u^{(\alpha)}_{(\bm{r}-\bm{t})_\mu,(\bm{r}'-\bm{t})_\nu}e^{i \sigma^3
 [t_2(r'_1-r_1)-t_3(r'_1-r_1-r'_2+r_2)]\chi_1};
\end{equation}
\item For space-group elements $\mathcal{O} \in
\{\overline{C}_6,S\}$, the singlet and the triplet parts transform
as
\begin{equation}\label{uuu}
\begin{aligned}
W_{\mathcal{O}}[\mathcal{O}(\bm{r}_\mu)]u^{(0)}_{\bm{r}_\mu,\bm{r}'_\nu}W^\dag_{\mathcal{O}}[\mathcal{O}(\bm{r}'_\nu)] &= u^{(0)}_{\mathcal{O}(\bm{r}_\mu),\mathcal{O}(\bm{r}'_\nu)},\\
W_{\mathcal{O}}[\mathcal{O}(\bm{r}_\mu)]u^{(i)}_{\bm{r}_\mu,\bm{r}'_\nu}\mathcal{R}^{\mathcal{O}}_{ij}
W^\dag_{\mathcal{O}}[\mathcal{O}(\bm{r}'_\nu)] &=
u^{(j)}_{\mathcal{O}(\bm{r}_\mu),\mathcal{O}(\bm{r}'_\nu)},
\end{aligned}
\end{equation}
 with
\begin{equation}
\mathcal{R}^{\overline{C}_6}=\left(\begin{array}{ccc}&&1\\1&&\\&1&\end{array}\right),\quad
\mathcal{R}^S =
\left(\begin{array}{ccc}&1&\\1&&\\&&-1\end{array}\right);
\end{equation}
\item For time reversal $\mathcal{T}$,
\begin{equation}\label{guguchow}
u^\alpha_{\bm{r}_\mu,\bm{r}'_\nu} = -
{W}_{\mathcal{T}}(\bm{r}_\mu)u^\alpha_{\bm{r}_\mu,\bm{r}'_\nu}
{W}_{\mathcal{T}}^\dag(\bm{r}'_\nu).
\end{equation}
\end{itemize}
By solving Eqs.~\eqref{uuu0}-\eqref{guguchow} for the sixteen real
parameters at each bond, we obtain the mean-field ans\"atze for the
PSG classes.

The pyrochlore bonds can be categorized into
{equivalence classes} (or orbits) of the space
group, where the bonds within each class are related by space group
transformations, while the bonds in different classes are unrelated.
In order to obtain the complete mean-field ansatz, it suffices to
obtain the mean-field solution for one representative bond of each
equivalence class. We choose and express these representative bonds
as follows:

\begin{itemize}
\item We use the Greek letters $\alpha$,$\beta$,$\gamma$,$\delta$ to parameterize the representative onsite bond $(0,0,0)_0\rightarrow (0,0,0)_0$, the Latin small letters $a$,$b$,$c$,$d$ to parameterize the representative nearest-neighbor bond $(0,0,0)_0\rightarrow (0,0,0)_1$, and the Latin capital letters $A$,$B$,$C$,$D$ to parameterize the representative next-nearest-neighbor (NNN) bond $(0,0,0)_1\rightarrow (0,-1,0)_2$. The choice of representative $i$th-nearest neighbor bonds and their parameterization for $3\leq i\leq 8$ can be found in App.~\ref{app:G}.
\item Take the representative NN bond for example. In an ansatz for the $\mathbb{Z}_2$ PSG without time reversal symmetry, all the 16 terms in Eq.~\eqref{bondsss} may be nonvanishing. We then use eight complex numbers $a_h,b_h,c_h,d_h,a_p,b_p,c_p,d_p$ to paramaterize the 16 real parameters in this bond: explicitly, we write (omitting the bond label $(0,0,0)_0\rightarrow (0,0,0)_1$)
\begin{equation}\label{u_gf_z2}
\begin{aligned}
u^{(0)} &= i\mathrm{Re}a_h1 -\mathrm{Re}a_p \sigma^1 - \mathrm{Im}a_p \sigma^2 - \mathrm{Im}a_h \sigma^3,\\
u^{(x)} &= \mathrm{Re}b_h1 + i(\mathrm{Re}b_p \sigma^1 + \mathrm{Im}b_p \sigma^2 + \mathrm{Im}b_h \sigma^3),\\
u^{(y)} &= \mathrm{Re}c_h1 + i(\mathrm{Re}c_p \sigma^1 + \mathrm{Im}c_p \sigma^2 + \mathrm{Im}c_h \sigma^3),\\
u^{(z)} &= \mathrm{Re}d_h1 + i(\mathrm{Re}d_p \sigma^1 +
\mathrm{Im}d_p \sigma^2 + \mathrm{Im}d_h \sigma^3).
\end{aligned}
\end{equation}
In the U(1) PSG, we only have hopping bilinears, therefore the
$\sigma^1$ and $\sigma^2$ terms in Eq.~\eqref{bondsss} vanish, and
we parameterize Eq.~\eqref{bondsss} as
\begin{equation}\label{u_gf_u1}
\begin{aligned}
u^{(0)} = i\mathrm{Re}a 1 -  \mathrm{Im}a \sigma^3,\quad
u^{(x)} = \mathrm{Re}b 1 + i \mathrm{Im}b \sigma^3,\\
u^{(y)} = \mathrm{Re}c 1 + i \mathrm{Im}c \sigma^3,\quad u^{(z)} =
\mathrm{Re}d 1 + i \mathrm{Im}d \sigma^3.
\end{aligned}
\end{equation}
\end{itemize}

The symmetry relations \eqref{uuu0}, \eqref{uuu}, and
\eqref{guguchow} impose constraints on these parameters, and the
numbers of independent real bond parameters are usually smaller than
16 or 8 in the $\mathbb{Z}_2$ and U(1)  ans\"atze, respectively. To
determine the independent bond parameters, one needs to find all the
symmetry operations (the so called ``stabilizers'' of the symmetry
group) that leave the representative bonds invariant. All such space
group symmetry operations for $i$th-nearest neighbor bonds ($i\leq
8$) have been listed in App.~\ref{app:G}. As an example, time
reversal symmetry leaves any bond invariant. In the U(1) case,
applying the time reversal PSG in Eq.~\eqref{t_gf_u1} to
Eq.~\eqref{guguchow} reduces the bond parameterization in
Eq.~\eqref{u_gf_u1} to
\begin{equation}\label{u_gf_u1tr}
\begin{aligned}
u^{(0)} = -\mathrm{Im}a\sigma^3,\ \ \
u^{(x)} = i\mathrm{Im}b\sigma^3,\\
u^{(y)} = i\mathrm{Im}c\sigma^3,\ \ \ u^{(z)} =
i\mathrm{Im}d\sigma^3.
\end{aligned}
\end{equation}
In the $\mathbb{Z}_2$ case, applying the time reversal PSG in
Eq.~\eqref{t_gf_z2} to Eq.~\eqref{guguchow} reduces the bond
parameterization in Eq.~\eqref{u_gf_z2} depending on the value of
$\eta_\mu$ for $\mu=0,1$:
\begin{equation}\label{u_gf_z2_1}
\begin{aligned}
u^{(0)} = -\mathrm{Re}a_p \sigma^1 -  \mathrm{Im}a_p \sigma^2,\quad
u^{(x)} = i(\mathrm{Re}b_p \sigma^1 + \mathrm{Im}b_p \sigma^2),\\
u^{(y)} = i(\mathrm{Re}c_p \sigma^1 + \mathrm{Im}c_p \sigma^2),\quad
u^{(z)} = i(\mathrm{Re}d_p \sigma^1 + \mathrm{Im}d_p \sigma^2)
\end{aligned}
\end{equation}
for $\eta_0=\eta_1$ and
\begin{equation}\label{u_gf_z2_2}
\begin{aligned}
u^{(0)} = i\mathrm{Re}a_h1  -  \mathrm{Im}a_h \sigma^3,\quad
u^{(x)} = \mathrm{Re}b_h1  + i\mathrm{Im}b_h \sigma^3,\\
u^{(y)} = \mathrm{Re}c_h1  + i\mathrm{Im}c_h \sigma^3,\quad u^{(z)}
= \mathrm{Re}d_h1  + i\mathrm{Im}d_h \sigma^3
\end{aligned}
\end{equation}
for $\eta_0=-\eta_1$. Note that the form of Eq.~\eqref{u_gf_z2_2}
coincides with that of a U(1) ansatz [see Eq.~\eqref{u_gf_u1}].
However, pairing terms (in which parameters have a subscript
``$p$'') do appear for other bonds (e.g., the representative NNN
bond), and this is generally not a U(1) ansatz.

The final result of the mean-field parameters for representative
bonds up to NNN are presented in Table \ref{MFTparameters_u1} for
the U(1) PSG and Table \ref{MFTparameters_z2} for the $\mathbb{Z}_2$
PSG. The effect of time reversal symmetry has also been addressed therein.

\renewcommand{\arraystretch}{1.0}
\begin{table*}
\centering
\caption{Independent mean-field parameters and constraints for the 18 U(1) PSG classes of pyrochlore space group symmetry only and 16 U(1) PSG classes of pyrochlore space group symmetry combined with time reversal symmetry. The tabulation is such that the latter are embedded in the former; and that daggered square brackets ``$\left[\,\,\,\right]^\dag$'' are used to enclose data that are specific to the former. The parameters not referenced are enforced to be zero. Note the zero- and $\pi$-flux PSG have identical free mean-field parameters up to NNN bonds, but this is no longer true when considering 3rd-nearest-neighbor bonds.}\label{MFTparameters_u1}
\begin{ruledtabular}
\begin{tabular}{l|lll|ll}
\qquad\qquad \;\;Class &\multicolumn{3}{c|}{Independent nonzero parameters}&\multicolumn{2}{c}{Constraints}\\
\cline{2-6}
$\chi_1$\---$(w_{\overline{C}_6}w_{S})$\---$(\chi_{ST_1}\chi_{\overline{C}_6S})$&Onsite&NN&NNN&NN&NNN\\
\hline
0\--- or $\pi$\---$(0\;0)$\---$(0\;0)$             &$\mathrm{Im}\alpha$    &$\mathrm{Im}a$, $\mathrm{Im}c$ &
$\mathrm{Im}A,\,{\left[\mathrm{Re}B\right]^\dag},\,\mathrm{Im}B,\,\mathrm{Im}D$ & $\mathrm{Im}c=-\mathrm{Im}d$ & $B=-C^*$
\\
0\---  or $\pi$\---$(0\;0)$\---$(0\;\pi)$             &$\mathrm{Im}\alpha$    &$\mathrm{Im}b${, $[\mathrm{Re}c]^\dag$} &
$\mathrm{Im}A,\,{\left[\mathrm{Re}B\right]^\dag},\,\mathrm{Im}B,\,\mathrm{Im}D$ & {$\left[\mathrm{Re}c=\mathrm{Re}d\right]^\dag$}  & $B=-C^*$    \\
%
0\---  or $\pi$\---$(0\;1)$\---$(0\;0)$&\NA &$\mathrm{Im}c$ &{${\left[\mathrm{Re}B\right]^\dag}$, }$\mathrm{Im}B$ & $\mathrm{Im}c=\mathrm{Im}d$  & $B=-C$ \\
0\---  or $\pi$\---$(0\;1)$\---$(0\;\pi)$&\NA &{${\left[\mathrm{Re}c\right]^\dag}$} &{${\left[\mathrm{Re}B\right]^\dag}$}, $\mathrm{Im}B$ & {${\left[\mathrm{Re}c=\mathrm{Re}d\right]^\dag}$}    & $B=-C$ \\
%
0\---  or $\pi$\---$(1\;0)$\---$(0\;0)$&\NA &$\mathrm{Im}c$ &$\mathrm{Im}A,\, \left[\mathrm{Re}B\right]^\dag,\,\mathrm{Im}B,\,\mathrm{Im}D$ & $\mathrm{Im}c=\mathrm{Im}d$  & $B=-C^*$ \\
0\---  or $\pi$\---$(1\;0)$\---$(0\;\pi)$&\NA &{$\left[\mathrm{Re}c\right]^\dag$} &$\mathrm{Im}A,\,{\left[\mathrm{Re}B\right]^\dag},\,\mathrm{Im}B,\,\mathrm{Im}D$ & ${\left[\mathrm{Re}c=\mathrm{Re}d\right]^\dag}$    & $B=-C^*$ \\
%
0\---  or $\pi$\---$(1\;1)$\---$(0\;0)$&\NA &$\mathrm{Im}a,\mathrm{Im}c$ &$\left[\mathrm{Re}B\right]^\dag,\,\mathrm{Im}B$ & $\mathrm{Im}c=-\mathrm{Im}d$  & $B=-C$ \\
0\---  or $\pi$\---$(1\;1)$\---$(0\;\pi)$&\NA &$\mathrm{Im}b,\,{\left[\mathrm{Re}c\right]^\dag}$ &${\left[\mathrm{Re}B\right]^\dag},\,\mathrm{Im}B$ & ${\left[\mathrm{Re}c=\mathrm{Re}d\right]^\dag}$   & $B=-C$ \\
{$\left[\frac{\pi}{2}\text{\---}(0\;1)\text{\---}(\pi\;0)\right]^\dag$}&{$\left[-\right]^\dag$} &{$\left[\mathrm{Im}c\right]^\dag$} &{$\left[\mathrm{Re}B,\,\mathrm{Im}B\right]^\dag$} & {$\left[\mathrm{Im}c=\mathrm{Im}d\right]^\dag$}  & {$\left[B=-C\right]^\dag$} \\
{$\left[\frac{\pi}{2}\text{\---}(0\;1)\text{\---}(\pi\;\pi)\right]^\dag$}&{$\left[-\right]^\dag$} &{$\left[\mathrm{Re}c\right]^\dag$} &{$\left[\mathrm{Re}B,\,\mathrm{Im}B\right]^\dag$} & {$\left[\mathrm{Re}c=\mathrm{Re}d\right]^\dag$}    & {$\left[B=-C\right]^\dag$} \\
\end{tabular}
\end{ruledtabular}
\end{table*}

\begin{table*}
\caption{Independent mean-field parameters and constraints for the 28 $\mathbb{Z}_2$ PSG classes of pyrochlore space group symmetry only and 48 $\mathbb{Z}_2$ PSG classes of pyrochlore space group symmetry combined with time reversal symmetry. The tabulation uses the double-daggered double-square brackets ``$\left\llbracket\,\,\,\right\rrbracket^\ddag$'' to enclose data that are specific to the latter, and whenever a class of the latter is derived from a class of the former then a symbol ``$\drsh$'' is used to specify their relation. Also we set $k=1$ (see Table \ref{table:z2_psg_table} for the definition of $k$). The parameters not referenced are enforced to be zero. We denoted $F_2 = \{\mathrm{Im}\alpha_h,\mathrm{Im}\alpha_p\}$, $F_4 = \{\mathrm{Im}a_h,\mathrm{Im}c_h,\mathrm{Im}a_p,\mathrm{Im}c_p\}$, $F_5 = \{{\mathrm{Re}B_h,\mathrm{Im}A_p,\mathrm{Im}B_p,\mathrm{Im}D_p}\}$, $F_6 = \{\mathrm{Im}A_h,\mathrm{Im}B_h,\mathrm{Im}D_h,\mathrm{Im}A_p,\,\mathrm{Im}B_p,\,\mathrm{Im}D_p\}$.}\label{MFTparameters_z2}
\begin{ruledtabular}
\begin{tabular}{l|lll|l}
Class ($\chi_1\chi_{ST_1}$)\---
&\multicolumn{3}{c|}{\multirow{ 1}{*}{Independent nonzero parameters}}&\multirow{ 2}{*}{Constraints (NN and NNN)}\\
\cline{2-4}
$(\chi_{\overline{C}_6S}\chi_{S\overline{C}_6}\chi_{\overline{C}_6})_j$
&Onsite&NN&NNN&\\
\hline
\multirow{2}{*}{\shortstack[l]{$(00)$\---  or $(\pi\pi)$\---$(000)$}}       & \multirow{2}{*}{$\mathrm{Im}\alpha_h,\alpha_p$} & \multirow{2}{*}{$\mathrm{Im}a_h,\mathrm{Im}c_h,a_p,c_p$} &\multirow{2}{*}{\shortstack[l]{$\mathrm{Im}A_h,B_h,\mathrm{Im}D_h$,\\$A_p,B_p,D_p$}} & \multirow{2}{*}{$\mathrm{Im}d_h = -\mathrm{Im}c_h,c_p = - d_p,C_h = -B^*_h,B_p=C_p$}\\
&&&& \\
 {$\drsh\left\llbracket(\chi_{\mathcal{T}\overline{C}_6}\chi_{\mathcal{T}S})=(00)\right\rrbracket^\ddag $}
& {${\left\llbracket F_2\right\rrbracket^\ddag}$}& {${\left\llbracket 
F_4\right\rrbracket^\ddag}$}& {${\left\llbracket 
F_6\right\rrbracket^\ddag}$}&$\left\llbracket \text{Constraints inherited from above}\right\rrbracket^\ddag$\\
\hline
\multirow{2}{*}{\shortstack[l]{$(00)$\--- or $(\pi\pi)$\---$(\pi00)$}}& \multirow{2}{*}{$\mathrm{Im}\alpha_h,\alpha_p$} &\multirow{2}{*}{$\mathrm{Im}b_h,\mathrm{Re}c_h,b_p$} & \multirow{2}{*}{\shortstack[l]{$\mathrm{Im}A_h,B_h,\mathrm{Im}D_h$,\\ $A_p,B_p,D_p$}} & \multirow{2}{*}{$\mathrm{Re}d_h = \mathrm{Re}c_h,C_h = -B^*_h,B_p=C_p$}\\
 &&&&\\
  {$\drsh\left\llbracket(\chi_{\mathcal{T}\overline{C}_6}\chi_{\mathcal{T}S})=(00)\right\rrbracket^\ddag $}
& {${\left\llbracket F_2\right\rrbracket^\ddag}$}& {${\left\llbracket \mathrm{Im}b_h,\mathrm{Im}b_p\right\rrbracket^\ddag}$}& {${\left\llbracket 
F_6\right\rrbracket^\ddag}$}&$\left\llbracket \text{Constraints inherited from above}\right\rrbracket^\ddag$\\
\hline
\multirow{2}{*}{\shortstack[l]{$(00)$\---  or $(\pi\pi)$\---$(0\pi\pi)$}}  & \multirow{ 2}{*}{$\mathrm{Re}\alpha_p$ }& \multirow{ 2}{*}{$\mathrm{Re}a_h,c_h,\mathrm{Im}c_p$} & \multirow{ 2}{*}{\shortstack[l]{$A_h,B_h,D_h$,\\$\mathrm{Im}A_p,B_p,\mathrm{Im}D_p$}} & \multirow{2}{*}{$d_h = - c^*_h,\mathrm{Im}d_p = \mathrm{Im}c_p,C_h = B_h,C_p = - B^*_p$}\\
&&&& \\
 {$\drsh\left\llbracket(\chi_{\mathcal{T}\overline{C}_6}\chi_{\mathcal{T}S})=(00)\right\rrbracket^\ddag $}
& {${\left\llbracket-\right\rrbracket^\ddag}$}& {${\left\llbracket \mathrm{Im}c_h,\mathrm{Im}c_p\right\rrbracket^\ddag}$}& {${\left\llbracket F_6\right\rrbracket^\ddag }$}&$\left\llbracket \text{Constraints inherited from above}\right\rrbracket^\ddag$\\
 {$\drsh\left\llbracket(\chi_{\mathcal{T}\overline{C}_6}\chi_{\mathcal{T}S})=(\pi0)\right\rrbracket^\ddag $}
& {$\left\llbracket \mathrm{Re}\alpha_p\right\rrbracket^\ddag$}& {$\left\llbracket \mathrm{Re}a_h,\,c_h\right\rrbracket^\ddag $}& {$\left\llbracket A_h,\,B_h,\,D_h\right\rrbracket^\ddag$}&$\left\llbracket \text{Constraints inherited from above}\right\rrbracket^\ddag$\\
\hline
\multirow{2}{*}{\shortstack[l]{$(00)$\--- or $(\pi\pi)$\---$(\pi\pi\pi)$}} & \multirow{2}{*}{$\mathrm{Re}\alpha_p$} &\multirow{2}{*}{$\mathrm{Re}b_h,\mathrm{Re}c_p$} & \multirow{2}{*}{\shortstack[l]{$A_h,B_h,D_h$,\\$\mathrm{Im}A_p,B_p,\mathrm{Im}D_p$}} & \multirow{2}{*}{$\mathrm{Re}d_p = \mathrm{Re}c_p,rC_h = B_h,C_p = - B^*_p$}\\
&&&& \\
 {$\drsh\left\llbracket(\chi_{\mathcal{T}\overline{C}_6}\chi_{\mathcal{T}S})=(00)\right\rrbracket^\ddag $}
& {${\left\llbracket-\right\rrbracket^\ddag}$}& {${\left\llbracket -\right\rrbracket^\ddag}$}& {${\left\llbracket F_6\right\rrbracket^\ddag} $}&$\left\llbracket \text{Constraints inherited from above}\right\rrbracket^\ddag$\\
 {$\drsh\left\llbracket(\chi_{\mathcal{T}\overline{C}_6}\chi_{\mathcal{T}S})=(\pi0)\right\rrbracket^\ddag $}
& {$\left\llbracket \mathrm{Re}\alpha_p\right\rrbracket^\ddag$}& {$\left\llbracket \mathrm{Re}b_h\right\rrbracket^\ddag$}& {$\left\llbracket  A_h,\,B_h,\,D_h\right\rrbracket^\ddag$}&$\left\llbracket \text{Constraints inherited from above}\right\rrbracket^\ddag$\\
\hline
\multirow{ 3}{*}{\shortstack[l]{$(0\pi)$\--- or $(\pi0)$\---$(0\pi0)_1$}}    & \multirow{ 3}{*}{\NA} & \multirow{ 3}{*}{$\mathrm{Re}b_h,\mathrm{Re}c_p$} & \multirow{ 3}{*}{$B_h,\mathrm{Re}A_p,B_p,\mathrm{Re}D_p$} & $\mathrm{Im}b_h = \sqrt{3}\mathrm{Re}b_h,\mathrm{Im}c_p = \sqrt{3}\mathrm{Re}c_p,c_p=d_p$, \\
&&&& $\mathrm{Im}A_p = -\sqrt{3}\mathrm{Re}A_p,C_h = -B_h$,\\ &&&& $\mathrm{Im}D_p = -\sqrt{3} \mathrm{Re}D_p,C_p = e^{-\frac{2\pi}{3}i}B_p^*$\\
 {$\drsh\left\llbracket(\chi_{\mathcal{T}\overline{C}_6}\chi_{\mathcal{T}S})=(0\pi)\right\rrbracket^\ddag $}
& {$\left\llbracket - \right\rrbracket^\ddag$}& {$\left\llbracket\mathrm{Re}b_h \right\rrbracket^\ddag$}& {$\left\llbracket \mathrm{Re}A_p,\,B_p,\,\mathrm{Re}D_p\right\rrbracket^\ddag $}&$\left\llbracket \text{Constraints inherited from above}\right\rrbracket^\ddag$\\
\hline
{\shortstack[l]{$(0\pi)$\--- or $(\pi0)$\---$(0\pi0)_3$}}    &{$\mathrm{Re}\alpha_p$} & {$\mathrm{Re}b_h,\mathrm{Re}c_p$} & {$B_h,\mathrm{Re}A_p,B_p,\mathrm{Re}D_p$} & {$\mathrm{Re}d_p = \mathrm{Re}c_p,C_h =-B_h,C_p = B^*_p$}\\
 {$\drsh\left\llbracket(\chi_{\mathcal{T}\overline{C}_6}\chi_{\mathcal{T}S})=(00)\right\rrbracket^\ddag $}
& {${\left\llbracket -\right\rrbracket^\ddag}$}& {${\left\llbracket -\right\rrbracket^\ddag}$}& {${\left\llbracket \mathrm{Im}B_h,\,\mathrm{Im}B_p\right\rrbracket^\ddag} $}&$\left\llbracket \text{Constraints inherited from above}\right\rrbracket^\ddag$\\
 {$\drsh\left\llbracket(\chi_{\mathcal{T}\overline{C}_6}\chi_{\mathcal{T}S})=(0\pi)\right\rrbracket^\ddag $}
& {$\left\llbracket \mathrm{Re}\alpha_p\right\rrbracket^\ddag$}& {$\left\llbracket \mathrm{Re}b_h\right\rrbracket^\ddag$}& {$\left\llbracket \mathrm{Re}A_p,\,B_p,\,\mathrm{Re}D_p\right\rrbracket^\ddag $}&$\left\llbracket \text{Constraints inherited from above}\right\rrbracket^\ddag$\\
\hline
{\shortstack[l]{$(0\pi)$\--- or $(\pi0)$\---$(\pi\pi0)_0$}} & {$\mathrm{Re}\alpha_p$} & {$\mathrm{Re}a_h,c_h,\mathrm{Im}c_p$} & {$B_h,\mathrm{Re}A_p,B_p,\mathrm{Re}D_p$} & {$d_h =-c^*_h,\mathrm{Im}d_p = \mathrm{Im}c_p,C_h =-B_h,C_p = B^*_p$}\\
 {$\drsh\left\llbracket(\chi_{\mathcal{T}\overline{C}_6}\chi_{\mathcal{T}S})=(00)\right\rrbracket^\ddag $}
& {${\left\llbracket -\right\rrbracket^\ddag}$}& {${\left\llbracket \mathrm{Im}c_h,\mathrm{Im}c_p\right\rrbracket^\ddag}$}& {${\left\llbracket \mathrm{Im}B_h,\,\mathrm{Im}B_p\right\rrbracket^\ddag} $}&$\left\llbracket \text{Constraints inherited from above}\right\rrbracket^\ddag$\\
 {$\drsh\left\llbracket(\chi_{\mathcal{T}\overline{C}_6}\chi_{\mathcal{T}S})=(0\pi)\right\rrbracket^\ddag $}
& {$\left\llbracket \mathrm{Re}\alpha_p\right\rrbracket^\ddag$}& {$\left\llbracket \mathrm{Re}a_h,\,c_h\right\rrbracket^\ddag $}& {$\left\llbracket \mathrm{Re}A_p,\,B_p,\,\mathrm{Re}D_p\right\rrbracket^\ddag $}&$\left\llbracket \text{Constraints inherited from above}\right\rrbracket^\ddag$\\
\hline
\multirow{ 3}{*}{\shortstack[l]{$(0\pi)$\--- or $(\pi0)$\---$(\pi\pi0)_2$}}  &\multirow{ 3}{*}{\NA} & \multirow{ 3}{*}{$\mathrm{Re}a_h,c_h,\mathrm{Re}c_p$} & \multirow{ 3}{*}{$B_h,\mathrm{Re}A_p,B_p,\mathrm{Re}D_p$} & $\mathrm{Im}a_h = - \sqrt{3} \mathrm{Re}a_h,d_h = e^{\frac{\pi}{3}i} c_h^*,\mathrm{Im}c_p = \frac{1}{\sqrt{3}} \mathrm{Re}c_p$,\\
&&&&$d_p = c_p,\mathrm{Im}A_p = \sqrt{3}\mathrm{Re} A_p,C_h = - B_h$,\\
&&&& $C_p = e^{\frac{2\pi}{3}i} B_p^*,\mathrm{Im}D_p = \sqrt{3}\mathrm{Re}D_p$\\
 {$\drsh\left\llbracket(\chi_{\mathcal{T}\overline{C}_6}\chi_{\mathcal{T}S})=(0\pi)\right\rrbracket^\ddag $}
& {$\left\llbracket - \right\rrbracket^\ddag$}& {$\left\llbracket \mathrm{Re}a_h,\,c_h\right\rrbracket^\ddag $}& {$\left\llbracket \mathrm{Re}A_p,\,B_p,\,\mathrm{Re}D_p\right\rrbracket^\ddag $}&$\left\llbracket \text{Constraints inherited from above}\right\rrbracket^\ddag$\\
\hline
{\shortstack[l]{$(0\pi)$\--- or $(\pi0)$\---$(00\pi)$}}      & {$\mathrm{Re}\alpha_p$} & {$\mathrm{Im}a_h,\,\mathrm{Im}c_h,\,a_p,\,c_p$} & { $\mathrm{Re}A_h,B_h,\mathrm{Re}D_h,B_p$} & {$\mathrm{Im}d_h = - \mathrm{Im}c_h,d_p = - c_p,C_h = B^*_h,C_p = - B_p$}\\
 {$\drsh\left\llbracket(\chi_{\mathcal{T}\overline{C}_6}\chi_{\mathcal{T}S})=(00)\right\rrbracket^\ddag $}
& {${\left\llbracket -\right\rrbracket^\ddag}$}& {${\left\llbracket F_4\right\rrbracket^\ddag} $}& {${\left\llbracket \mathrm{Im}B_h,\,\mathrm{Im}B_p\right\rrbracket^\ddag} $}&$\left\llbracket \text{Constraints inherited from above}\right\rrbracket^\ddag$\\
 {$\drsh\left\llbracket(\chi_{\mathcal{T}\overline{C}_6}\chi_{\mathcal{T}S})=(\pi\pi)\right\rrbracket^\ddag $}
& {$\left\llbracket \mathrm{Re}\alpha_p\right\rrbracket^\ddag $}& {$\left\llbracket a_p,\,c_p\right\rrbracket^\ddag $}& {$\left\llbracket \mathrm{Re}A_h,\,B_h,\,\mathrm{Re}D_h\right\rrbracket^\ddag $}&$\left\llbracket \text{Constraints inherited from above}\right\rrbracket^\ddag$\\
\hline
{\shortstack[l]{$(0\pi)$\--- or $(\pi0)$\---$(\pi0\pi)$}}    & {$\mathrm{Re}\alpha_p$} & {$\mathrm{Im}b_h,\mathrm{Re}c_h,b_p$} & {$\mathrm{Re}A_h,B_h,\mathrm{Re}D_h,B_p$} & {$\mathrm{Re}d_h = \mathrm{Re}c_h,C_h = B^*_h,B_p = - C_p$}\\
 {$\drsh\left\llbracket(\chi_{\mathcal{T}\overline{C}_6}\chi_{\mathcal{T}S})=(00)\right\rrbracket^\ddag $}
& {${\left\llbracket -\right\rrbracket^\ddag}$}& {${\left\llbracket \mathrm{Im}b_h,\mathrm{Im}b_p\right\rrbracket^\ddag} $}& {${\left\llbracket \mathrm{Im}B_h,\,\mathrm{Im}B_p\right\rrbracket^\ddag} $}&$\left\llbracket \text{Constraints inherited from above}\right\rrbracket^\ddag$\\
 {$\drsh\left\llbracket(\chi_{\mathcal{T}\overline{C}_6}\chi_{\mathcal{T}S})=(\pi\pi)\right\rrbracket^\ddag $}
& {$\left\llbracket \mathrm{Re}\alpha_p\right\rrbracket^\ddag $}& {$\left\llbracket b_p\right\rrbracket^\ddag $}& {$ \left\llbracket \mathrm{Re}A_h,\,B_h,\,\mathrm{Re}D_h\right\rrbracket^\ddag $}&$\left\llbracket \text{Constraints inherited from above}\right\rrbracket^\ddag$\\
\hline
\multirow{ 3}{*}{\shortstack[l]{$(0\pi)$\--- or $(\pi0)$\---$(0\pi\pi)_0$}}  & \multirow{ 3}{*}{\NA} & \multirow{ 3}{*}{$\mathrm{Re}b_h,\mathrm{Re}c_p$
} & \multirow{ 3}{*}{$B_h,\mathrm{Re}A_p,B_p,\mathrm{Re}D_p$} & $\mathrm{Im}b_h = -\sqrt{3}\mathrm{Re}b_h,\mathrm{Im}c_p = \frac{1}{\sqrt{3}}\mathrm{Re}c_p,c_p=d_p$, \\
&&&& $\mathrm{Im}A_p = -\frac{1}{\sqrt{3}}\mathrm{Re}A_p,C_h =-B_h$,\\
&&&&  $C_p = e^{-\frac{\pi}{3}i} B_p^*$,
$\mathrm{Im}D_p = -\frac{1}{\sqrt{3}} \mathrm{Re}D_p$\\
 {$\drsh\left\llbracket(\chi_{\mathcal{T}\overline{C}_6}\chi_{\mathcal{T}S})=(0\pi)\right\rrbracket^\ddag $}
& {$\left\llbracket - \right\rrbracket^\ddag$}& {$\left\llbracket \mathrm{Re}b_h\right\rrbracket^\ddag $}& {$\left\llbracket \mathrm{Re}A_p,\,B_p,\,\mathrm{Re}D_p\right\rrbracket^\ddag $}&$\left\llbracket \text{Constraints inherited from above}\right\rrbracket^\ddag$\\
\hline
\multirow{ 3}{*}{\shortstack[l]{$(0\pi)$\--- or $(\pi0)$\---$(\pi\pi\pi)_0$}}& \multirow{ 3}{*}{\NA} &\multirow{ 3}{*}{$\mathrm{Re}a_h,c_h,\mathrm{Re}c_p$} & \multirow{ 3}{*}{${B_h,\mathrm{Re}A_p, B_p,\mathrm{Re}D_p}$} & $\mathrm{Im}a_h = - \sqrt{3}\mathrm{Re} a_h,d_h = e^{i\frac{\pi}{3}} c_h^*$,\\
&&&& $\mathrm{Im}c_p = - \sqrt{3}\mathrm{Re}c_p,c_p=d_p,\mathrm{Im}A_p = -\frac{1}{\sqrt{3}}\mathrm{Re}A_p$, \\
&&&& $C_h=-B_h,C_p = e^{-\frac{\pi}{3}i} B_p^*$,
$\mathrm{Im}D_p = - \frac{1}{\sqrt{3}}\mathrm{Re}D_p$\\
 {$\drsh\left\llbracket(\chi_{\mathcal{T}\overline{C}_6}\chi_{\mathcal{T}S})=(0\pi)\right\rrbracket^\ddag $}
& {$\left\llbracket - \right\rrbracket^\ddag$}& {${\left\llbracket \mathrm{Re}a_h,c_h\right\rrbracket^\ddag }$}& {$\left\llbracket {\,\mathrm{Re}A_p,B_p,\,\mathrm{Re}D_p}\right\rrbracket^\ddag $}&$\left\llbracket \text{Constraints inherited from above}\right\rrbracket^\ddag$\\
\hline
{\shortstack[l]{$(0\pi)$\--- or $(\pi0)$\---$(0\pi\pi)_1$}} &{\NA} &{$\mathrm{Re}b_h,\mathrm{Im}c_p$} & {$B_h,\mathrm{Im}A_p,B_p,\mathrm{Im}D_p$} & {$\mathrm{Im}d_p = \mathrm{Im}c_p,C_h=-B_h,C_p = - B^*_p$}\\
{$\drsh\left\llbracket(\chi_{\mathcal{T}\overline{C}_6}\chi_{\mathcal{T}S})=(0\pi)\right\rrbracket^\ddag $}
& {$\left\llbracket - \right\rrbracket^\ddag$}& {$\left\llbracket \mathrm{Re}b_h\right\rrbracket^\ddag $}& {$\left\llbracket \mathrm{Im}A_p,\,B_p,\,\mathrm{Im}D_p\right\rrbracket^\ddag $}&$\left\llbracket \text{Constraints inherited from above}\right\rrbracket^\ddag$\\
 {$\drsh\left\llbracket(\chi_{\mathcal{T}\overline{C}_6}\chi_{\mathcal{T}S})=(\pi0)\right\rrbracket^\ddag $}
& {$\left\llbracket - \right\rrbracket^\ddag$}& {$\left\llbracket \mathrm{Re}b_h\right\rrbracket^\ddag $}& {$\left\llbracket {\mathrm{Re}B_h,\mathrm{Re}B_p}\right\rrbracket^\ddag $}&$\left\llbracket \text{Constraints inherited from above}\right\rrbracket^\ddag$\\
 {$\drsh\left\llbracket(\chi_{\mathcal{T}\overline{C}_6}\chi_{\mathcal{T}S})=(\pi\pi)\right\rrbracket^\ddag $}
& {$\left\llbracket - \right\rrbracket^\ddag$}& {$\left\llbracket {-}\right\rrbracket^\ddag $}& {$\left\llbracket F_5\right\rrbracket^\ddag $}&$\left\llbracket \text{Constraints inherited from above}\right\rrbracket^\ddag$\\
\hline
{\shortstack[l]{$(0\pi)$\--- or $(\pi0)$\---$(\pi\pi\pi)_1$}}& {\NA} &{$\mathrm{Re}a_h,c_h,\mathrm{Re}c_p$} & {$B_h,\mathrm{Im}A_p,B_p,\mathrm{Im}D_p$} &{$d_h=-c^*_h,\mathrm{Re}d_p=\mathrm{Re}c_p$,  $C_h=-B_h,C_p = -B^*_p$}\\
 {$\drsh\left\llbracket(\chi_{\mathcal{T}\overline{C}_6}\chi_{\mathcal{T}S})=(0\pi)\right\rrbracket^\ddag $}
& {$\left\llbracket - \right\rrbracket^\ddag$}& {$\left\llbracket \mathrm{Re}a_h,\,c_h\right\rrbracket^\ddag $}& {$\left\llbracket \mathrm{Im}A_p,\,B_p,\,\mathrm{Im}D_p\right\rrbracket^\ddag $}&$\left\llbracket \text{Constraints inherited from above}\right\rrbracket^\ddag$\\
 {$\drsh\left\llbracket(\chi_{\mathcal{T}\overline{C}_6}\chi_{\mathcal{T}S})=(\pi0)\right\rrbracket^\ddag $}
& {$\left\llbracket - \right\rrbracket^\ddag$}& {$\left\llbracket \mathrm{Re}a_h,{\mathrm{Re}c_h,\mathrm{Re}c_p}\right\rrbracket^\ddag $}& {$\left\llbracket {\mathrm{Re}B_h,\mathrm{Re}B_p}\right\rrbracket^\ddag $}&$\left\llbracket \text{Constraints inherited from above}\right\rrbracket^\ddag$\\
 {$\drsh\left\llbracket(\chi_{\mathcal{T}\overline{C}_6}\chi_{\mathcal{T}S})=(\pi\pi)\right\rrbracket^\ddag $}
& {$\left\llbracket - \right\rrbracket^\ddag$}& {$\left\llbracket \mathrm{Re}c_p, {\mathrm{Im}c_h}\right\rrbracket^\ddag $}& {$\left\llbracket F_5 \right\rrbracket^\ddag $}&$\left\llbracket \text{Constraints inherited from above}\right\rrbracket^\ddag$\\
\end{tabular}
\end{ruledtabular}
\end{table*}

\subsection{Symmetry properties of the 0-flux ans\"atze}

The symmetry constraints imposed by the PSG given in the last
subsection are formulated in real space. For analyzing the
properties of the mean-field ans\"atze, it is more helpful to see
how the projective symmetry transformations apply in momentum space.
In the 0-flux case ($\chi_1=0$), the action of each projective
symmetry is simple and can be explicitly given. To start with, we
define the Bogoliubov-de Gennes (BdG) basis (where the spin indices
are suppressed):
$$\Phi_{\bm{k}} = \left(
f^{\phantom{\dag}}_{\bm{k},0},f^{\phantom{\dag}}_{\bm{k},1},f^{\phantom{\dag}}_{\bm{k},2},f^{\phantom{\dag}}_{\bm{k},3},
f^\dag_{-\bm{k},0},f^\dag_{-\bm{k},1},f^\dag_{-\bm{k},2},f^\dag_{-\bm{k},3}\right)^T.$$
The Hamiltonian is then written as
\begin{equation}\label{H}
H = \sum\limits_{\bm{k}\in \text{BZ}^+} \Phi^\dag_{\bm{k}}
\mathcal{H}_{\text{BdG}}(\bm{k}) ^{\vphantom\dagger}\Phi_{\bm{k}},
\end{equation}
where the momentum sum is over half of the Brillouin zone (BZ) with,
say $k_3>0$. This Hamiltonian has the standard Bogoliubov form
\begin{equation}\label{HH}
\mathcal{H}_{\text{BdG}}(\bm{k}) = \left(\begin{array}{cc}
\mathcal{H}_{\text{U(1)}}(\bm{k})& \mathcal{H}_p(\bm{k})\\
\mathcal{H}^\dag_p & -\mathcal{H}_{\text{U(1)}}^T(-\bm{k})
\end{array}\right),
\end{equation}
where $\mathcal{H}_{\text{U(1)}}$ and $\mathcal{H}_p$ correspond to
the hopping and pairing terms in Eq.~\eqref{bondsss} for the
$\mathbb{Z}_2$ PSG. For the U(1) PSG, the pairing terms vanish and
the BdG form corresponds to two copies of the U(1) Hamiltonian
$\mathcal{H}_{\text{U(1)}}$.

In terms of the BdG Hamiltonian matrix
$\mathcal{H}_{\text{BdG}}(\bm{k})$ describing each 0-flux ansatz, we
have the following symmetry constraints:
\begin{equation}
\begin{aligned}
W^\dag_{\overline{C}_6,w_{\overline{C}_6}}(\bm{k})
\mathcal{H}_{\text{BdG}}(\bm{k})
&W_{\overline{C}_6,w_{\overline{C}_6}}(\bm{k})\\
&= \mathcal{H}_{\text{BdG}}\left(\overline{C}_6(\bm{k})+\mu^3
\phi_{ST_1} {\bm{b}}_1\right),
\end{aligned}
\end{equation}
\begin{equation}
\begin{aligned}
W^\dag_{S,w_S}(\bm{k}) \mathcal{H}_{\text{BdG}}(\bm{k})
&W_{S,w_S}(\bm{k})\\
&= \mathcal{H}_{\text{BdG}}\left(S(\bm{k})+\mu^3 \phi_{ST_1}
({\bm{b}}_1+3{\bm{b}}_3)\right),
\end{aligned}
\end{equation}
where $\bm{b}_i$ (with $i=1,2,3$) are the three reciprocal lattice
vectors in units of $a^{-1}$, while $\mu^i$ (with $i=1,2,3$) are the
Pauli matrices acting in the particle-hole space $(f,f^\dag)^T$. The
derivation of these equations together with the forms of
$W_{\overline{C}_6,w_{\overline{C}_6}}(\bm{k})$ and
$W_{S,w_S}(\bm{k})$ can be found in App.~\ref{app:F}. We point out
that, due to the projective nature of the symmetry transformation,
an annihilation operation $f$ can be mapped to either an
annihilation operator $f$ or a creation operator $f^\dag$.
Therefore, the unitary matrices
$W_{\mathcal{O},w_{\mathcal{O}}}(\bm{k})$ with
$\mathcal{O}=\overline{C}_6,S$ are either off-diagonal
(corresponding to $w_{\mathcal{O}}=1$) or diagonal (corresponding to
$w_{\mathcal{O}}=0$) in the particle-hole space. Let us understand
the implications of this property for a U(1) 0-flux ansatz with
$w_{\overline{C}_6}=1$ by considering the action of
$I=\overline{C}_6^3$:
\begin{equation}
w_{\overline{C}_6}=1:\quad W_I^\dag
\mathcal{H}_{\text{U(1)}}(\bm{k}) W_I
=-\mathcal{H}^*_{\text{U(1)}}(\bm{k})\quad \forall \bm{k} \in
\text{BZ}. \label{wc6}
\end{equation}
Therefore, projective inversion acts like the product of faithful
inversion and charge conjugation, which ensures that the energies
come in $\pm E(\bm{k})$ pairs for the entire BZ.

\subsection{0-flux U(1)  ans\"atze with $w_S=1$: projective symmetry protected gapless nodal star\label{sec:nodalstar_c}}

\begin{figure}
\centering
\includegraphics[width=0.25\textwidth]{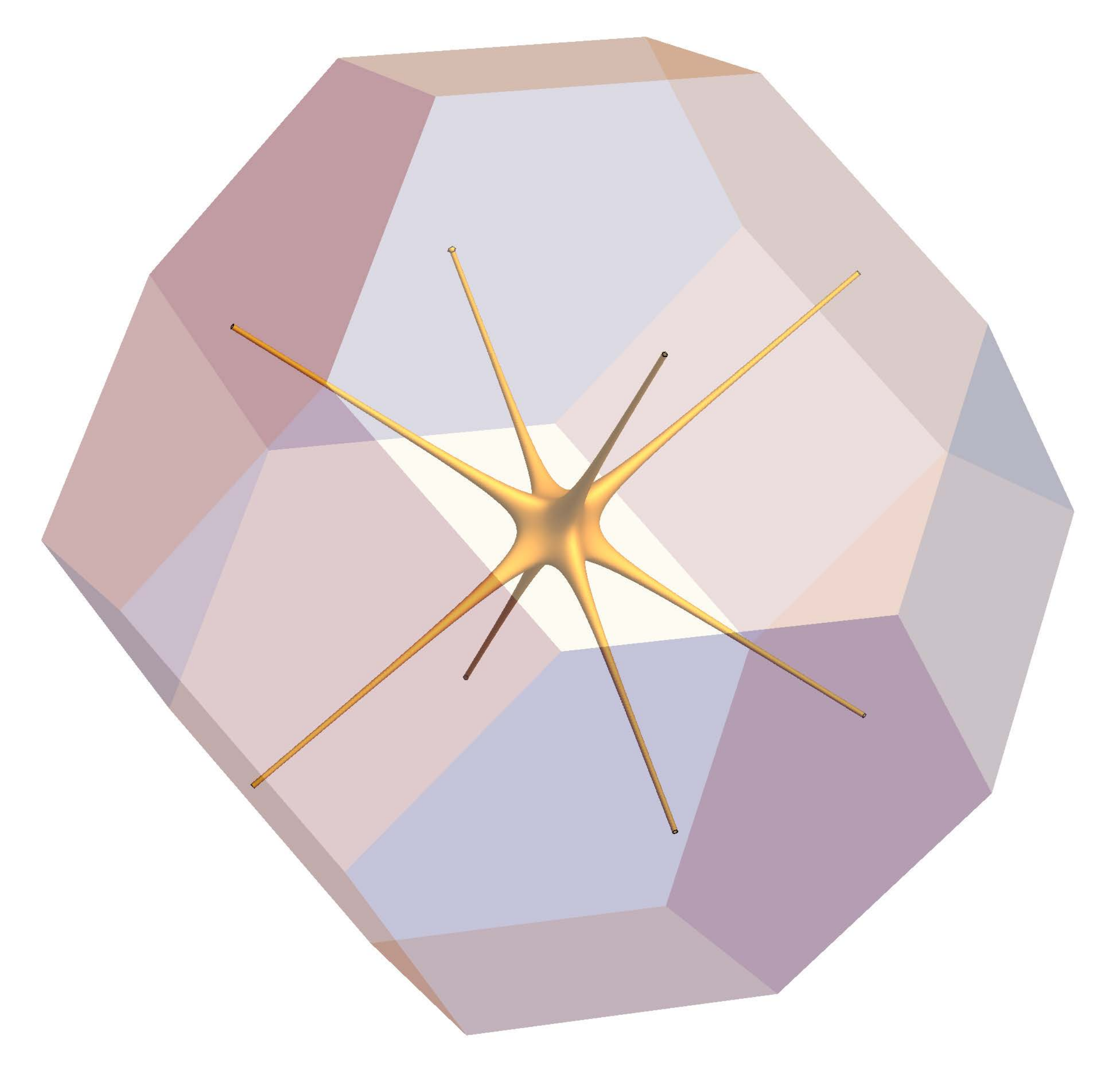}
\caption{{Illustration of the nodal star Fermi surface; the contour corresponds to an energy infinitesimally above the Fermi level.}}\label{nodal_star}
\end{figure}

Unlike the $w_{\overline{C}_6}=1$ classes, a 0-flux U(1) ansatz with
$w_{\overline{C}_6}=0$ does not necessarily have energy levels
coming in $\pm E(\bm{k})$ pairs at each momentum $\bm{k}$. However,
in the case of $w_S=1$ (regardless of $w_{\overline{C}_6}$), there
exists a one-dimensional (1D) submanifold in the BZ along which this
is true. In fact, along this submanifold,
\begin{equation}
\Lambda = \{(\varsigma_1,\varsigma_2,\varsigma_3)k|\varsigma_{1,2,3}
= \pm\},
\end{equation}
the energy levels not only come in $\pm E(\bm{k})$ pairs, but are
always \emph{gapless} at $E=0$, i.e., there are always two
degenerate modes sitting at the Fermi energy. We call this 1D
submanifold $\Lambda$ the \emph{star} manifold, and refer to the
gapless modes as the \emph{nodal star} zero modes; see
Fig.~\ref{nodal_star} for an illustration. A direct check by adding
up to 8th-nearest neighbor bonds (given in App.~\ref{app:G}) with
generic bond parameters shows that the zero modes are robust as long
as the projective symmetries are intact. This is strong evidence
that the nodal star zero modes are not accidental but are protected
by the projective symmetries.

Such a nodal star structure has been reported in other contexts. In
Ref.~\cite{burnell2009monopole}, Burnell \emph{et al.}~studied an
SU(2) invariant ansatz with purely imaginary NN hoppings forming
$\pi/2$ fluxes on all pyrochlore faces: the so-called ``Monopole
Flux'' state. Such an ansatz preserves charge conjugation, the
product of inversion and time reversal $I\circ\mathcal{T}$, and the
24 ``proper elements'' of the point group $O_h$, but it breaks the
individual time reversal and inversion symmetries. The 24 ``proper
elements'' and the composed symmetry $I\circ\mathcal{T}$ allow the
mapping of one NN bond to all the other NN bonds, while charge
conjugation and $I\circ\mathcal{T}$ ensure that the Hamiltonian is
real and an odd function of $\bm{k}$. An algebraic proof was then
given to show that the matrix structure on the star submanifold
leads to nodal star zero modes and that the symmetries forbid a
chemical potential which could otherwise gap out the nodal star.
However, the proof relies on the restriction of the hopping to the
nearest neighbors, and it is not clear in the proof if the nodal star is truly
symmetry protected, i.e., whether further neighbor symmetry allowed
hoppings can gap out the nodal star. A similar nodal star state also
appears in the bosonic description of a $\mathbb{Z}_2$ spin liquid
\cite{liu2019competing} at the NN level. In that case, it was
explicitly shown that an infinitesimal NNN bond amplitude is enough
to gap out the nodal line to discrete points.

Here, we provide a proof that the nodal star zero modes appearing in
our $w_S=1$ classes are indeed protected by the projective
symmetries, see App.~\ref{app:H}. The proof is algebraic, and can be
viewed as a generalized version of that given in
Ref.~\cite{burnell2009monopole}. The proof relies on the following
observation: while an unprojective screw symmetry relates the
Hamiltonian at momentum $\bm{k} = (k_x,k_y,k_z)$ with that at
momentum $(k_y,k_x,-k_z)$, a projective screw symmetry relates the
Hamiltonian at momentum $\bm{k}=(k_x,k_y,k_z)$ with that at momentum
$(-k_y,-k_x,k_z)$. {Therefore, focusing on the
$(1,1,1)$ nodal line without loss of generality, the symmetry
operation
\begin{equation}
R\equiv S\circ C_3\circ C_3\circ S\circ C_3\circ S
\end{equation}
leaves the momenta along the nodal line unchanged in the case of a
projective $S$. This symmetry constrains the Hamiltonian along the
nodal line [cf.~Eq.~(\ref{wc6})],
\begin{equation}
W_R^\dag(k,k,k) \mathcal{H}_{\text{U(1)}}(k,k,k) W_R(k,k,k)
=-\mathcal{H}^*_{\text{U(1)}}(k,k,k),
\end{equation}
and implies that the energy levels come in $\pm E(\bm{k})$ pairs.
Considering the analogous action of the threefold rotation $C_3$
[which also maps $(k,k,k)$ to itself],
\begin{equation}
W_{C_3}^\dag(k,k,k) \mathcal{H}_{\text{U(1)}}(k,k,k) W_{C_3}(k,k,k)
=\mathcal{H}_{\text{U(1)}}(k,k,k),
\end{equation}
and the specific forms of $W_R(k,k,k)$ and $W_{C_3}(k,k,k)$, it can
then be shown (see App.~\ref{app:H}) that the rank of the matrix
$\mathcal{H}_{\text{U(1)}}(k,k,k)$ is at most $6$, which implies
that it has at least two zero eigenvalues.} Since our proof only
relies on the symmetry properties of the spinon Hamiltonian, the
result universally applies to any fully symmetric mean-field ansatz
with a projective screw symmetry ($w_S=1$), even beyond the NN (or
NNN) level. In turn, this suggests that a pyrochlore spin liquid
with a nodal star Fermi surface may be commonplace.

\section{Nodal star U(1) spin liquid\label{sec:exp}}

In this section, we restrict ourselves to the study of U(1) spin
liquids with a gapless nodal star structure as was put forward in
the last subsection. Our goal is to develop a full-fledged low
energy theory whose degrees of freedom include both the nodal star
spinons and the U(1) gauge field. The gauge field has physical
consequences and may lead to observable effects. Such effects have
been explored in U(1) spin liquids with a spinon Fermi surface where
the U(1) gauge fluctuations lead to $T^{2/3}$ and $T\ln(1/T)$
scaling in the specific heat for two and three spatial dimensions,
respectively \cite{senthil2004weak,motrunich2005variational}. These
non-Fermi liquid behaviors have been important experimental
teststones for the discovery of spinon Fermi surface U(1) spin
liquids. In this regard, it is interesting to ask how gauge
fluctuations affect the thermodynamic properties of the nodal star
U(1) spin liquid. We hope that the answer to this question provided
in this section can serve as a primer for the more interesting
properties of the nodal star U(1) spin liquid.

\subsection{Low energy effective model for spinon nodal bands}

In this subsection, we take a specific class of nodal star spin
liquid and study its low energy properties in detail. We choose the
U(1) class $0$\---$(1\;1)$\---$(0\;\pi)$ and only keep the NN
mean-field parameters $b= i b_i$ and $c = c_r$. Although the energy
at arbitrary momentum cannot be written in a closed form, the
energies along the star $(\varsigma_1 k,\varsigma_2 k,\varsigma_3 k)
\in \Lambda$ have a simple expression:
\begin{equation}
\begin{aligned}
&E_{1,2}= 0, \quad E_3 = -E_4 = 4 \sqrt{2} c_r,\\
&E_{5,6} = -E_{7,8} = \sqrt{6 b_i^2+20 c_r^2-6(b_i^2-2c_r^2) \cos
k}.
\end{aligned}
\end{equation}
For simplicity, we set $b_i = \sqrt{2} c_r$; this specific ansatz
should be continuously connected to those at other parameter
regions. The low energy dispersion along the nodal lines and in
vicinity of the $\Gamma$ point is then well described by the
following effective Hamiltonian:
\begin{subequations}
\begin{align}\label{heffh}
\mathcal{H}(\bm{k}) &= \bm{d}_{\bm{k}} \cdot \bm{\sigma},\\
\bm{d}_{\bm{k}} &= \left(\begin{array}{c}\cos k_3 - \cos k_2\\
\cos k_1 - \cos k_3\\
\cos k_2 - \cos k_1
\end{array}\right).
\end{align}
\end{subequations}
Along each momentum section perpendicular to the star lines, the
spinon field has the dispersion of a 2+1D Dirac field, where the
Dirac velocity $v= \sqrt{3} \sin k$ is a sinusoidal function of the
star momentum $(\varsigma_1 k,\varsigma_2 k,\varsigma_3 k)$. In the
vicinity of the $\Gamma$ point, the model can be further simplified
by expanding $\bm{d}_{\bm{k}}$:
\begin{equation}\label{expdk}
\bm{d}_{\bm{k}} = (k_2^2-k_3^2,k_3^2-k_1^2,k_1^2- k_2^2) + O(k^4).
\end{equation}
The spinon nodal star creates an interesting instance of U(1) gauge
fields interacting with gapless matter. This is in contrast with the
quantum spin ice model established in
Ref.~\cite{hermele2004pyrochlore} where the matter fields are gapped
and the low energy description is the Maxwell theory.

\subsection{Nodal star spinons with U(1) gauge field}

We now assume that the nodal star spinons are coupled to a U(1)
gauge field. The low energy effective Hamiltonian describing the
spinon nodal bands in Eq.~\eqref{heffh} corresponds to a Lagrangian
\begin{equation}\label{toyH}
\mathcal{L}_0 = \psi^\dag_k(-i k_0 + \mathcal{H}(\bm{k}))\psi_k,
\end{equation}
where we use the imaginary time formulation and denote $k=(k_0,
\bm{k})$ as the four-momentum in Euclidean spacetime. The U(1) gauge
field has a Maxwell term
\begin{equation}\label{max}
\mathcal{L}_M = \frac{1}{2g^2}A^\mu({k})(k^2 \delta_{\mu\nu}-k_\mu
k_\nu) A^\nu(-{k}),
\end{equation}
which emerges in this low energy effective theory by integrating out
the high energy spinon bands. Finally, the gauge field couples to
the spinon fields in the form of
\begin{equation}
\begin{aligned}
\mathcal{L}_1 =  & \sum_q i A^0(-{q}) \psi^\dag_{{k}-{q}/2}\psi_{{k} + {q}/2} \\
& \,~\quad + A^i(-{q}) \psi^\dag_{{k}-{q}/2}\frac{\partial \mathcal{H}(\bm{k})}{\partial k_i} \psi_{{k} + {q}/2}\label{vertex1}\\
& + \sum_{q,q'} A^i({q})A^j({q}')
\psi^\dag_{{k}+{q}'}\frac{\partial^2 \mathcal{H}(\bm{k})}{\partial
k_j\partial k_i} \psi_{{k} - {q}}+O(A^3),
\end{aligned}
\end{equation}
where the first two lines are the usual minimal coupling terms and
the third line is a diamagnetic coupling term. The complete theory
describing the low energy nodal star spinons and the U(1) gauge
field is thus
\begin{equation}
\mathcal{L} = \mathcal{L}_0 + \mathcal{L}_M + \mathcal{L}_1+
\mathcal{L}_{gf}+\mathcal{L}_{gh},
\end{equation}
where we have also included a gauge fixing term and a ghost term for
later use \cite{peskin2018introduction}:
\begin{subequations}
\begin{align}
\mathcal{L}_{gf} &= \frac{1}{2 \xi} k_\mu k_\nu A^\mu(k) A^\nu(-k), \label{Lgf} \\
\mathcal{L}_{gh} &= \frac{1}{g^2}\bar{\eta}_k k^2  \eta_k
\label{Lgh}.
\end{align}
\end{subequations}
It is the goal of the next subsection to derive an effective theory
for the photon field.

\subsection{Vacuum polarization for the emergent photons}

\begin{figure}[t]
\centering
\includegraphics[width=0.4\textwidth]{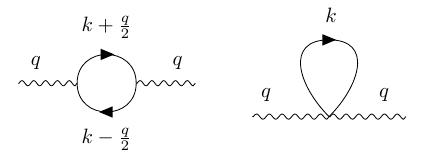}
\caption{The two diagrams for the photon self-energy at one loop
level: the ``vacuum polarization bubble'' (left) and the ``tadpole''
(right). Solid (wavy) lines denote spinon (photon)
propagators.}\label{feynman}
\end{figure}

We follow the usual perturbative approach to calculate the photon
effective action within the random phase approximation (RPA). The
validity of this calculation will be commented below.

At one-loop level, the spinon-gauge coupling $\mathcal{L}_1$
produces two diagrams for the photon self-energy, as shown in
Fig.~\ref{feynman}. {Due to gauge invariance, the
photon self-energy must (i) vanish when $q\rightarrow 0$ and (ii)
satisfy the Ward identity at small $q$. We prove these properties at
one-loop level in Appendix \ref{app:H}. We show that, at $q=0$, the
two diagrams of Fig.~\ref{feynman}, the ``vacuum polarization
bubble'' and the ``tadpole'', cancel each other. Furthermore, at $q
\neq 0$, the corrections to the tadpole are only $O(q^2)$, while, as
we will show below, there are corrections to the vacuum polarization
bubble at a lower order of $q$. Therefore, at leading order in $q$,
the photon self-energy can be identified as the $q$-dependent part
of the vacuum polarization bubble. Ignoring a minus sign resulting
from the fermion loop, the vacuum polarization bubble reads}
\begin{equation}\label{po}
\Pi^{\mu\nu}(q) = \int \frac{d^4 k}{(2\pi)^4} \mathrm{Tr}\left[
\Gamma^\mu(\bm{k}) G_0(k+\frac{q}{2}) \Gamma^\nu(\bm{k})
G_0(k-\frac{q}{2})\right],
\end{equation}
where
$G_0(k) = \frac{1}{ik_0 - \mathcal{H}(\bm{k})}$ is the Green's
function for the bare spinon Lagrangian $\mathcal{L}_0$, and the
vertex $\Gamma^\mu$ obtained from Eq.~\eqref{vertex1} has the
following form:
\begin{equation}\label{vert}
\begin{aligned}
\Gamma^0(\bm{k}) = i 1_{2\times 2}, &\quad \Gamma^1(\bm{k}) =\sin k_1 (\sigma^2-\sigma^3),\\
\Gamma^2(\bm{k}) =\sin k_2 (\sigma^3-\sigma^1),&\quad
\Gamma^3(\bm{k}) =\sin k_3 (\sigma^1 -\sigma^2).
\end{aligned}
\end{equation}
While the anisotropic forms of the Hamiltonian [Eq.~\eqref{toyH}]
and the vertices [Eq.~\eqref{vert}] make it difficult to evaluate
the polarization in Eq.~\eqref{po} exactly, it is physically clear
that there are two distinct momentum regions in the BZ. These two
regions, the star region and the gapped region away from the star,
result in different scalings of the photon self energy. The
contribution from the gapped region away from the nodal lines,
$\Pi_1$, is $O(q^2)$, as can be directly seen from expanding the
polarization in Eq.~\eqref{po}. The final result for $\Pi_1$ is
constrained by the Ward identity (see Appendix \ref{app:H}) to have
the form of the bare Maxwell term in Eq.~\eqref{max}, hence it
simply renormalizes the corresponding coupling constant $g$. In
contrast, the contribution from the star region, $\Pi_0$, is linear
in $q$ \emph{with logarithmic corrections}. This means that $\Pi_0$
completely dominates over $\Pi_1$ at low energy and small momentum.
In the following, we focus on the calculation of $\Pi_0$.

We first provide an intuitive understanding for the linear scaling
$\Pi_0\sim |\bm{q}|$ and the existence of logarithmic corrections.
In the BZ, each plane perpendicular to the nodal direction has a
Dirac dispersion $E = v|\bm{k}_\perp|$ with a Dirac velocity $v=
\sqrt{3}\sin k_\parallel$ that is a sinusoidal function of the nodal
line momentum $(\varsigma_1,\varsigma_2,\varsigma_3)k_\parallel$.
Restricted to such a plane, the spinon-gauge coupled system can be
viewed as a QED3. The vacuum polarization diagram in QED3 scales
linearly in $q_\perp=(q_0,v\bm{q}_\perp)$ as
\begin{equation}
\Pi^{ab}_{\text{QED3}}= \sqrt{q_\perp^2}\left(  \delta^{ab} -
\frac{q_\perp^a q_\perp^ b}{q_\perp^2}\right),\quad a=0,1,2.
\end{equation}
The polarization of one nodal line branch can therefore be obtained
by considering copies of QED3 interacting with each other. To the
leading order of $q$, these QED3 copies are decoupled, and we find
that the polarization in the star region, $\Pi_0$, summed over these
QED3 copies and over different line branches, scales linearly with
$q$. Note, however, that this picture is oversimplified as the
vanishing Dirac velocity at the $\Gamma$ point would lead to
unphysical divergence in the limit of $q_0/|\bm{q}|\rightarrow 0$
when integrating over copies of QED3:
\begin{equation}\label{pidiverg}
\begin{aligned}
\int^\pi_0 d k_\parallel \Pi^{00}_{\text{QED3}} &\sim
\int^\pi_0 d k_\parallel \frac{1}{\sqrt{q_0^2 + \bm{q}^2_\perp \sin^2 k_\parallel}}\\
&\xrightarrow{q_0/|\bm{q}_\perp|\rightarrow 0} \mathrm{divergent!}
\end{aligned}
\end{equation}
In reality, the quadratic dispersion near the $\Gamma$ point takes
over as the Dirac dispersion flattens, which removes the unphysical
divergence and introduces a small momentum cutoff $\theta_0$ for the
nodal line momentum $k_\parallel$. The cutoff is determined by the
criterion that the Dirac dispersion becomes comparable to the
quadratic dispersion around the $\Gamma$ point, $v|\bm{k}_\perp|\sim
\bm{k}^2$, which gives $\theta_0\sim |\bm{k}|$ and thus changes the
integration range in Eq.~\eqref{pidiverg} as
\begin{equation}\label{cutofftheta}
\int^\pi_0 d k_\parallel \rightarrow \int^{\pi-|\bm{q}|}_{|\bm{q}|}
d k_\parallel.
\end{equation}
This cutoff introduces a logarithmic correction to the $00$, $0i$
and $i0$ components of the polarization tensor.

To understand the scaling behavior of $\Pi_0$ in a more rigorous
manner, we provide in Appendix \ref{app:I} the scaling analysis of
the vacuum polarization tensor in Eq.~\eqref{po} for $ q_0/|\bm{q}|
\ll 1$. We find that
\begin{subequations}\label{scalinggeneralform}
\begin{align}
\Pi_0^{00} &\sim - |\bm{q}|\ln\left(\frac{1}{\bm{q}^2+\omega^2/\bm{q}^2}\right)+|\bm{q}|f_{00}(q_0/\bm{q}^2),\\
\Pi_0^{0i} &\sim  i q_0 \ln\left(\frac{1}{\bm{q}^2+\omega^2/\bm{q}^2}\right)+\bm{q}^2f_{0i}(q_0/\bm{q}^2),\\
\Pi_0^{ij} &\sim |\bm{q}|+|\bm{q}|^3f_{ij}(q_0/\bm{q}^2),
\end{align}
\end{subequations}
where, in each expression, the first and the second terms denote the
contributions from the nodal lines and the region near the $\Gamma$
point, respectively, while $f_{00}(x)$, $f_{0i}(x)$, and $f_{ij}(x)$
are regular functions for any $0<|x|<1$. We see that, in all
components of the polarization tensor, the contribution from the
$\Gamma$ region is subdominant compared to that from the nodal lines
themselves.

The analysis in the preceding two paragraphs allows us to obtain the
analytic form of the dominant contribution to the photon self-energy
by performing a ``QED3 type'' calculation for the nodal lines. The
detailed calculation can be found in Appendix \ref{app:I}. Here we
stress that the ``QED3 type'' calculation performed here must be
understood with caution. In the usual perturbative calculation of
QED3, a large parameter $N$ (the number of fermion flavors) is
introduced to ensure the validity of the RPA and the convergence in
the infrared (IR) limit. While we introduce no explicit large
parameter $N$ in our calculation, such a large $N$ should be
understood to be present whenever needed, and we hope that the
result can be analytically continued to the $N=1$ case.

With this caution in mind, we present the final result here: at the
leading order of $q$, the photon self-energy is
\begin{equation}
\Pi(iq_0,\bm{q}) = \sum_{\varsigma_1,\varsigma_2,\varsigma_3 = \pm1}
\Pi_{\varsigma_1,\varsigma_2,\varsigma_3}(iq_0,\bm{q}),
\end{equation}
where $\Pi_{\varsigma_1,\varsigma_2,\varsigma_3}(iq_0,\bm{q})$ is
the contribution from the nodal line branch
$(\varsigma_1,\varsigma_2,\varsigma_3)$. The individual
contributions are
\begin{widetext}
\begin{equation}\label{retardedpi}
\begin{aligned}
&\Pi_{\varsigma_1,\varsigma_2,\varsigma_3}(iq_0,\bm{q})
=\frac{\sqrt{q_0^2}}{16 \sqrt{3} \pi}
 \times\\
&\left(\begin{array}{cccc}
\frac{Q^2}{q^2_0}F{}& -\varsigma_1\frac{Q_1}{q_0} F{}&-\varsigma_2\frac{Q_2}{q_0} F{}&-\varsigma_3\frac{Q_3}{q_0} F{}\\
-\varsigma_1\frac{Q_1}{q_0}F{}& \quad\quad\; (1+
\frac{Q^2_{11}}{Q^2}) F{} +(1- \frac{Q^2_{11}}{Q^2})
E{}&-\varsigma_1\varsigma_2\left[ ( \frac{1}{2}-
\frac{Q^2_{33}}{Q^2}) F{} +(\frac{1}{2}+ \frac{Q^2_{33}}{Q^2})
E{}\right] &-\varsigma_1\varsigma_3\left[ (\frac{1}{2}-
\frac{Q^2_{11}}{Q^2}) F{} +(\frac{1}{2}+ \frac{Q^2_{11}}{Q^2})
E{}\right]
\\
-\varsigma_2\frac{Q_2}{q_0} F{}&-\varsigma_1\varsigma_2\left[ (\frac{1}{2}- \frac{Q^2_{33}}{Q^2}) F{} +(\frac{1}{2}+ \frac{Q^2_{33}}{Q^2})  E{}\right]& \quad\quad\; (1+ \frac{Q^2_{22}}{Q^2}) F{} +(1- \frac{Q^2_{22}}{Q^2})  E{}&-\varsigma_2\varsigma_2\left[ (\frac{1}{2}- \frac{Q^2_{11}}{Q^2}) F{} +(\frac{1}{2}+ \frac{Q^2_{11}}{Q^2})  E{}\right]\\
-\varsigma_3\frac{Q_3}{q_0} F{}&-\varsigma_1\varsigma_3\left[
(\frac{1}{2}- \frac{Q^2_{22}}{Q^2}) F{} +(\frac{1}{2}+
\frac{Q^2_{22}}{Q^2})  E{}\right]&-\varsigma_2\varsigma_3\left[
(\frac{1}{2}- \frac{Q^2_{11}}{Q^2}) F{} +(\frac{1}{2}+
\frac{Q^2_{11}}{Q^2})  E{}\right]& \quad\quad\; (1+
\frac{Q^2_{33}}{Q^2}) F{} +(1- \frac{Q^2_{33}}{Q^2})  E{}
\end{array}\right),
\end{aligned}
\end{equation}
\end{widetext}
where
\begin{subequations}
\begin{align}
F &= F(\pi-|\bm{q}|,-Q^2/q_0^2) - F(|\bm{q}|,-Q^2/q_0^2),\\
E &= E(\pi-|\bm{q}|,-Q^2/q_0^2) - E(|\bm{q}|,-Q^2/q_0^2)
\end{align}
\end{subequations}
are the incomplete elliptic integrals of the first and second kinds,
respectively, with elliptic modulus $Q/q_0$, and
\begin{subequations}\label{qqqqqqq}
\begin{align}
Q_1 &= 2 \varsigma_1 q_1 -\varsigma_2 q_2 - \varsigma_3 q_3,\\
Q_2 &= -\varsigma_1 q_1 +2\varsigma_2 q_2 - \varsigma_3 q_3,\\
Q_3 &= -\varsigma_1 q_1 -\varsigma_2 q_2 +2 \varsigma_3 q_3,\\
Q^2 &= \frac{1}{3}(Q_1^2+Q_2^2+Q_3^2),\\
Q_{11}^2&= Q^2 - 3 (\varsigma_2 q_2 - \varsigma_3 q_3)^2 ,\\
Q_{22}^2&= Q^2 - 3 (\varsigma_3 q_3 - \varsigma_1 q_1)^2,\\
Q_{33}^2&= Q^2 - 3 (\varsigma_1 q_1 - \varsigma_2 q_2)^2.
\end{align}
\end{subequations}
For each branch $(\varsigma_1,\varsigma_2,\varsigma_3)$,
$\Pi_{\varsigma_1,\varsigma_2,\varsigma_3}(iq_0,\bm{q})$ has two
zero eigenvalues corresponding to eigenvectors $(q_0,q_1,q_2,q_3)$
and $(0,\varsigma_1,\varsigma_2,\varsigma_3)$; the former one is the
longitudinal four-momentum vector. The remaining two nonzero
eigenvalues, $\frac{\sqrt{3}}{32 \pi}\sqrt{q_0^2} E$ and
$\frac{\sqrt{3}}{32\pi}\frac{q_0^2+\frac{1}{3}Q^2}{\sqrt{q_0^2}} F$,
are nondegenerate; they correspond to one ``transverse'' eigenvector
$\left(0,\varsigma_1(\varsigma_2 q_2-\varsigma_3
q_3),\varsigma_2(\varsigma_3
q_3-\varsigma_1q_1),\varsigma_3(\varsigma_1 q_1-\varsigma_2
q_2)\right)$ and one ``longitudinal'' eigenvector $(-
\frac{Q^2}{q_0},\zeta_1 Q_1,\zeta_2 Q_2,\zeta_3 Q_3)$, where
``transverse'' and ``longitudinal'' are understood with respect to
the spatial three-momentum $(Q_1,Q_2,Q_3)$. The nondegeneracy here
implies that the boost symmetry of the QED3 is broken in our theory,
which can be traced back to the ``boost symmetry breaking'' of the
vertices in Eq.~\eqref{vert}.

The photon self-energy $\Pi(iq_0,\bm{q})$ contains the longitudinal
four-momentum vector as an eigenvector corresponding to eigenvalue
zero, which ensures that the Ward identity $q_\mu \Pi^{\mu\nu} = 0$
is preserved. However, since the four star branches have different
three-momenta $(Q_1,Q_2,Q_3)$, $\Pi(iq_0,\bm{q})$ can no longer be
decomposed into longitudinal and transverse modes.

Suppose $e_i(i q_0,\bm{q})$ with $i=1,2,3$ are the three nonzero
eigenvalues of $\Pi(i q_0,\bm{q})$ that correspond to the
eigenvectors $v_i(i q_0,\bm{q})$. The dressed photon Green's
function is then
\begin{equation}
D^{\mu\nu}(iq_0,\bm{q}) = \sum_{i=1}^3
\frac{P^{\mu\nu}_i}{\frac{1}{g^2}q^2 + e_i(q)} +  \xi \frac{q^\mu
q^\nu}{q^4}, \label{Dmn}
\end{equation}
where $P^{\mu\nu}_i= v_i^\mu v_i^\nu$ are the projectors for the
$i=1,2,3$ modes. The last term results from the gauge fixing term
$\mathcal{L}_{gf}$ in Eq.~\eqref{Lgf}.

We remind the reader that the photon self-energy calculated here is
for zero temperature. A finite temperature calculation can also be
considered following Ref.~\cite{PhysRevD.100.073009}. We leave such
a calculation to future work.

\subsection{Photon contribution to thermodynamics}

We now proceed to calculate the photon contribution to the
thermodynamics. The photon free energy reads
\begin{equation}
F = -\frac{1}{2\beta} \sum_n \int \frac{d^3\bm{q}}{(2\pi)^3}
\left(\ln \det D(i\omega_n,\bm{q}) + 2\ln
\beta^2(\omega^2_n+\bm{q}^2)\right),
\end{equation}
where $\omega_n$ is the Matsubara frequency and $D$ is the photon
Matsubara Green's function. The second term of $F$ comes from the
fictitious free energy for the ghost fields $\eta_k$ in
Eq.~\eqref{Lgh}.
Half of this fictitious term will cancel the contribution from the
gauge fixing term $\xi q^\mu q^\nu/q^4$ in the photon Green's
function, while the other half will contribute a positive term
$\propto T^4$ to the free energy. Such a term would cancel out the
longitudinal mode in the free gauge theory, however, as we will see,
this is no longer the case in the full theory when the photons are
coupled to the spinons. Converting the Matsubara sum to a contour
integral, we then obtain 
\begin{equation}
\begin{aligned}
F = & \frac{\pi^2}{90}T^4 + \sum_{i=1}^3 \int ^{+\infty}_{-\infty} \frac{d\omega}{2\pi} \frac{1}{e^{\beta \omega}-1} \times\\
&\,\,\int \frac{d^3 q}{(2\pi)^3} \tan^{-1} \left( \frac{-\omega
0^++\mathrm{Im} e^R_i(\omega,\bm{q})}{-\omega^2+\bm{q}^2 +
\mathrm{Re}e^R_i(\omega,\bm{q})}\right),
\end{aligned}
\end{equation}
where $e_i^R(\omega,\bm{q})$ are the eigenvalues of the retarded
polarization $\Pi^R(\omega,\bm{q}) = \Pi(iq_0\rightarrow
\omega^+,\bm{q})$ with the notation $\omega^+ = \omega+i0^+$. Note
that the dressed photon Green's function corresponds to zero
temperature and that the temperature dependence of the free energy
comes entirely from the Boltzmann function. The momentum-frequency
integral in the free energy can be separated into two regions that
give different scaling behaviors.
\\
\emph{The ``dynamic'' region: $|\omega|/|\bm{q}|\geq 2$}.  \; In this region, we have
$\frac{Q^2}{\omega^2}\leq \frac{4\bm{q}^2}{\omega^2} \leq 1$, and the elliptic functions $E$ and
$F$ are both real. The eigenvalues $e^R_{1,2,3}$ are then purely
imaginary due to the prefactor $\sqrt{q_0^2} \rightarrow -i \omega$
on the upper half plane in Eq.~\eqref{retardedpi}. Furthermore, the
eigenvalues are of the same amplitude:
\begin{equation}
e^R_{1,2,3}\rightarrow-\frac{i \mathrm{sgn}(\omega)}{4\sqrt{3}
\pi}\text{\,\, \,\,\,when\,\, }|\omega|/|\bm{q}|\rightarrow \infty,
\end{equation}
which is much larger than $-\omega^2+\bm{q}^2$ at small frequency.
We then have $\tan^{-1} \left( \frac{ \mathrm{Im}
e^R_i(\omega,\bm{q})}{-\omega^2+\bm{q}^2 }\right) \sim -
\frac{\pi}{2} \mathrm{sgn}(\omega)$ and, hence, the free energy
scales with temperature as 
\begin{equation}
 F_{\text{dyn}}(T) \sim  3 \int^{+\infty}_{-\infty} \frac{d \omega}{2\pi} \frac{1}{e^{\beta \omega}-1}\frac{\frac{4\pi}{3}}{(2\pi)^3} {\frac{\omega^3}{8}} \left(-\frac{\pi}{2}\right) = - \frac{\pi^2}{{480} } T^4,
\end{equation}
{where we have dropped an unphysical part $F_{\text{dyn}}(0)$ which is divergent and temperature independent.} One notices that the free energy (and other thermodynamic
properties, such as the entropy and the specific heat) of each
dressed photon is {a fraction} of that of a free photon, which can be viewed as being contributed by a fractional degree of freedom. Such a
phenomenon also appears at infinite coupling of large $N$ QED3
\cite{PhysRevLett.123.241602}.
\vspace{0.2cm}
\\
\emph{The ``static'' region: $|\omega|/|\bm{q}|\ll 1$.}\; In this
region, by using the asymptotic forms of the elliptic functions, the
polarization $\Pi^{\mu\nu}$ in Eq.~\eqref{retardedpi} agrees with
the general scaling form in Eq.~\eqref{scalinggeneralform}. The Ward
identity $\omega \Pi^{00}+q_i \Pi^{i0}=0$ suggests that
$\Pi^{0i}\sim \frac{|\omega|}{|\bm{q}|}\Pi^{00}\ll \Pi^{00}$.
Therefore, the $\Pi^{00}$ component is decoupled from the $3\times
3$ block of the polarization tensor with spatial indices and is
identified with one of the eigenvalues, $e^R_3$. The remaining
eigenvalues $e^R_{1,2}$ then correspond to the two transverse
polarization modes. All three eigenvalues $e^R_{1,2,3}$ are
generally complex, and we find the following scaling form for them:
\begin{subequations}
\begin{align}
e^R_{1,2} & \sim   -i\omega  \left[E\left(0,\frac{2\bm{q}^2}{(\omega^+)^2}\right)-F\left(0,\frac{2\bm{q}^2}{(\omega^+)^2}\right)\right]\notag\\
 & \sim \  |\bm{q}| - i\frac{\omega^2}{|\bm{q}|}\text{sgn}(\omega),\\
e^R_3 & \sim  i\frac{\bm{q}^2}{\omega}\left[F\left(\pi-\theta_0,\frac{2\bm{q}^2}{(\omega^+)^2}\right)-F\left(\theta_0,\frac{2\bm{q}^2}{(\omega^+)^2}\right)\right]\notag\\
&\xrightarrow{\theta_0\sim |\bm{q}|}|\bm{q}|  \ln \left(\frac{1}{\max \{\frac{|\omega|}{|\bm{q}|},|\bm{q}|\}}\right) \notag\\
&\qquad\quad  + i
|\bm{q}|u\left(\frac{|\omega|}{|\bm{q}|}\right)\text{sgn}(\omega)
\theta\left( \frac{|\omega|}{|\bm{q}|}-|\bm{q}|\right),
\end{align}
\end{subequations}
where $u(\frac{|\omega|}{|\bm{q}|})$ scales as $u(x)\sim
\mathrm{const.} + x^2$. The eigenvalues of the transverse modes,
$e^R_{1,2}$, do not diverge near $\Gamma$ or $\mathrm{L}$, therefore
the cutoff $\theta_0$ has been set to zero.

In writing these asymptotic expressions, we have neglected the
angular dependence in the momentum. The validity of this
approximation has been numerically verified. For
$|\omega|>\bm{q}^2$, the physical modes $i=1,2,3$ lead to a dressed
photon Green's function of the form
\begin{equation}
\begin{aligned}
D^{\mu\nu} =& \frac{P^{\mu\nu}_1+P^{\mu\nu}_2}{\frac{1}{g^2}(\bm{q}^2c^2 - \omega^2) + \frac{1}{2 \sqrt{3} \pi} (|\bm{q}| - i \text{sgn}(\omega)\frac{\omega^2}{|\bm{q}|})}\\
&+
\frac{P^{\mu\nu}_3}{\frac{1}{g^2}(\bm{q}^2c^2-\omega^2)+\frac{1}{2
\sqrt{3} \pi} (|\bm{q}|\ln \frac{|\bm{q}|}{|\omega|} + i
\text{sgn}(\omega)|\bm{q}|)}.
\end{aligned}
\end{equation}
Note that the imaginary parts have opposite signs and that the
$P^{\mu\nu}_3$ term will contribute negatively to the specific heat.
The photon ``bare'' velocity $c$ and the coupling $g$ come from
integrating out the gapped spinon bands at higher energies. The
``bare'' velocity $c$ should be comparable to the mean Dirac
velocity of the nodal line spinons, $c \sim 1$, and we take
$g\lesssim 1$ following Ref.~\cite{gan1993non}. Therefore, at small
$\bm{q}$, we have $|\bm{q}|>\bm{q}^2> \omega^2$, and we only keep
$|\bm{q}|$ in the real part. The free energy can then be written as
$F_{\text{sta}} = F^{1,2}_{\text{sta}}+F^3_{\text{sta}}$, where the
contributions $F^{1,2}_{\text{sta}}$ and $F^3_{\text{sta}}$
correspond to the eigenvalues $e^R_{1,2}$ and $e^R_3$, respectively:
\begin{subequations}
\begin{align}
F^{1,2}_{\text{sta}} &= -2 \int_{0}^{\Lambda} \frac{q^2 dq}{2\pi^2} \int^{|\bm{q}|}_{0} \frac{d \omega}{2\pi} \frac{1}{e^{\beta \omega } - 1} 2\tan^{-1} \frac{\omega^2}{\bm{q}^2},\\
F^3_{\text{sta}} &= -2 \int_{0}^{\Lambda} \frac{q^2 dq}{2\pi^2}
\int^{<|\bm{q}|}_{\bm{q}^2} \frac{d \omega}{2\pi} \frac{1}{e^{\beta
\omega } - 1} \tan^{-1} \frac{1}{\ln \frac{|\omega|}{|\bm{q}|}},
\end{align}
\end{subequations}
where the dimensionless parameter $\Lambda$ is the upper momentum
cutoff at which the nodal line approximation becomes invalid.
Setting $x=\omega/|\bm{q}|$ in $F^{1,2}_{\text{sta}}$, we obtain
\begin{equation}
F^{1,2}_{\text{sta}} \sim - \int^{\Lambda}_0 q^3 d q \int^1_0 d x
\frac{x^2}{e^{\beta |\bm{q}| x}-1} \sim -\int^{\Lambda}_0 dq q^3
F\left(\frac{|\bm{q}|}{T}\right),
\end{equation}
where $F(y) = \int^1_0 dx \frac{x^2}{e^{ yx}-1}$. For $y\rightarrow
0$, we have $F(y)= \frac{1}{2y}+O(1)$, while for $y\rightarrow
+\infty$, the upper limit can be extended to $+\infty$, and we have
$F(y)\rightarrow 2\zeta(3)/y^3$. Therefore, at low temperatures,
$T<\Lambda$, we have $F^{1,2}_{\text{sta}}\sim - 2 (\int^T_0 dq q^3
\frac{T}{q}  + \int^\Lambda_T dq q^3 \frac{T^3}{q^3}) \sim -\Lambda
T^3 +O(T^4)$, and the leading contribution to free energy is
$-\Lambda T^3$.

For the remaining contribution $F^3_{\text{sta}}$, we perform the
following transformation:
\begin{equation}\label{Fu2d2}
\begin{aligned}
F^3_{\text{sta}} &\xrightarrow{\omega \equiv q^{\alpha+1}} -2 \int^\Lambda_0 q^2 dq \int^{\alpha_0>0}_{1} \frac{q^{\alpha+1}}{\alpha(e^{\beta q^{\alpha+1}}-1)} d\alpha\\
&\xrightarrow{z \equiv \beta q^{\alpha+1}}
2\int^1_{\alpha_0>0}\frac{d\alpha}{\alpha(\alpha+1)}
T^{\frac{\alpha+4}{\alpha+1}}\int^{\frac{\Lambda^{\alpha+1}}{T}}_0
\frac{z^{\frac{3}{\alpha+1}}}{e^z-1} dz.
\end{aligned}
\end{equation}
Numerics show that $F^3_{\text{sta}}$ is independent of $\alpha_0$
whenever $\alpha_0<0.5$. Note also that, when $z\gg 1$, we can
approximate $e^z-1\sim e^z$, meaning that, at large $z$, the
integral will contribute to the free energy with an exponentially
small term $e^{-\frac{1}{T}}$. Therefore, we can safely extend the
upper limit to infinity, and the integral in $z$ then gives
\begin{equation}
\int^\infty_0 \frac{z^{\frac{3}{\alpha+1}}}{e^z-1} dz =
\Gamma\left(\frac{4+\alpha}{1+\alpha}\right)\mathrm{Li}_{\frac{4+\alpha}{1+\alpha}}(1).
\end{equation}
{Since it is approximately true for $0.5<\alpha < 1$
that
\begin{equation}
\frac{\Gamma\left(\frac{4+\alpha}{1+\alpha}\right)\mathrm{Li}_{\frac{4+\alpha}{1+\alpha}}(1)}{\alpha(\alpha+1)}\sim
\frac{0.84}{\alpha^2},
\end{equation}
the contribution $F^3_{\text{sta}}$ takes the leading-order form
\begin{equation}
F^3_{\text{sta}} \sim \int^1_{\alpha_0} \frac{1}{\alpha^2}
T^{\frac{\alpha+4}{\alpha+1}} d\alpha \sim
-\frac{T^{\frac{5}{2}}}{\ln T} + O\left(\frac{T^{\frac{5}{2}}}{
\ln^2 T}\right).
\end{equation}
From all the analysis above, we conclude that the contribution
$F^3_{\text{sta}}$ dominates the dressed photon free energy at low
temperature. Note that this dominant term contributes negatively to
the specific heat.}

\subsection{Final result for the specific heat}

At the non-interacting level, the temperature scaling of the spinon
free energy can be obtained from the spinon density of states by a
simple power counting. The spinons near the nodal lines and the
$\Gamma$ point have densities of states $g(\epsilon)\propto
\epsilon$ and $g(\epsilon)\propto  \sqrt{\epsilon}$, and contribute
to the specific heat as $c_v\propto T^2$ and $c_v \propto
T^{\frac{3}{2}}$, respectively. The final result for specific heat
is then
\begin{equation}
c_v \sim T^{\frac{3}{2}} + \frac{T^{\frac{3}{2}}}{\ln T} +
\text{subleading terms}.
\end{equation}
Compared to a U(1) QSL with gapped matter fields, the leading term
in the specific heat has a lower power law exponent,
{$c_v \propto T^{3/2}$}, while the subleading term
has a negative contribution at low temperature. Since the Dirac
velocity $v$ is related to the pyrochlore spin exchange $J$ by
$v\sim J a/\hbar$, the small value of $J$ (typically a few meV)
indicates that this $T^{3/2}$ scaling {likely
dominates at low temperature over non-magnetic contributions} and
may serve as strong evidence for the observation of a nodal
star spin liquid.

\section{Discussion and outlook\label{sec:discussion}
}

\subsection{Summary}
\label{sec:summary}

In this paper, we obtain the complete classification of
spin-orbit-coupled spin liquids with either $\mathbb{Z}_2$ or U(1)
gauge structure on the pyrochlore lattice, within the PSG framework
for Abrikosov fermions. We find that there are at most 18 U(1) and 28
$\mathbb{Z}_2$ PSG classes with full pyrochlore space group symmetry,
and that the number of classes reduces to 16 for U(1) and increases to
48 for $\mathbb{Z}_2$ if time reversal symmetry is further imposed. We
present the explicit form of the mean-field Hamiltonian for each PSG
class upon gauge fixing. We also show that, in the U(1) case, several
classes of mean-field ans\"atze possess robust spinon zero modes along
high symmetry lines in the Brillouin zone and that these nodal lines
are protected by the projective screw symmetry. A low energy effective
theory for the nodal line spinons coupled to U(1) gauge fields is
given. Finally, we calculate the spinon contribution to the photon
self energy at one loop level and study the thermodynamics of the
dressed photon within the RPA approximation. {We find that the most
  dominant contributions to the specific heat are $T^{3/2}$ from the
  bare spinons and $T^{3/2} / \ln T$ from the dressed photons.}

\subsection{$\mathbb{Z}_2$ PSGs from fermionic and bosonic partons}

We now discuss the relationship between the $\mathbb{Z}_2$ PSGs
obtained from fermionic and bosonic partons. In a previous work
\cite{liu2019competing}, we employed Schwinger bosons to classfy
$\mathbb{Z}_2$ spin liquids on the pyrochlore lattice with full
lattice and time reversal symmetries. There, we found 16 distinct
PSG classes and labeled them by four $\mathbb{Z}_2$ parameters,
$n_1$, $n_{ST_1}$, $n_{\overline{C}_6S}$, and $n_{\overline{C}_6}$,
that are the bosonic counterparts of the $\chi$'s in the current
work. The other IGG parameters are all related to these four
parameters, for example, we have $n_{S\overline{C}_6} = n_1 +
n_{\overline{C}_6} + n_{ST_1}$, $n_{\mathcal{T}\overline{C}_6} =
n_{\overline{C}_6}$, and $n_{\mathcal{T}S} = n_1 + n_{ST_1}$.

By comparing these bosonic quantum numbers with the fermionic ones,
we see that each bosonic class corresponds to an appropriate class
in the fermionic PSG with lattice and time reversal symmetries. In
fact, the bosonic classes all have their counterparts already in the
fermionic PSG with only pyrochlore space group symmetry through the
corresponding $\mathbb{Z}_2$ quantum numbers
($\eta_{S\overline{C}_6} = \eta_1
\eta_{ST_1}\eta_{\overline{C}_6}$). The fermionic PSG, however, has
a larger number of classes, some of which do not have bosonic
counterparts. From Appendix \ref{app:z2_SG}, it is seen that the
additional fermionic classes exist due to the violation of the
condition $\eta_{S\overline{C}_6} = \eta_1
\eta_{ST_1}\eta_{\overline{C}_6}$ {or as a result of
multiple solutions to the SU(2) equation (which are distinguished by
an additional discrete parameter $j$)}. Upon imposing time reversal
symmetry, the number of fermionic classes increases from 28 to 48,
and the 16 bosonic classes still have 16 counterparts among them.

Physically, the bosonic and fermionic PSGs are supposed to describe
fractionalized excitations with bosonic and fermionic statistics,
respectively: the bosonic and fermionic spinons. This has been well
understood for 2D $\mathbb{Z}_2$ QSLs with topological order. In the
2D case, the elementary (fractionalized) excitations are bosonic
spinons, fermionic spinons, and visons. A fermionic spinon can be
viewed as a bound state of a bosonic spinon and a vison, which
induces a corresponding product rule between the vison, boson, and
fermion PSGs. It was argued in Ref.~\cite{PhysRevB.97.094422} on
general grounds that the classes common in bosonic and fermionic
PSGs realize gapped $\mathbb{Z}_2$ symmetric spin liquids, while the
additional classes in the fermionic PSG realize symmetry protected
gapless $\mathbb{Z}_2$ spin liquids. However, it is not clear
whether the claim directly applies to our case. Concretely, one can
compare the independent nonzero mean-field parameters at a given
bond level between the corresponding bosonic and fermionic classes,
and the numbers do not always match. The possible reasons are that
(i) the required assumption of U(1) spin symmetry for the proof of
Ref.~\cite{PhysRevB.97.094422} is absent in our case, and (ii) the
dimensionality is different in our case. Indeed, the dimensional
augmentation to 3D may fundamentally change the correspondence
between the bosonic and fermionic PSGs since the visons are now
line-like objects {and do not straightforwardly
relate bosonic and fermionic spinons to each other.}

Understanding the relation between the fermionic and bosonic PSG
classifications is an important goal. In addition to extracting the
statistics of the fractionalized excitations discussed above, it can
be used to map out the phases proximate to a QSL and the possible
transition types. Indeed, the bosonic representation has the
fundamental advantage that it can describe a transition to a
magnetically ordered state via the condensation of bosonic spinons,
while such a transition cannot be easily described in the fermionic
representation. Therefore, we hope to establish a clearer
understanding of this important relation in a future work.

\subsection{$\mathbb{Z}_2$ and U(1) PSGs using fermionic partons}

The $\mathbb{Z}_2$ mean-field  ans\"atze have spinon pairing terms
which manifestly break the U(1) symmetry down to its subgroup
$\mathbb{Z}_2$. However, in special cases, when some of the
mean-field parameters are switched off, the $\mathbb{Z}_2$ ans\"atze
may possess an enlarged symmetry. If this enlarged symmetry is (or
contains) U(1), the ansatz with parameters switched off belongs to
some ``root'' U(1) PSG class and the $\mathbb{Z}_2$ PSG can be
viewed as being derived from this U(1) PSG class by ``gauge symmetry
breaking'' via the Higgs mechanism. The simplest way one can enlarge $\mathbb{Z}_2$ symmetry to U(1) is by switching off all the pairing parameters in a
$\mathbb{Z}_2$ ansatz. If this can be consistently done without
violating the PSG, we obtain an explicit U(1) ansatz with only
hopping terms. 
For example, if we take the non-projective $\mathbb{Z}_2$ class $(00)$\---$(000)$ and naively switch off the pairings, we get exactly the U(1) non-projective mean-field state $0$\---(00)\---(00).

However, the correspondence between a $\mathbb{Z}_2$ and a U(1) ansatz may not always be
apparent and may be masked by the different gauge fixing conventions
used for the $\mathbb{Z}_2$ and the U(1)  ans\"atze. For example, a
mean-field Hamiltonian with only singlet pairing terms ($a_p$ and
its equivalents at further neighbor bonds) also has a U(1) symmetry.
This pairing U(1) symmetry can be converted to the usual hopping
U(1) symmetry by an appropriate gauge transformation. For
$\mathbb{Z}_2$ classes with time reversal and
$(\chi_{\mathcal{T}\overline{C}_6},\chi_{\mathcal{T}S},k) = (0,0,3)$
(see Table \ref{table:z2_psg_table}), such a gauge transformation
can be chosen as
\begin{equation}
W = e^{-i\frac{\pi}{4} \sigma^2},
\end{equation}
which transforms the time reversal PSG according to
$W_{\mathcal{T}}(\bm{r}_\mu) = i \sigma^3\rightarrow
WW_{\mathcal{T}}(\bm{r}_\mu)W^\dag = i \sigma^1$. The mean-field
parameters therefore transform as
$$i(\mathrm{Re}a \sigma^1+ \mathrm{Im} b \sigma^2)\rightarrow
i(-\mathrm{Re}a \sigma^3+ \mathrm{Im} b \sigma^2),$$ i.e., the real
part of the pairing term is transformed into a hopping term, which
is consistent with Eq.~\eqref{u_gf_u1tr}. In this case, keeping only
the real part of the singlet pairing will recover a U(1) ansatz. 

In closing this subsection, we point out that a general mapping between $\mathbb{Z}_2$ and U(1) pyrochlore PSG classes is still lacking. Understanding such a relation will be important in mapping out phase diagrams containing various spin liquids and magnetic orders.

\subsection{The non-projective U(1) class: topological insulator\label{subsec:TI}}

Let us examine the topological properties of the U(1) PSG class $0$\---$(00)$\---$(00)$: this is the ``trivial'' class in which symmetries are realized linearly (i.e., nonprojectively). This symmetry structure also applies to physical electrons instead of spinons:  it can describe ordinary, non-fractionalized itinerant electrons on the pyrochlore lattice. Such systems have been intensely studied in the context of pyrochlore iridates \cite{doi:10.1143/JPSJ.80.044708,PhysRevLett.103.206805,PhysRevB.87.155101}. There, the most striking prediction from theory is the existence of a topological insulator phase, which occupies a finite volume in the phase space spanned by spin-orbit couplings up to NNN \cite{doi:10.1143/JPSJ.80.044708,PhysRevLett.103.206805,PhysRevB.87.155101}.

The most complete of these prior works \cite{doi:10.1143/JPSJ.80.044708,PhysRevLett.103.206805,PhysRevB.87.155101} is Ref.~\cite{PhysRevB.87.155101}, which determined the general form of the Hamiltonian up to second neighbor hoppings. Here, we provide an explicit mapping between the parameters used there and those used in Table \ref{MFTparameters_u1}: there are in total two real independent parameters $(t_1,t_2)$ for the NN bonds and three real independent parameters $(t'_1,t'_2,t'_3)$ for the NNN bonds \cite{PhysRevB.87.155101}, which are related to our PSG results by $(t_1,t_2,t'_1,t'_2,t'_3)=(\mathrm{Im}a,\mathrm{Im}c,\mathrm{Im}A,\mathrm{Im}{B}+\mathrm{Im}{D},\mathrm{Im}{B}-\mathrm{Im}{D})$. This serves as a partial check of our classification and allows the results in Refs~\cite{doi:10.1143/JPSJ.80.044708,PhysRevLett.103.206805,PhysRevB.87.155101} to directly apply in the PSG context. 

Given the existence of a topological insulator phase in the ``trivial'' U(1) PSG class, it is reasonable to believe that other classes may also support nontrivial topological phases. Among them, it would be of specific interest to identify those that are protected by the \emph{projectiveness} of the symmetry and would appear only in systems with fractionalized degrees of freedom.

\subsection{Future directions}

The mean-field ans\"atze for the PSG classes listed in this work
provide abundant ground state candidates for model spin Hamiltonians
on the pyrochlore lattice. In the works reported so far, the
monopole flux state \cite{burnell2009monopole} has the lowest energy
as a variational mean-field ansatz for the pyrochlore Heisenberg
model. The monopole flux state does not belong to our classification
with the full pyrochlore lattice symmetry since it spontaneously
breaks lattice inversion. An interesting question is then whether
any of the fully symmetric states may be energetically favored by
the Heisenberg model, and if not, what is the physical reason for
the energy being lowered by spontaneous symmetry breaking. In this regard, it is interesting to study the PSG classification of chiral spin liquids on the pyrochlore lattice, in which certain space group symmetries are replaced by them composed with time reversal symmetries. Furthermore, since rare-earth pyrochlore materials are intrinsically
spin-orbit coupled, it is also natural to take our PSG ans\"atze as
variational states for the full spin-orbit coupled pyrochlore
exchange model \cite{ross2011quantum}. These questions are addressed
in an ongoing work that will be reported elsewhere.

One outcome of this work is the realization, in several PSG classes, of the nodal star U(1) spin
liquid, which represents a new family of pyrochlore U(1) spin
liquids beyond the known prototypes, whose low energy nodal structure is protected by the pyrochlore space group symmetries. We point out that the proof of the symmetry protected nature of the nodal lines also applies to several classes of chiral spin liquids, including the monopole flux state \cite{burnell2009monopole}. In the present work, our main focus has been on the spinon corrections to the gauge field. Subsequent questions \--- such as how the gauge field feeds back into the spinons and how the vertices receive corrections \--- have been
outside the scope of this work and require a more  involved
calculation. These calculations may reveal additional contributions
to the thermodynamic properties and will provide insights for
another important observable, the spin susceptibility. Even at the
non-interacting level, the nodal line spinons will lead to spectral
features that should be observable in, e.g., neutron scattering
experiments. For example, a broad low energy continuum should be
seen along {appropriate high-symmetry planes of the
Brillouin zone}. The observation of such signatures may serve as
direct evidence for a pyrochlore spin liquid state. We leave the
study of these aspects of the nodal star U(1) spin liquid to a
future work.

There are also several more broad directions to be explored. The pyrochlore PSG may be studied from the perspective of symmetry protected
crystalline insulators and symmetry enriched topological orders. Apart from the case of the non-projective U(1) ans\"atze mentioned in the preceding subsection, the topological aspects of the spinon bands have not been investigated in this work and deserve further study.  Another future direction is the stability of the gapless states beyond the mean field approximation.  While at the latter level, we have proven that the nodal line is symmetry protected, it is not clear if it remains robust against symmetry allowed spinon interactions. In this regard, it would be interesting to look for criteria that forbid a single gapped many-body ground state of the spinon Hamiltonian from appearing in the presence of spinon interactions and full pyrochlore symmetry, similar to the proposals of Ref.~\cite{PhysRevB.97.094422}. In the context of $\mathbb{Z}_2$
QSLs, a related theme would be to generalize the Lieb-Schultz-Mattis
theorem to the pyrochlore lattice and other geometrically
frustrated 3D lattice types. The screw symmetry, which is crucial to
the nodal star spin liquids, is an instance of nonsymmorphic
symmetry and, in this regard, it is compelling to connect our
results with recent works, such as Refs.~\cite{2015PNAS..11214551W,PhysRevB.101.224437} and especially Ref.~\cite{PhysRevX.8.011040} (considering that the classification result for the pyrochlore space group there is missing).

\begin{acknowledgments}
We acknowledge Yasir Iqbal, Federico Becca, Francesco Ferrari, and Mengxing Ye  for helpful discussions.
  C.L. and L.B. were supported by the DOE, Office of
  Science, Basic Energy Sciences under Award No. DE-FG02-08ER46524.
  This research benefitted from the facilities of the Kavli Institute
  for Theoretical Physics, supported in part by the National Science
  Foundation under Grant No. NSF PHY-1748958.
  The work of G.B.H.~was supported by the U.S.~Department of Energy, Office of Science, National Quantum Information Science Research Centers, Quantum Science Center.

\end{acknowledgments}

\appendix

\section{\label{app:u1_SG}Solving U(1) PSG equations: space group part}

The space group part of the PSG equations are
\begin{subequations}\label{simpfiedgrprelations}
\begin{align}
&(G_{T_i}T_i)(G_{T_{i+1}}T_{i+1})(G_{T_i}T_i)^{-1} (G_{T_{i+1}}T_{i+1})^{-1}\in \text{IGG},\label{PSG123}\\
&(G_{\overline{C}_6} \overline{C}_6)^6 \in \text{IGG},\label{PSGc6}\\
&(G_SS)^2 (G_{T_3} T_3)^{-1} \in \text{IGG},\label{PSGs}\\
&(G_{\overline{C}_6}\overline{C}_6)(G_{T_i}T_i)(G_{\overline{C}_6}\overline{C}_6)^{-1} (G_{T_{i+1}}T_{i+1})\in \text{IGG},\label{PSG789}\\
&(G_SS)(G_{T_i}T_i)(G_SS)^{-1}(G_{T_3}T_3)^{-1} (G_{T_i}T_i)\in\text{IGG},\label{PSG1314}\\
&(G_SS)(G_{T_3}T_3)(G_S S)^{-1}(G_{T_3}T_3)^{-1} \in \text{IGG},\label{PSG15}\\
&{}[(G_{\overline{C}_6}\overline{C}_6)(G_SS)]^4 \in \text{IGG},\label{PSG18}\\
&{}[(G_{\overline{C}_6}\overline{C}_6)^3(G_SS)]^2 \in \text{IGG}.\label{PSGhashtag}
\end{align}
\end{subequations}
The corresponding SU(2) equations are
\begin{subequations}\label{chi}
\begin{align}
W_{T_i}(\bm{r}_\mu)W_{T_{i+1}}[T^{-1}_i(\bm{r}_\mu)]\cdot& \nonumber \\
W^{-1}_{T_i}[T^{-1}_{i+1}(\bm{r}_\mu)]W^{-1}_{T_{i+1}}(\bm{r}_\mu)& = e^{i\sigma^3\chi_i},\label{tttt}\\
W_{\overline{C}_6}(\bm{r}_\mu)W_{\overline{C}_6}[\overline{C}_6^{-1}(\bm{r}_\mu)]\cdot&\\
W_{\overline{C}_6}[\overline{C}_6^{-2}(\bm{r}_\mu)]W_{\overline{C}_6}[\overline{C}_6^{-3}(\bm{r}_\mu)]\cdot&\nonumber\\
 W_{\overline{C}_6}[\overline{C}_6^{-4}(\bm{r}_\mu)]W_{\overline{C}_6}[\overline{C}_6^{-5}(\bm{r}_\mu)]&= e^{i\sigma^3\chi_{\overline{C}_6}}\label{cccccc}\\
W_S(\bm{r}_\mu) W_S[S^{-1}(\bm{r}_\mu)]W^{-1}_{T_3}(\bm{r}_\mu) &= e^{i\sigma^3\chi_S},\label{sst}\\
W_{\overline{C}_6}(\bm{r}_\mu)W_{T_i}[\overline{C}^{-1}_6(\bm{r}_\mu)]\cdot&\nonumber\\
W^{-1}_{\overline{C}_6}[T_{i+1}(\bm{r}_\mu)]W_{T_{i+1}}[T_{i+1}(\bm{r}_\mu)]& = e^{i\sigma^3\chi_{\overline{C}_6T_i}},\label{ctct}\\
W_S(\bm{r}_\mu)W_{T_i}[S^{-1}(\bm{r}_\mu)]\cdot \nonumber\\
W^{-1}_S[T^{-1}_3 T_i(\bm{r}_\mu)]\cdot&\nonumber\\
W^{-1}_{T_3}[T_i(\bm{r}_\mu)] W_{T_i}[T_i(\bm{r}_\mu)] &= e^{i\sigma^3\chi_{ST_i}},\label{ststt}\\
W_S(\bm{r}_\mu)W_{T_3}[S^{-1}(\bm{r}_\mu)]\cdot&\nonumber\\
W^{-1}_S[T^{-1}_3(\bm{r}_\mu)]W^{-1}_{T_3}(\bm{r}_\mu)&= e^{i\sigma^3\chi_{ST_3}},\label{stst}\\
W_{\overline{C}_6}(\bm{r}_\mu)W_S[\overline{C}_6^{-1}(\bm{r}_\mu)]\cdot & \nonumber\\
W_{\overline{C}_6}[(\overline{C}_6 S)^{-1}(\bm{r}_\mu)]\cdot&\nonumber\\
W_S[(\overline{C}_6S \overline{C}_6)^{-1}(\bm{r}_\mu)]\cdot & \nonumber\\
W_{\overline{C}_6}[(\overline{C}_6S \overline{C}_6S)^{-1}(\bm{r}_\mu)]\cdot&\nonumber\\
W_S[(\overline{C}_6S \overline{C}_6S\overline{C}_6)^{-1}(\bm{r}_\mu)]\cdot&\nonumber\\
W_{\overline{C}_6}[(\overline{C}_6S \overline{C}_6S\overline{C}_6S)^{-1}(\bm{r}_\mu)]\cdot&\nonumber\\
W_S[(\overline{C}_6S \overline{C}_6S\overline{C}_6S\overline{C}_6)^{-1}(\bm{r}_\mu)] &= e^{i\sigma^3\chi_{\overline{C}_6S}},\label{cscscscs}\\
W_{\overline{C}_6}[\overline{C}_6^{-2}(\bm{r}_\mu)]W_S[\overline{C}_6^{-3}(\bm{r}_\mu)]\cdot&\nonumber\\
W_{\overline{C}_6}[(\overline{C}_6^3S)^{-1}(\bm{r}_\mu)]\cdot & \nonumber\\
W_{\overline{C}_6}[(\overline{C}_6^3S\overline{C}_6)^{-1}(\bm{r}_\mu)]\cdot& \nonumber\\
W_{\overline{C}_6}[(\overline{C}_6^3S\overline{C}^2_6)^{-1}(\bm{r}_\mu)]W_S[S(\bm{r}_\mu)]&= e^{i\sigma^3\chi_{S\overline{C}_6}},\label{cccscccs}
 \end{align}
 \end{subequations}
 where all the $\chi \in [0,2 \pi)$ for U(1) IGG, and $\chi \in \{0,\pi\}$ for $\mathbb{Z}_2$ IGG.

The general form for $W_{\mathcal{O}}(\bm{r})$ is $W_{\mathcal{O}}(\bm{r}) = (i \sigma^1)^{w_{\mathcal{O}}}e^{i\sigma^3\phi_{\mathcal{O}}(\bm{r})}$, where $w_{\mathcal{O}} = 0$ or $1$, $\mathcal{O}\in \{T_1,T_2,T_3,\overline{C}_6,S\}$. From Eq.~\eqref{ststt} we see we must have $w_{T_3}=0$, further from Eq.~\eqref{ctct} we have $w_{T_1}=w_{T_2}=0$. Therefore Eq.~\eqref{tttt} becomes a pure phase equation
\begin{equation}
\phi_{T_i}(\bm{r}_\mu)+\phi_{T_{i+1}}[T_i^{-1}(\bm{r}_\mu)]-\phi_{T_i}[T_{i+1}^{-1}(\bm{r}_\mu)]-\phi_{T_{i+1}}(\bm{r}_\mu)=\chi_i,
\end{equation}
using gauge freedom to set $\phi_{T_1}(\bm{r}_\mu)=0$, $\phi_{T_2}(r_1,0,0)_\mu=0$, and $\phi_{T_3}(r_1,r_2,0)=0$, we have
\begin{equation}\label{sol123}
\phi_{T_1}(\bm{r}_\mu)=0,\ \ \ \phi_{T_2}(\bm{r}_\mu) = -\chi_1 r_1,\ \ \
\phi_{T_3}(\bm{r}_\mu) = \chi_3r_1 - \chi_2 r_2.
\end{equation}
Eq.~\eqref{ctct} gives
\begin{equation}
\begin{aligned}
(-1)^{w_{\overline{C}_6}}\Big(\phi_{\overline{C}_6}(\bm{r}_\mu) + \phi_{T_i}[\overline{C}^{-1}_6(\bm{r}_\mu)]-\phi_{\overline{C}_6}[T_{i+1}&(\bm{r}_\mu)]\Big)\\
+\phi_{T_{i+1}}[T_{i+1}(\bm{r}_\mu)] &= \chi_{\overline{C}_6T_i},
\end{aligned}
\end{equation}
consistency condition
\begin{equation}\label{consisc}
\begin{aligned}
&\Delta_i \phi_{\overline{C}_6}(\bm{r}_\mu)+\Delta_{i+1}\phi_{\overline{C}_6}[T_i^{-1}(\bm{r}_\mu)]\\
&=\Delta_{i+1} \phi_{\overline{C}_6}(\bm{r}_\mu)+\Delta_i \phi_{\overline{C}_6}[T_{i+1}^{-1}(\bm{r}_\mu)], \quad i=1,2,3
\end{aligned}
\end{equation}
where we defined $\Delta_i \phi(\bm{r}_\mu) = \phi(\bm{r}_\mu)-\phi[T_i^{-1}(\bm{r}_\mu)]$, requires
\begin{equation}
\begin{aligned}
(-1)^{w_{\overline{C}_6}} \chi_1 = \chi_3,\ \ \ &-\chi_1+(-1)^{w_{\overline{C}_6}}\chi_2 = 0,\ \ \ \\
\chi_2 &= (-1)^{w_{\overline{C}_6}}\chi_3,
\end{aligned}
\end{equation}
which means that, if $w_{\overline{C}_6} = 0$, then $\chi_1=\chi_2 = \chi_3$ and no quantization condition is imposed on $\chi_{1,2,3}$; if $w_{\overline{C}_6} = 1$, then $\chi_1=\chi_2 = \chi_3 = 0$ or $\pi$. Combining the two cases we can write
\begin{equation}\label{phic6}
\begin{aligned}
\phi_{\overline{C}_6}(\bm{r}_\mu)
=& -r_1(r_2-r_3)\chi_1\\
& -(\delta_{\mu,2}-\delta_{\mu,3})\chi_1r_1+\delta_{\mu,2}\chi_1r_3\\
&-(-1)^{w_{\overline{C}_6}}(\chi_{\overline{C}_6T_3}r_1+\chi_{\overline{C}_6T_1}r_2+\chi_{\overline{C}_6T_2}r_3)\\
& + \rho_\mu,
\end{aligned}
\end{equation}
where we abbreviated $\rho_\mu \equiv \phi_{\overline{C}_6}(0,0,0)_\mu$. Plug the expression \eqref{phic6} in Eq.~\eqref{cccccc}, we get the condition
\begin{itemize}
\item When $w_{\overline{C}_6}=0$,
\begin{equation}
\begin{aligned}
6 \rho_0 &= 2(\rho_1+\rho_2+\rho_3) +\chi_{\overline{C}_6T_1}+ \chi_{\overline{C}_6T_2}+ \chi_{\overline{C}_6T_3}\\
 &= \chi_{\overline{C}_6};
 \end{aligned}
\end{equation}
\item when $w_{\overline{C}_6}=1$,
\begin{equation}
\chi_{\overline{C}_6} = \chi_{\overline{C}_6T_1}+ \chi_{\overline{C}_6T_2}+ \chi_{\overline{C}_6T_3}=0.
\end{equation}
\end{itemize}
Plug the result \eqref{sol123} in Eqs.~\eqref{ststt} and Eq.~\eqref{stst}, we get
\begin{subequations}
\begin{align}
\Delta_1 \phi_S(\bm{r}_\mu)&=
\chi_1(r_1-r_2+\delta_{\mu,1}-\delta_{\mu,2})\nonumber\\
&\quad+(-1)^{w_S}(-\chi_{ST_1}+\chi_{ST_3}),\\
\Delta_2 \phi_S(\bm{r}_\mu)&=
\chi_1(2r_1-(-1)^{w_S}r_1-r_2+2\delta_{\mu,1}-\delta_{\mu,2})\nonumber\\
&\quad+(-1)^{s_W}(-\chi_{ST_2}+\chi_{ST_3}),\\
\Delta_3 \phi_S(\bm{r}_\mu)&=\chi_1((1+(-1)^{w_S})(r_1-r_2)+\delta_{\mu,1}-\delta_{\mu,2})\nonumber\\
&\quad+(-1)^{w_S}\chi_{ST_3}.
\end{align}
\end{subequations}
Consistency conditions for $\phi_S$ similar to \eqref {consisc} give
\begin{equation}
\begin{aligned}
(-2+(-1)^{w_S})\chi_1 = \chi_1,\quad (1+(-1)^{w_S})\chi_1 = 0,\\
\quad 0 = -(1+(-1)^{w_S})\chi_1,\qquad\quad\quad
\end{aligned}
\end{equation}
therefore when $w_S=0$, $\chi_1=0$ or $\pi$; when $w_S=1$, $\chi_1 = 0$, $\frac{\pi}{2}$, $\pi$ or $\frac{3\pi}{2}$. And
\begin{equation}\label{phis}
\begin{aligned}
\phi_S (\vec{r}_\mu)
&=\left(\frac{(r_1+1)r_1}{2}-\frac{(r_2+1)r_2}{2}-r_1r_2\right)\chi_1\\
&+\left[(\delta_{\mu,1}-\delta_{\mu,2})\chi_1+(-1)^{w_S}(\chi_{ST_3}-\chi_{ST_1})\right]r_1\\
&+\left[(2\delta_{\mu,1}-\delta_{\mu,2})\chi_1+(-1)^{w_S}(\chi_{ST_3}-\chi_{ST_2})\right]r_2\\
&+\left[(\delta_{\mu,1}-\delta_{\mu,2})\chi_1+(-1)^{w_S}\chi_{ST_3}\right]r_3\\
&+\theta_\mu,
\end{aligned}
\end{equation}
where we abbreviated $\theta_\mu=\phi_S(0,0,0)_\mu$. Plug the result \eqref{phis} and \eqref{sol123} into Eq.~\eqref{sst}, we get
\begin{itemize}
\item When $w_S = 0$,
\begin{subequations}
\begin{align}
\chi_{ST_3}&=0,\\
\theta_0+\theta_3 &= 2\theta_1-\chi_1+\chi_{ST_1}\nonumber\\
&=2\theta_2+\chi_1+\chi_{ST_2}\nonumber\\
&=\chi_S;
\end{align}
\end{subequations}
\item
When $w_S = 1$,
\begin{subequations}
\begin{align}
2\chi_1+2\chi_{ST_1}-\chi_{ST_3} &= 2\chi_1-2\chi_{ST_2} +\chi_{ST_3}\nonumber\\
 &= 0,\\
-\theta_0+\theta_3+\chi_{ST_3} &= -\chi_1-\chi_{ST_1}+\chi_{ST_3}\nonumber\\
&=\chi_1-\chi_{ST_2}+\chi_{ST_3}\nonumber\\
&=\theta_0-\theta_3 \nonumber\\
&= \chi_S.
\end{align}
\end{subequations}
\end{itemize}
At this point we are left with two equations, \eqref{cscscscs} and \eqref{cccscccs}. Depending on the $\mathbb{Z}_2$ value of $w_{\overline{C}_6}$ and $w_S$ we have the following four cases:
\begin{itemize}
\item $(w_{\overline{C}_6},w_S)=(0,0)$:
Eq.~\eqref{cscscscs} gives
\begin{equation}\label{A2gA2hbegin}
\chi_{\overline{C}_6T_1}+\chi_{\overline{C}_6T_2}+\chi_{\overline{C}_6T_3}+\sum_{\mu=0}^3(\rho_\mu+\theta_\mu) = \chi_{\overline{C}_6S},
\end{equation}
and Eq.~\eqref{cccscccs} gives
\begin{subequations}
\begin{align}
&2\chi_{ST_1}-3\chi_{\overline{C}_6T_1}+3\chi_{\overline{C}_6T_2}-\chi_{\overline{C}_6T_3}\nonumber\\
&\quad= 2\chi_{ST_2}-3\chi_{\overline{C}_6T_1}-\chi_{\overline{C}_6T_2}+3\chi_{\overline{C}_6T_3}\nonumber\\
&\quad=0,\\
&\theta_0+\theta_3+3\rho_0+\rho_1+\rho_2+\rho_3+\chi_{\overline{C}_6T_2}+\chi_{\overline{C}_6T_3}\nonumber\\
&\quad=2\theta_1+2(\rho_1+\rho_2+\rho_3)+2\chi_{\overline{C}_6T_1}+2\chi_{\overline{C}_6T_3}\nonumber\\
&\quad=2\theta_2+2(\rho_1+\rho_2+\rho_3)+2\chi_{\overline{C}_6T_1}+2\chi_{\overline{C}_6T_2}\nonumber\\
&\quad=\chi_{S\overline{C}_6}.
\end{align}
\end{subequations}
\item $(w_{\overline{C}_6},w_S)=(0,1)$:
Eq.~\eqref{cscscscs} gives
\begin{subequations}
\begin{align}
&-\theta_0-\theta_1+\theta_2+\theta_3+\rho_0-\rho_1+\rho_2-\rho_3\nonumber\\
&\quad\quad\qquad +\chi_{\overline{C}_6T_1}-\chi_{\overline{C}_6T_2}-\chi_{\overline{C}_6T_3}=\chi_{\overline{C}_6S},\\
&\theta_0+\theta_1-\theta_2-\theta_3-\rho_0+\rho_1-\rho_2+\rho_3\nonumber\\
&\quad\quad\qquad-2(\chi_{ST_1}-\chi_{ST_2}+\chi_{ST_3})\nonumber\\
&\quad\quad\qquad+\chi_{\overline{C}_6T_1}-\chi_{\overline{C}_6T_2}-\chi_{\overline{C}_6T_3}=\chi_{\overline{C}_6S},\\
&2(-\chi_{ST_1}+\chi_{ST_2}-\chi_{ST_3})\nonumber\\
&\quad\quad\quad +2(\chi_{\overline{C}_6T_1}-\chi_{\overline{C}_6T_2}-\chi_{\overline{C}_6T_3})=0,
\end{align}
\end{subequations}
where the second equation can be obtained from the first and the third.
Eq.~\eqref{cccscccs} gives
\begin{subequations}
\begin{align}
&-\chi_{ST_3}+\chi_{\overline{C}_6T_1}-\chi_{\overline{C}_6T_2}-\chi_{\overline{C}_6T_3}=0,\\
&-\theta_0+\theta_3+3\rho_0-\rho_1-\rho_2-\rho_3-\chi_{\overline{C}_6T_2}-\chi_{\overline{C}_6T_3}\nonumber\\
&\qquad=-\chi_{ST_3}+\chi_{\overline{C}_6T_1}-\chi_{\overline{C}_6T_2}-\chi_{\overline{C}_6T_3}\nonumber\\
&\qquad=\theta_0-\theta_3-3\rho_0+\rho_1+\rho_2+\rho_3+\chi_{\overline{C}_6T_2}+\chi_{\overline{C}_6T_3}\nonumber\\
&\qquad=\chi_{S\overline{C}_6},
\end{align}
\end{subequations}

\item $(w_{\overline{C}_6},w_S)=(1,0)$:
Eq.~\eqref{cscscscs} gives
\begin{subequations}
\begin{align}
&2(\chi_{ST_1}-\chi_{ST_2}+\chi_{ST_3})\nonumber\\
&\qquad +2(\chi_{\overline{C}_6T_1}-\chi_{\overline{C}_6T_2}-\chi_{\overline{C}_6T_3})=0,\\
&-\theta_0-\theta_1+\theta_2+\theta_3-\rho_0+\rho_1-\rho_2+\rho_3\nonumber\\
&\qquad\qquad\qquad +\chi_{\overline{C}_6T_1}-\chi_{\overline{C}_6T_2}-\chi_{\overline{C}_6T_3}\nonumber\\
&\quad =\theta_0+\theta_1-\theta_2-\theta_3+\rho_0-\rho_1+\rho_2-\rho_3\nonumber\\
&\qquad \qquad\qquad -\chi_{\overline{C}_6T_1}+\chi_{\overline{C}_6T_2}+\chi_{\overline{C}_6T_3}\nonumber\\
&\quad =\chi_{\overline{C}_6S},
\end{align}
\end{subequations}
this gives $2\chi_{\overline{C}_6S} = 0$ so $\chi_{\overline{C}_6S} = 0$ or $\pi$. Eq.~\eqref{cccscccs} gives
\begin{subequations}
\begin{align}
&\chi_{ST_3}-\chi_{\overline{C}_6T_1}-\chi_{\overline{C}_6T_2}-\chi_{\overline{C}_6T_3}=0,\\
&-\theta_0+\theta_3-\rho_0+\rho_1-\rho_2+\rho_3-\chi_{\overline{C}_6T_2}-\chi_{\overline{C}_6T_3}\nonumber\\
&\quad =0=\chi_{S\overline{C}_6}.
\end{align}
\end{subequations}

\item $(w_{\overline{C}_6},w_S)=(1,1)$:
Eq.~\eqref{cscscscs} gives
\begin{equation}
\sum\limits_{\mu=0}^3(\theta_\mu-\rho_\mu) = \chi_{\overline{C}_6S},
\end{equation}
and Eq.~\eqref{cccscccs} gives
\begin{subequations}\label{A2gA2hend}
\begin{align}
&-2\chi_{ST_1}+\chi_{ST_3}=-2\chi_{ST_2}+\chi_{ST_3} = 0,\\
&\theta_0+\theta_3-\rho_0-\rho_1+\rho_2-\rho_3+\chi_{\overline{C}_6T_2}+\chi_{\overline{C}_6T_3}\nonumber\\
&\quad\qquad =2\theta_1-2\rho_1-2\rho_2+2\rho_3-2\chi_{\overline{C}_6T_2}\nonumber\\
&\quad\qquad =2\theta_2+2\rho_1-2\rho_2-2\rho_3-2\chi_{\overline{C}_6T_3}\nonumber\\
&\quad\qquad =\chi_{S\overline{C}_6}.
\end{align}
\end{subequations}

\end{itemize}

We now choose a gauge to fix some of the phases. This gauge applies to all four classes above. First, under gauge transformation $W(\bm{r}_\mu)=1$ for $\mu=0$, $W(\bm{r}_i) = e^{i \sigma^3\psi_i r_i }$ for $i=1,2,3$, where $\psi_i$ is any constant phase, the values of $\chi_{\overline{C}_6T_1}$, $\chi_{\overline{C}_6T_2}$ and $\chi_{ST_2}$ change. This means they are ineffective in labeling the PSG classes, and by properly choosing $\psi_i$ we can set them to be
\begin{equation}\label{eq:gaugefx1}
\chi_{\overline{C}_6T_1}=0,\ \ \ \chi_{\overline{C}_6T_2}=0, \ \ \ \chi_{ST_2}=0.
\end{equation}
Then, we can use the IGG freedom (the freedom of choosing a global U(1) phase) to set
\begin{equation}\label{eq:gaugefx2}
\rho_0=\theta_2=0\ \ \Rightarrow \ \ \chi_{\overline{C}_6} = 0.
\end{equation}
Finally, using the ``sublattice'' gauge transformation
\begin{equation}\label{eq:gaugetr}
W(\bm{r}_\mu) = e^{i \phi_\mu}
\end{equation}
where $\phi_\mu$ is any constant phase, $W_{\overline{C}_6}(\bm{r}_\mu)$ for $\mu=1,2$ and $W_S(\bm{r}_\mu)$ for $\mu=0$ will transform as
\begin{subequations}
\begin{align}
(i\sigma^1)^{w_{\overline{C}_6}} e^{i \phi_{\overline{C}_6}(\bm{r}_1)\sigma^3}&\rightarrow
e^{i\phi_1\sigma^3} (i\sigma_1)^{w_{\overline{C}_6}} e^{i \phi_{\overline{C}_6}(\bm{r}_2)\sigma^3}
e^{-i\phi_3 \sigma^3}\nonumber\\
&=  (i\sigma^1)^{w_{\overline{C}_6}} e^{i ((-1)^{w_{\overline{C}_6}}\phi_1+\phi_{\overline{C}_6}(\bm{r}_1)-\phi_3)\sigma^3},\\
(i\sigma^1)^{w_{\overline{C}_6}} e^{i \phi_{\overline{C}_6}(\bm{r}_1)\sigma^3}&\rightarrow
e^{i\chi_1\sigma^3} (i\sigma^1)^{w_{\overline{C}_6}} e^{i \phi_{\overline{C}_6}(\bm{r}_2)\sigma^3} e^{-i\phi_1\sigma^3}\nonumber\\
 &=  (i\sigma_1)^{w_{\overline{C}_6}} e^{i ((-1)^{w_{\overline{C}_6}}\phi_2+\phi_{\overline{C}_6}(\bm{r}_2)-\phi_1)\sigma^3},\\
(i\sigma_1)^{w_S} e^{i \phi_S(\bm{r}_0)\sigma^3}&\rightarrow
e^{i\phi_0\sigma^3} (i\sigma^1)^{w_S} e^{i \phi_S(\bm{r}_0)\sigma^3}
e^{-i\phi_3\sigma^3}\nonumber\\
&=  (i\sigma^1)^{w_S} e^{i ((-1)^{w_S}\phi_0+\phi_S(\bm{r}_0)-\phi_3)\sigma^3}.
\end{align}
\end{subequations}
Then, by properly choosing $\phi_{0,1,2,3}$ we are able to set
\begin{equation}\label{eq:gaugefx3}
\rho_1=\rho_2=\theta_0=0.
\end{equation}
Eqs.~\eqref{eq:gaugefx1}, \eqref{eq:gaugefx2} and \eqref{eq:gaugefx3} significantly simply Eqs.~\eqref{A2gA2hbegin}\---\eqref{A2gA2hend}, and furthermore allow them to be solved without any ambiguity. The final result is presented in Table \ref{table:u1_psg_table}.

\section{\label{app:u1_TR} Solving U(1) PSG equations: adding time reversal}

There is one complication when considering time reversal $\mathcal{T}$. Acting on spins, it is the anti-unitary operator
\begin{equation}
\mathcal{T}\colon \hat{\bm{S}}\rightarrow i\sigma^y \mathcal{K} \hat{\bm{S}} (-i \sigma^y \mathcal{K}),
\end{equation}
where $\mathcal{K}$ is the complex conjugation operator that complex conjugates everything on its right. This induces an action on $\Psi$ as
\begin{equation}
\mathcal{T}\colon \Psi\rightarrow (i\sigma^y) \mathcal{K} \Psi.
\end{equation}
Introducing a gauge field associated to $\mathcal{T}$, we have
\begin{equation}\label{originaldeftimereversal}
G_{\mathcal{T}}\circ \mathcal{T}\colon
\Psi(\bm{r}_\mu)\rightarrow  (i\sigma^y)\mathcal{K}\Psi(\bm{r}_\mu)W_{\mathcal{T}}(\bm{r}_\mu).
\end{equation}

Now we apply this to a mean-field bond:
\begin{equation}
\begin{aligned}
&G_{\mathcal{T}}\circ \mathcal{T}\colon H^{\alpha}_{\bm{r}_\mu,\bm{r}'_\nu}
\rightarrow \\
& \mathrm{Tr}\left[ (\sigma^\alpha)^*
(i\sigma^y)\mathcal{K}\Psi(\bm{r}_\mu)W_{\mathcal{T}}(\bm{r}_\mu) u^\alpha_{\bm{r}_\mu,\bm{r}'_\nu} W^\dag_{\mathcal{T}}(\bm{r}'_\nu) \Psi^\dag(\bm{r}'_\nu)(-i \sigma^y) \mathcal{K}\right]\\
 &=\mathrm{Tr}\left[ (\sigma^\alpha)^* \sigma^y \Psi(\bm{r}_\mu) W^*_{\mathcal{T}}(\bm{r}_\mu) \left(u^\alpha_{\bm{r}_\mu,\bm{r}'_\nu}\right)^* W^T_{\mathcal{T}}(\bm{r}'_\nu) \Psi^\dag(\bm{r}'_\nu) \sigma^y\right],
\end{aligned}
\end{equation}
where we have noted that both the bond $u^\alpha_{\bm{r}_\mu,\bm{r}'_\nu}$ and $\sigma^\alpha$ are complex-conjugated by $\mathcal{K}$. However we can get rid of this complex conjugation by defining a new gauge field $\widetilde{W}_{\mathcal{T}}(\bm{r}_\mu)$ by
\begin{equation}\label{redefinedww}
W_{\mathcal{T}}(\bm{r}_\mu) \equiv i \sigma^2  \widetilde{W}_{\mathcal{T}}(\bm{r}_\mu).
\end{equation}
Using the identity $\sigma^2 u \sigma^2 = u^*$ for $u \in \mathrm{SU}(2)$, we have $W^*_{\mathcal{T}}(\bm{r}_\mu) = \widetilde{W}_{\mathcal{T}}(\bm{r}_\mu) i\sigma^2$ and
\begin{equation}
\begin{aligned}
\sigma^2 \left(u^0_{\bm{r}_\mu,\bm{r}'_\nu}\right)^* \sigma^2
&=-u^0_{\bm{r}_\mu,\bm{r}'_\nu},\\
\sigma^2 \left(u^{(i)}_{\bm{r}_\mu,\bm{r}'_\nu}\right)^* \sigma^2
&=u^{(i)}_{\bm{r}_\mu,\bm{r}'_\nu},\quad i=x,y,z,\\
\end{aligned}
\end{equation}
and
\begin{equation}
\begin{aligned}
\sigma^2 (\sigma^0)^*\sigma^2 &= \sigma^0,\\
\sigma^2 (\sigma^i)^*\sigma^2 &= -\sigma^i,\quad i=x,y,z,\\
\end{aligned}
\end{equation}
we see that the form \eqref{bondsss} is specially designed so that
\begin{equation}
\begin{aligned}
\label{TTTT}
&G_{\mathcal{T}}\circ \mathcal{T}\colon
H^\alpha_{\bm{r}_\mu,\bm{r}'_\nu}=\mathrm{Tr}\left[\sigma^\alpha \Psi_{\bm{r}_\mu} u^\alpha_{\bm{r}_\mu,\bm{r}'_\nu} \Psi^\dag_{\bm{r}'_\nu}\right]\\
&\rightarrow
-\mathrm{Tr}\left[\sigma^\alpha \Psi_{\bm{r}_\mu} \widetilde{W}_{\mathcal{T}}(\bm{r}_\mu)u^\alpha_{\bm{r}_\mu,\bm{r}'_\nu}  \widetilde{W}_{\mathcal{T}}^\dag(\bm{r}'_\nu)\Psi^\dag_{\bm{r}'_\nu}\right].
\end{aligned}
\end{equation}
Therefore, with the redefined gauge $\widetilde{W}_{\mathcal{T}}$, $\widetilde{\mathcal{T}}=G_{\mathcal{T}}\circ \mathcal{T}$ can be regarded as a unitary operation with an additional sign flip for the mean-field parameters (this sign flip keeps track of the anti-unitarity of $\mathcal{T}$).

For the rest of the appendices and in the main text, we will remove the ``$\widetilde{\phantom{..}}$'' in $\widetilde{W}_{\mathcal{T}}$ and  call it $W_{\mathcal{T}}$ for simplicity.  It is then easy to see that in a time reversal symmetric ansatz Eq.~\eqref{TTTT} leads to Eq.~\eqref{guguchow}.

The SU(2) equations associated with Eqs.~\eqref{4j} and \eqref{4i} are
\begin{subequations}\label{PSGtimet}
\begin{align}
&{W}_{\mathcal{T}}(\bm{r}_\mu) W_{T_i}(\bm{r}_\mu){W}_{\mathcal{T}}^{-1}[T_i^{-1}(\bm{r}_\mu)]W_{T_i}^{-1}(\bm{r}_\mu) =e^{i\sigma^3\chi_{\mathcal{T}T_i}},\label{eq:timet}\\
&{W}_{\mathcal{T}}(\bm{r}_\mu) W_{\overline{C}_6}(\bm{r}_\mu){W}_{\mathcal{T}}^{-1}[\overline{C}_6^{-1}(\bm{r}_\mu)]W_{\overline{C}_6}^{-1}(\bm{r}_\mu) =e^{i\sigma^3\chi_{\mathcal{T}\overline{C}_6}},\label{eq:timec}\\
&{W}_{\mathcal{T}}(\bm{r}_\mu) W_S(\bm{r}_\mu){W}_{\mathcal{T}}^{-1}[S^{-1}(\bm{r}_\mu)]W_S^{-1}(\bm{r}_\mu) =e^{i\sigma^3\chi_{\mathcal{T}S}},\label{eq:times}\\
&W^2_{\mathcal{T}}(\bm{r}_\mu) = e^{i \sigma^3 \chi_{\mathcal{T}}},\label{eq:timestimes}
\end{align}
\end{subequations}
where all the $\chi \in [0,2\pi)$ for U(1) IGG, and $\chi \in \{0,\pi\}$ for $\mathbb{Z}_2$ IGG.

The above analysis explains the general strategy of treating the projective time reversal operation. Below we specialize to case of a U(1) gauge group and solve the corresponding PSG equations \eqref{PSGtimet}. The general form of $W_{\mathcal{T}}$ is
\begin{equation}
W_{\mathcal{T}}(\bm{r}_\mu) = (i \sigma^1)^{w_{\mathcal{T}}} e^{i \phi_{\mathcal{T}}(\bm{r}_\mu) \sigma^3},
\end{equation}
where $w_{\mathcal{T}} = 0$ or $1$. We now discuss these two cases separately.

When $w_{\mathcal{T}} = 0$, Eq.~\eqref{eq:timet} gives
\begin{equation}\label{eq:trttt}
\phi_{\mathcal{T}}(\bm{r}_\mu) = \phi_{\mathcal{T}}(\bm{0}_\mu) + \sum_{i=1}^3 \chi_{\mathcal{T}T_i} r_i.
\end{equation}
Then look at Eq.~\eqref{eq:timestimes}. We can use the IGG freedom to set $\chi_{\mathcal{T}} = 0$. This requires
\begin{equation}\label{eq:trttt1}
2\phi_{\mathcal{T}}(\bm{0}_\mu) = 2 \chi_{\mathcal{T}T_i}=0,\quad \mu=0,1,2,3\text{ and } i=1,2,3.
\end{equation}
Plug the form \eqref{eq:trttt} in Eqs.~\eqref{eq:timec} and \eqref{eq:times}, we get
\begin{subequations}
\begin{align}
&(-1)^{w_{\overline{C}_6}}\chi_{\mathcal{T}T_3}+\chi_{\mathcal{T}T_1}\nonumber\\
&=
(-1)^{w_{\overline{C}_6}}\chi_{\mathcal{T}T_1}+\chi_{\mathcal{T}T_2}\nonumber\\
&=
(-1)^{w_{\overline{C}_6}}\chi_{\mathcal{T}T_2}+\chi_{\mathcal{T}T_3}
=0,\label{eq:u1tt1}\\
&\phi_{\mathcal{T}}(\bm{0}_0)((-1)^{w_{\overline{C}_6}}-1)\nonumber\\
&=(-1)^{w_{\overline{C}_6}}\phi_{\mathcal{T}}(\bm{0}_3)-\phi_{\mathcal{T}}(\bm{0}_1)\nonumber\\
&=(-1)^{w_{\overline{C}_6}}\phi_{\mathcal{T}}(\bm{0}_1)-\phi_{\mathcal{T}}(\bm{0}_2)\nonumber\\
&=(-1)^{w_{\overline{C}_6}}\phi_{\mathcal{T}}(\bm{0}_2)-\phi_{\mathcal{T}}(\bm{0}_3)\nonumber\\
&=-\chi_{\mathcal{T}\overline{C}_6},\label{eq:u1tt2}\\
&-\chi_{\mathcal{T}T_1}((-1)^{w_S}+1)+(-1)^{w_S}\chi_{\mathcal{T}T_3}\nonumber\\
&=
-\chi_{\mathcal{T}T_2}((-1)^{w_S}+1)+(-1)^{w_S}\chi_{\mathcal{T}T_3}\nonumber\\
&=
\chi_{\mathcal{T}T_3}((-1)^{w_S}-1)=0,\label{eq:u1tt3}\\
&\phi_{\mathcal{T}}(\bm{0}_3)(-1)^{w_S}-\phi_{\mathcal{T}}(\bm{0}_0)
-\chi_{\mathcal{T}T_3}(-1)^{w_S}\nonumber \\
&=
\phi_{\mathcal{T}}(\bm{0}_1)((-1)^{w_S}-1)-\chi_{\mathcal{T}T_1}(-1)^{w_S}\nonumber\\
&=
\phi_{\mathcal{T}}(\bm{0}_1)((-1)^{w_S}-1)-\chi_{\mathcal{T}T_1}(-1)^{w_S}\nonumber\\
&=
\phi_{\mathcal{T}}(\bm{0}_0)(-1)^{w_S}-\phi_{\mathcal{T}}(\bm{0}_3)\nonumber\\
&= - \chi_{\mathcal{T}S},\label{eq:u1tt4}
\end{align}
\end{subequations}
Eqs.~\eqref{eq:u1tt1} and \eqref{eq:u1tt3} only give zero solution $\chi_{\mathcal{T}T_1}=\chi_{\mathcal{T}T_2}=\chi_{\mathcal{T}T_3}=0$ for any combinations of $w_{\overline{C}_6}=0,1$ and $w_S=0,1$. Then, Eqs.~\eqref{eq:u1tt2} and \eqref{eq:u1tt4} have the only solution $\phi_{\mathcal{T}}(\bm{0}_0)=\phi_{\mathcal{T}}(\bm{0}_1)=\phi_{\mathcal{T}}(\bm{0}_2)=\phi_{\mathcal{T}}(\bm{0}_3)$; and by further using the IGG freedom of time reversal we can set this phase to zero. Therefore the final solution for $w_{\mathcal{T}}=0$ is $W_{\mathcal{T}}(\bm{r}_\mu)=1$. However, this implies $u^\alpha_{\bm{r}_\mu,\bm{r}'_\nu} = - u^\alpha_{\bm{r}_\mu,\bm{r}'_\nu}$ according to Eq.~\eqref{guguchow}, which gives vanishing mean-field  ans\"atze. This indicates that $w_{\mathcal{T}}=0$ is not physical.

Next consider the case $w_{\mathcal{T}}=1$. We again solve Eq.~\eqref{eq:timet} first:
consistency condition requires $-2\chi_1 = 0$, therefore $\chi_1 = 0$ or $\pi$, and we have
\begin{equation}
\phi_{\mathcal{T}}(\bm{r}_\mu) = \phi_{\mathcal{T}}(\bm{0}_\mu) - \sum_{i=1}^3 \chi_{\mathcal{T}T_i} r_i.
\end{equation}
Then we solve Eqs.~\eqref{eq:timec} and \eqref{eq:times}.
In all four cases given by $w_{\overline{C}_6}=0$ or $1$ and $w_S = 0$ or $1$, we have $2\chi_{\overline{C}_6T_1}=2\chi_{\overline{C}_6T_2}=2\chi_{\overline{C}_6T_3}=2\chi_{ST_1}=2\chi_{ST_2} = 2\chi_{ST_3} = 0$. Then, since $\chi_1 = 0$ or $\pi$ and $\chi_{ST_1}=0$ or $2\pi/3$ are the only possible values in the space group PSG solution,  we see that
\begin{equation}
\chi_{\overline{C}_6T_1}=\chi_{\overline{C}_6T_2}=\chi_{\overline{C}_6T_3}=\chi_{ST_1}=\chi_{ST_2} = \chi_{ST_3} = 0.
\end{equation}
Furthermore, analogous to the $w_{\mathcal{T}}=0$ case, we obtain
$\chi_{\mathcal{T}T_1}=\chi_{\mathcal{T}T_2}=\chi_{\mathcal{T}T_3}=0$ and $\phi_{\mathcal{T}}(\bm{0}_0)=\phi_{\mathcal{T}}(\bm{0}_1)=\phi_{\mathcal{T}}(\bm{0}_2)=\phi_{\mathcal{T}}(\bm{0}_3)$. Then we can always set this phase to zero using the IGG freedom of time reversal. This further implies
\begin{equation}\label{tstc}
\chi_{\mathcal{T}\overline{C}_6} = \chi_{\mathcal{T}S}=0.
\end{equation}
In conclusion, in the U(1) PSG case, adding time reversal symmetry does not introduce additional PSG parameters but further restricts $\chi_1 = 0$ or $\pi$, and $\chi_{ST_1} = 0$. The time reversal gauge part has the form (in our choice of gauge fixing)
\begin{equation}
W_{\mathcal{T}} (\bm{r}_\mu) = i \sigma^1.
\end{equation}

The final result is presented in Table \ref{table:u1_psg_table}.

\section{\label{app:z2_SG}Solving $\mathbb{Z}_2$ PSG equations: space group part}

In solving the $\mathbb{Z}_2$ PSG equations \eqref{chi}, all the $\chi \in \{0,\pi\}$, therefore we introduce a short hand notation
\begin{equation}
\eta = e^{i \sigma^3 \chi} = \pm 1\quad \text{for }\chi\text{'s in Eq.~\eqref{chi}}
\end{equation}
The general form for $W_{\overline{C}_6}(\bm{r}_\mu)$ and $W_S(\bm{r}_\mu)$ is given in Eqs.~\eqref{eq:wc6} and \eqref{eq:ws}. Here, in order to clearly distinguish different notations, we will rewrite the SU(2) matrices at the origin $W_{\mathcal{O},\mu}$ using a different symbol
\begin{equation}
g_{\mathcal{O},\mu} \equiv W_{\mathcal{O},\mu}\quad \text{for } \mathcal{O} = \overline{C}_6,S.
\end{equation}
The solution is in complete parallel to the U(1) PSG case which we briefly review below.

First solve Eq.~\eqref{tttt}: using gauge freedom as in the U(1) case, we get
\begin{equation}\label{z2solt}
W_{T_1}(\bm{r}_\mu) =1,\quad W_{T_2}(\bm{r}_\mu) = \eta_1^{r_1},\quad W_{T_3}(\bm{r}_\mu) = \eta_3^{r_1}\eta_2^{r_2}.
\end{equation}

Then solve Eq.~\eqref{ctct}: plugging in the solution \eqref{z2solt}, we have
\begin{subequations}
\begin{align}
&W_{\overline{C}_6}(\bm{r}_\mu) W^{-1}_{\overline{C}_6}[T_2(\bm{r}_\mu)]\eta_1^{r_1}= \eta_{\overline{C}_6T_1},\\
&W_{\overline{C}_6}(\bm{r}_\mu) \eta_1^{-(r_2+\delta_{\mu,2})} W^{-1}_{\overline{C}_6}[T_3(\bm{r}_\mu)] \eta_3^{r_1}\eta_2^{r_2} = \eta_{\overline{C}_6T_2},\\
&W_{\overline{C}_6}(\bm{r}_\mu)\eta_3^{-(r_2+\delta_{\mu,2})}\eta_2^{-( r_3+\delta_{\mu,3})}W^{-1}_{\overline{C}_6}[T_1(\bm{r}_\mu)] = \eta_{\overline{C}_6T_3}.
\end{align}
\end{subequations}
Consistency condition requires $\eta_1=\eta_2=\eta_3$ and we get
\begin{equation}\label{fC6}
\begin{aligned}
W_{\overline{C}_6}(\bm{r}_\mu) &= g_{\overline{C}_6,\mu}\eta_1^{r_1(r_2+r_3)}\\
  &\qquad\cdot(\eta_{\overline{C}_6T_3}\eta_1^{\delta_{\mu,2}+\delta_{\mu,3}})^{r_1}\eta_{\overline{C}_6T_1}^{r_2}(\eta_{\overline{C}_6T_2}\eta_1^{\delta_{\mu,2}})^{r_3}.
  \end{aligned}
\end{equation}

Then solve Eq.~\eqref{ststt} and Eq.~\eqref{stst}: plugging in the solution \eqref{z2solt}, we have
\begin{subequations}
\begin{align}
&W_S(\bm{r}_\mu) W^{-1}_S[T_1T_3^{-1}(\bm{r}_\mu)] \eta_1^{-(r_1+r_2+1)}= \eta_{ST_1},\label{fSSS1}\\
&W_S(\bm{r}_\mu) \eta_1^{-(r_1+\delta_{\mu,1})} W^{-1}_S[T_2T_3^{-1}(\bm{r}_\mu)] \eta_1^{-(r_1+r_2+1)}\eta_1^{r_1} = \eta_{ST_2},\label{fSSS2}\\
&W_S(\bm{r}_\mu)\eta_1^{-(r_1+r_2+\delta_{\mu,1}+\delta_{\mu,2})}  W^{-1}_S[T_3^{-1}(\bm{r}_\mu)] \eta_1^{-(r_1+r_2)} =\eta_{ST_3}. \label{fSSS3}
\end{align}
\end{subequations}
The consistency condition is always satisfied, and we have
\begin{equation}
\begin{aligned}
W_S(\bm{r}_\mu) &=g_{S,\mu}\eta_1^{\frac{1}{2}(r_1+r_2)(r_1+r_2+1)}(\eta_{ST_1}\eta_{ST_3}\eta_1^{\delta_{\mu,1}+\delta_{\mu,2}})^{r_1}\\
&\qquad\cdot(\eta_{ST_2}\eta_{ST_3}\eta_1^{\delta_{\mu,2}})^{r_2}(\eta_{ST_3}\eta_1^{\delta_{\mu,1}+\delta_{\mu,2}})^{r_3}.
\end{aligned}
\end{equation}
Now we are left with Eqs.~\eqref{cccccc}, \eqref{sst}, \eqref{cscscscs}, and \eqref{cccscccs}. Plugging the solution for $W_{T_{1,2,3},\overline{C}_6,S}$ that we just obtained in these equations, we are led to the following constraints
\begin{subequations}
\begin{align}
\eta_{\overline{C}_6T_1}\eta_{\overline{C}_6T_2}\eta_{\overline{C}_6T_3} &= 1,\label{eta789}\\
\eta_{ST_3}&=1,
\end{align}
\end{subequations}
and the following equations to solve
\begin{subequations}\label{z2_all_intra_cell_constraints}
\begin{align}
g^6_{\overline{C}_6,0}=(g_{\overline{C}_6,1}g_{\overline{C}_6,3}g_{\overline{C}_6,2})^2 &=\eta_{\overline{C}_6},\\
g^2_{S,1}\eta_{ST_1}\eta_1=g^2_{S,2}\eta_{ST_2}\eta_1=g_{S,3}g_{S,0}&=\eta_S,\\
g_{\overline{C}_6,0}g_{S,0}g_{\overline{C}_6,3}g_{S,2}g_{\overline{C}_6,2}g_{S,1}g_{\overline{C}_6,1}g_{S,3} &= \eta_{\overline{C}_6S},\\
 \eta_{\overline{C}_6T_1}g^3_{\overline{C}_6,0}g_{S,0}g_{\overline{C}_6,3}g_{\overline{C}_6,2}g_{\overline{C}_6,1}g_{S,3} &= \eta_{S\overline{C}_6},\\
g_{\overline{C}_6,1}g_{\overline{C}_6,3}g_{\overline{C}_6,2}g_{S,1}
g_{\overline{C}_6,1}g_{\overline{C}_6,3}g_{\overline{C}_6,2}g_{S,1}  &= \eta_{S\overline{C}_6},\\
g_{\overline{C}_6,2}g_{\overline{C}_6,1}g_{\overline{C}_6,3}g_{S,2}
g_{\overline{C}_6,2}g_{\overline{C}_6,1}g_{\overline{C}_6,3}g_{S,2}  &= \eta_{S\overline{C}_6},\\
g_{\overline{C}_6,3}g_{\overline{C}_6,2}g_{\overline{C}_6,1}g_{S,3}g^3_{\overline{C}_6,0}g_{S,0}
\eta_{\overline{C}_6T_1}&=\eta_{S\overline{C}_6}.
\end{align}
\end{subequations}

First let us use the  IGG gauge freedom to simplify these equations. We can always set $\eta_S=\eta_{\overline{C}_6T_1}=\eta_{\overline{C}_6T_2}=1$, and by Eq.~\eqref{eta789} we also have $\eta_{\overline{C}_6T_3}=1$. Then, under the ``sublattice'' gauge transformation \eqref{eq:gaugetr}, we can fix $g_{\overline{C}_6,1}=1$,  $g_{\overline{C}_6,2}=1$, and $g_{S,0}=1$. Note this also implies $g_{S,3}=1$. To summarize, gauge fixing gives
\begin{equation}
g_{\overline{C}_6,1}=g_{\overline{C}_6,2}=g_{S,0}=g_{S,3}=1.
\end{equation}
Now Eq.~\eqref{z2_all_intra_cell_constraints} is simplified to
\begin{subequations}\label{z2_all_intra_cell_constraints111}
\begin{align}
g^6_{\overline{C}_6,0}=g^2_{\overline{C}_6,3}&=\eta_{\overline{C}_6},\label{c03}\\
g^2_{S,1}\eta_{ST_1}\eta_1=g^2_{S,2}\eta_{ST_2}\eta_1&=1,\label{s12}\\
g_{\overline{C}_6,0}g_{\overline{C}_6,3}g_{S,2}g_{S,1}&= \eta_{\overline{C}_6S},\label{c03s21}\\
g^3_{\overline{C}_6,0}g_{\overline{C}_6,3} &= \eta_{S\overline{C}_6},\label{c0c3}\\
g_{\overline{C}_6,3}g_{S,1}g_{\overline{C}_6,3}g_{S,1}  &= \eta_{S\overline{C}_6},\label{cscs1}\\
g_{\overline{C}_6,3}g_{S,2}g_{\overline{C}_6,3}g_{S,2}  &= \eta_{S\overline{C}_6}.\label{cscs2}
\end{align}
\end{subequations}

Next we claim that
\begin{equation}\label{st1st2}
\eta_{ST_1}=\eta_{ST_2}.
\end{equation}
The proof proceeds as follows: If $\eta_{\overline{C}_6}=1$ then Eq.~\eqref{c03} gives $g_{\overline{C}_6,3}=\pm 1$, which together with Eqs.~\eqref{cscs1} and \eqref{cscs2} proves \eqref{st1st2} in Eq.~\eqref{s12}. If $\eta_{\overline{C}_6}=-1$, then $g_{\overline{C}_6,3}=i\bm{a}\cdot\bm{\sigma}$ for some unit vector $\bm{a}$, and we proceed to prove $\eta_{ST_1}=\eta_{ST_2}$ by contradiction: without loss of generality we assume $\eta_{ST_1}\eta_1=1=-\eta_{ST_2}\eta_1$, then $g_{S,1}\equiv \eta_{S,1}=\pm 1$ and $g_{S,2}=i\bm{b}\cdot\bm{\sigma}$ for some unit vector $\bm{b}$. Therefore Eq.~\eqref{cscs1} gives $\eta_{S\overline{C}_6}=-1$, and Eq.~\eqref{cscs2} gives $[(i \bm{a}\cdot \bm{\sigma})(i \bm{b}\cdot \bm{\sigma})]^2=-1$, which implies $\bm{a}\perp\bm{b}$ and that $g_{\overline{C}_6,3}g_{S,2} = -i \bm{c}\cdot \bm{\sigma}$ with $\bm{c} = \bm{b}\times \bm{a}$. Then from Eq.~\eqref{c03s21} we get $g_{\overline{C}_6,0} = i \bm{c}\cdot \bm{\sigma} \eta_{S\overline{C}_6} \eta_{S,1}$, which contradicts Eq.~\eqref{c0c3} given that $\bm{c}\perp \bm{a}$. Therefore Eq.~\eqref{st1st2} holds.

Next, depending on the value of $\eta_{S\overline{C}_6}$, $\eta_{\overline{C}_6}$, $\eta_{ST_1}$ and $\eta_1$, we have the following cases:
\begin{itemize}
\item If $\eta_1\eta_{ST_1}=1$, then $g_{S,1}=\pm 1$ and $g_{S,2}=\pm 1$, then we have
\begin{equation}\label{sstts}
\eta_{S\overline{C}_6} = g^2_{\overline{C}_6,3} = \eta_{\overline{C}_6},
\end{equation}
meaning that Eq.~\eqref{c03s21} to the cubic power gives $g_{\overline{C}_6,0}^3 = g_{\overline{C}_6,3}^{-3} \eta_{S\overline{C}_6}g_{S,2}g_{S,1}$; together with Eq.~\eqref{sstts} we see that
\begin{equation}
g_{S,1}g_{S,2} = \eta_{\overline{C}_6S}.
\end{equation}
We have the following two cases after gauge fixing:
\begin{itemize}
\item When $\eta_{S\overline{C}_6}=\eta_{\overline{C}_6}=1$,
\begin{equation}
(g_{\overline{C}_6,0},g_{\overline{C}_6,3},g_{S,1},g_{S_2}) = (1,1,1,\eta_{\overline{C}_6S});
\end{equation}
\item When $\eta_{S\overline{C}_6}=\eta_{\overline{C}_6}=-1$,
\begin{equation}
(g_{\overline{C}_6,0},g_{\overline{C}_6,3},g_{S,1},g_{S_2}) = (-i \sigma^k,i \sigma^k,1,\eta_{\overline{C}_6S}),
\end{equation}
where $\sigma^k$ can be any of the three Pauli matrices.
\end{itemize}
\item If $\eta_1\eta_{ST_1}=-1$, we have $g_{S,1} = i \bm{n}_1\cdot \bm{\sigma}$ and $g_{S,2} = i \bm{n}_2\cdot \bm{\sigma}$ for some unit vectors $\bm{n}_1$ and $\bm{n}_2$. We have:
\begin{itemize}
\item If $\eta_{\overline{C}_6}=1$, then $g_{\overline{C}_6,3}= \pm 1$ therefore $\eta_{S\overline{C}_6}=-1$.  Combine \eqref{c03s21} and \eqref{c0c3} we get $(g_{S,2}g_{S,1})^{-3} = -\eta_{S\overline{C}_6}$, which gives $g_{S,2}g_{S,1} = e^{i\frac{ \pi}{3} j \bm{n}\cdot \bm{\sigma}}$ for some unit vector $\bm{n}$, and $j=1,3$ for $\eta_{\overline{C}_6S}=1$ while $j=0,2$ for $\eta_{\overline{C}_6S}=-1$. This then implies $\bm{n}_2\cdot \bm{n}_1 + i(\bm{n}_2\times \bm{n}_1)\cdot \bm{\sigma} = - e^{i\frac{\pi}{3}j \bm{n}\cdot \bm{\sigma}}$. After gauge fixing, we get
\begin{equation}
\begin{aligned}
&(g_{\overline{C}_6,0},g_{\overline{C}_6,3},g_{S,1},g_{S,2})\\
&=
(-\sigma^k(\cos \frac{\pi j}{3}\sigma^k+\sin \frac{\pi j}{3} \sigma^{k+1}),
1,\\
& \quad \quad  i \sigma^k,i(\cos \frac{\pi j}{3}\sigma^k+\sin \frac{\pi j}{3} \sigma^{k+1})),
\end{aligned}
\end{equation}
where $\sigma^k$ can be any of the three Pauli matrices. This gives eight classes depending on the values of $\eta_1$, $\eta_{\overline{C}_6}$ and $j$.
\item if $\eta_{\overline{C}_6}=-1$, we have $g_{\overline{C}_6,3} = i \bm{n}_3 \cdot \bm{\sigma}$ for some $\bm{n}_3$, and $g^3_{\overline{C}_6,0}=-i \bm{n}_3\cdot \bm{\sigma}\eta_{S\overline{C}_6}$. Define $g_{\overline{C}_6,0} = e^{i \theta \bm{n}_0\cdot \bm{\sigma}}$, then we have $\cos 3\theta=0$, $\theta= \frac{\pi(2j+1)}{6}$, which gives $g_{\overline{C}_6,0} = e^{-i\frac{\pi(2j+1)}{6}(-1)^j \eta_{S\overline{C}_6} \bm{n}_3\cdot \bm{\sigma}}$, with independent $j=0,1,2$. Choose gauge fixing such that $g_{S,1} = i \sigma^k$ where $\sigma^k$ is any of the three Pauli matrices (we denote the corresponding $\bm{n}_1$ as $\bm{x}_k$) and $g_{S,2} = e^{i \phi \sigma^{k-1}}i \sigma^k$ for some $\phi$, then equation \eqref{cscs1} requires that $\bm{n}_3$ be either parallel (anti-parallel) or perpendicular to $\bm{x}_k$, depending on the value of $\eta_{S\overline{C}_6}$. After gauge fixing, we obtain the following form:
\begin{itemize}
\item If  $\eta_{S\overline{C}_6}=1$, we have
\begin{equation}
(g_{\overline{C}_6,0},g_{\overline{C}_6,3},g_{S,1},g_{S,2}) = (i \sigma^k,i \sigma^k, i \sigma^k, \eta_{\overline{C}_6S} i \sigma^k),
\end{equation}
the parameters $\eta_{\overline{C}_6S}$ and $\eta_1$ give four independent classes.
\item If $\eta_{S\overline{C}_6}=-1$, we have
\begin{equation}
\begin{aligned}
&(g_{\overline{C}_6,0},g_{\overline{C}_6,3},g_{S,1},g_{S,2}) =\\
& ( e^{i\frac{\pi(2j+1)}{6}(-1)^j\sigma^{k-1}},i \sigma^{k-1},i \sigma^k,\\
& \qquad -\eta_{\overline{C}_6S}i \sigma^k e^{i\left(\frac{\pi}{2}+\frac{\pi(2j+1)}{6}(-1)^j\right) \sigma^{k-1}}),
\end{aligned}
\end{equation}
where $j=0,1$ together with $\eta_{\overline{C}_6S}$ and $\eta_1$ gives eight independent classes.
\end{itemize}
\end{itemize}
\end{itemize}

The final result is presented in Table \ref{table:z2_psg_table}.

\section{\label{app:E}Solving $\mathbb{Z}_2$ PSG equations: adding time reversal}
In this subsection we assume that all the space group PSG equations have been solved (and gauge-fixed). We then add time reversal to the PSG and solve all the PSG equations containing time reversal: just as in the U(1) case, we are solving Eqs.~\eqref{PSGtimet}, but now the right-hand sides of these equations are $\pm 1$. From Eq.~\eqref{eq:timet} we get
\begin{equation}\label{eq:Wt123}
W_{\mathcal{T}}(\bm{r}_\mu) = g_{\mathcal{T},\mu} \eta_{\mathcal{T}T_1}^{r_1}\eta_{\mathcal{T}T_2}^{r_2}\eta_{\mathcal{T}T_3}^{r_3},
\end{equation}
where we followed the notation in the last appendix: $g_{\mathcal{T},\mu} \equiv W_{\mathcal{T},\mu}$. Plug the form \eqref{eq:Wt123} in Eqs.~\eqref{eq:timec}\---\eqref{eq:timestimes}, we obtain
\begin{equation}\label{eq:timetnotrans}
\eta_{\mathcal{T}T_1} = \eta_{\mathcal{T}T_2} = \eta_{\mathcal{T}T_3} =0
\end{equation}
and
\begin{subequations}\label{lasttrconstraints1}
\begin{align}
g_{\mathcal{T},0}g_{\overline{C}_6,0}g^{-1}_{\mathcal{T},0}g^{-1}_{\overline{C}_6,0}
&=g_{\mathcal{T},1}g_{\overline{C}_6,1}g^{-1}_{\mathcal{T},3}g^{-1}_{\overline{C}_6,1}\nonumber\\
&=g_{\mathcal{T},2}g_{\overline{C}_6,2}g^{-1}_{\mathcal{T},1}g^{-1}_{\overline{C}_6,2}\nonumber\\
&=g_{\mathcal{T},3}g_{\overline{C}_6,3}g^{-1}_{\mathcal{T},2}g^{-1}_{\overline{C}_6,3}\nonumber\\
&=\eta_{\mathcal{T}\overline{C}_6},\label{eq:lastonetothink1}\\
g_{\mathcal{T},0}g_{S,0}g^{-1}_{\mathcal{T},3}g^{-1}_{S,0}
&=g_{\mathcal{T},1}g_{S,1}g^{-1}_{\mathcal{T},1}g^{-1}_{S,1}\nonumber\\
&=g_{\mathcal{T},2}g_{S,2}g^{-1}_{\mathcal{T},2}g^{-1}_{S,2}\nonumber\\
&=g_{\mathcal{T},3}g_{S,3}g^{-1}_{\mathcal{T},0}g^{-1}_{S,3}\nonumber\\
&=\eta_{\mathcal{T}S},\label{eq:lastonetothink2}\\
g^2_{\mathcal{T},\mu} &= \eta_{\mathcal{T}}.\label{eq:lastonetothink}
\end{align}
\end{subequations}
Eq.~\eqref{eq:timetnotrans} means that $W_{\mathcal{T}}(\bm{r}_\mu) = g_{\mathcal{T},\mu}$, which only depends on the sublattice indices. We now claim that $\eta_{\mathcal{T}}=-1$. Otherwise $\eta_{\mathcal{T}}=1$, then $g_{\mathcal{T},\mu}$ has the form of a sign factor $\eta_{\mathcal{T},\mu} = \pm 1$ times the identity matrix. Plugging this form in Eqs.~\eqref{eq:lastonetothink1} and ~\eqref{eq:lastonetothink2} gives $1 = \eta_{\mathcal{T},1}\eta_{\mathcal{T},3} = \eta_{\mathcal{T},2}\eta_{\mathcal{T},1} = \eta_{\mathcal{T},3}\eta_{\mathcal{T},2} = \eta_{\mathcal{T}\overline{C}_6}$ and $\eta_{\mathcal{T},0}\eta_{\mathcal{T},3} = 1=\eta_{\mathcal{T}S}$, meaning that all the sublattice signs $\eta_{\mathcal{T},\mu}$ must be the same. As was argued in Appendix \ref{app:u1_TR} this will lead to a vanishing mean-field  ans\"atze. Therefore we must have $\eta_{\mathcal{T}} = -1$.

Recall that in classifying the space group PSG in Appendix \ref{app:z2_SG} gauge fixing already gives $g_{\overline{C}_6,1}=g_{\overline{C}_6,2}=g_{S,0}=g_{S,3} = 1$. Then Eqs.~\eqref{eq:lastonetothink1} and \eqref{eq:lastonetothink1} enforce
\begin{equation}
g_{\mathcal{T},1} = \eta_{\mathcal{T}\overline{C}_6}\eta_{\mathcal{T}S}g_{\mathcal{T},0},\quad
g_{\mathcal{T},2} = \eta_{\mathcal{T}S}g_{\mathcal{T},0}, \quad
g_{\mathcal{T},3} = \eta_{\mathcal{T}S}g_{\mathcal{T},0},
\end{equation}
and the two equations reduce to
\begin{equation}
\begin{aligned}
g_{\mathcal{T},0}g_{\overline{C}_6,0} = \eta_{\mathcal{T}\overline{C}_6} g_{\overline{C}_6,0}g_{\mathcal{T},0},&\quad
g_{\mathcal{T},0}g_{\overline{C}_6,3} =  \eta_{\mathcal{T}\overline{C}_6} g_{\overline{C}_6,3}g_{\mathcal{T},0},\\
g_{\mathcal{T},0}g_{S,1}=\eta_{\mathcal{T}S}g_{S,1}g_{\mathcal{T},0},&\quad g_{\mathcal{T},0}g_{S,2}=\eta_{\mathcal{T}S} g_{S,2}g_{\mathcal{T},0}.
\end{aligned}
\end{equation}
$g_{\mathcal{T},0}$ is then determined by $g_{\overline{C}_6,0},g_{\overline{C}_6,3},g_{S,1},g_{S,2}$ and there are five cases listed below.
\begin{itemize}
\item In the case $(\eta_1\eta_{ST_1},\eta_{\overline{C}_6},\eta_{S\overline{C}_6}) = (1,1,1)$, we have $(g_{\overline{C}_6,0},g_{\overline{C}_6,3},g_{S,1},g_{S,2})
= (1,1,1,\eta_{\overline{C}_6S})$, we see we must have $\eta_{\mathcal{T}\overline{C}_6}=\eta_{\mathcal{T}S}=1$, and it is easy to use the remaining global SU(2) gauge freedom to set e.g. $g_{\mathcal{T},0}=i\sigma^k$, where $\sigma^k$ can be any of the three Pauli matrices.
\item In the case $(\eta_1\eta_{ST_1},\eta_{\overline{C}_6},\eta_{S\overline{C}_6}) = (1,-1,-1)$, we have $(g_{\overline{C}_6,0},g_{\overline{C}_6,3},g_{S,1},g_{S,2})= (- i \sigma^k,i \sigma^k,1,\eta_{\overline{C}_6S})$, it is easy to see from the expression of $g_{S,1}$ and $g_{S,2}$ that $\eta_{\mathcal{T}S}=1$. Then time reversal is either $g_{\mathcal{T},0} = i\sigma^k$ which gives $\eta_{\mathcal{T}\overline{C}_6}=1$, or $g_{\mathcal{T},0}=i\sigma^{k-1}$ which gives $\eta_{\mathcal{T}\overline{C}_6}=-1$. Note we have used the gauge freedom (rotating along $\sigma^k$ axis) to set $g_{\mathcal{T},0}=i\sigma^{k-1}$.
\item In the case $(\eta_1\eta_{ST_1},\eta_{\overline{C}_6},\eta_{S\overline{C}_6}) = (-1,1,-1)$, we have $(g_{\overline{C}_6,0},g_{\overline{C}_6,3},g_{S,1},g_{S,2})
=(-\eta_{\overline{C}_6S} e^{i\frac{\pi j}{3} \sigma^{k-1}},1, i \sigma^k, i \sigma^k e^{i\frac{\pi j}{3} \sigma^{k-1}})$. When $\eta_{\overline{C}_6S} = 1$, we get $j=1,3$,  and when $\eta_{\overline{C}_6S}=-1$, we get $j=0,2$. We always have $\eta_{\mathcal{T}\overline{C}_6}=1$. If $j=1$ or $2$, then we must have $g_{\mathcal{T},0}=i\sigma^{k-1}$, and $\eta_{\mathcal{T}S}=-1$; if $j=0$ or $3$, then we can have $g_{\mathcal{T},0} = i \sigma^k$ or $g_{\mathcal{T},0}=i\sigma^{k-1}$ (after gauge fixing), which gives $g_{\mathcal{T}S}=1$ or $g_{\mathcal{T}S} = -1$, respectively.

\item In the case $(\eta_1\eta_{ST_1},\eta_{\overline{C}_6},\eta_{S\overline{C}_6}) = (-,-,+)$, $g_{\overline{C}_6,0} =  i \sigma^k$, $g_{\overline{C}_6,3} = i \sigma^k$, $g_{S,1} = i \sigma^k$, $g_{S,2} = \eta_{\overline{C}_6S} i \sigma^k$. Two solutions exist: we can have either $g_{\mathcal{T},0}=i\sigma^k$ corresponding to $\eta_{\mathcal{T}\overline{C}_6}=\eta_{\mathcal{T}S}=1$, or $g_{\mathcal{T},0}=i\sigma^{k-1}$ (after gauge fixing), corresponding to $\eta_{\mathcal{T}\overline{C}_6} = \eta_{\mathcal{T}S} = -1$.

\item Lastly, in the case $(\eta_1\eta_{ST_1},\eta_{\overline{C}_6},\eta_{S\overline{C}_6}) = (-1,-1,-1)$, $g_{\overline{C}_6,0} = e^{i\frac{\pi(2j+1)}{6}(-1)^j\sigma^{k-1}}$, $g_{\overline{C}_6,3} = i \sigma^{k-1}$, $g_{S,1} = i \sigma^k$, $g_{S,2} = \eta_{\overline{C}_6S} i\sigma^{k+1}e^{i\frac{\pi(2j+1)}{6}(-1)^j\sigma^{k-1}}$.
When $j=0$ we must have $\eta_{\mathcal{T}\overline{C}_6}=1$ and $g= i\sigma^{k-1}$, which gives $\eta_{\mathcal{T}S} = -1$; when $j=1$ three solutions exist: we can have $g= i\sigma^{k-1}$ corresponding to $(\eta_{\mathcal{T}\overline{C}_6},\eta_{\mathcal{T}S})=(1,-1)$, or $g= i\sigma^k$ corresponding to $(\eta_{\mathcal{T}\overline{C}_6},\eta_{\mathcal{T}S}) = (-1,1)$, or $g=i\sigma^{k+1}$ corresponding to $(\eta_{\mathcal{T}\overline{C}_6},\eta_{\mathcal{T}S}) = (-1,-1)$.
\end{itemize}

The final result is presented in Table \ref{table:z2_psg_table}.

\section{\label{app:F}0-flux symmetry properties}

In this Appendix we study the symmetry transformation of the Hamiltonian and operators in the 0-flux  ans\"atze in more detail. Under an arbitrary symmetry operation $\mathcal{O}$, following Eq.~\eqref{b_under_o} we have
\begin{equation}
G_{\mathcal{O}}\circ\mathcal{O}\colon
 \Psi_{\bm{r}_\mu}\rightarrow U^\dag_{\mathcal{O}} \Psi_{\mathcal{O}(\bm{r}_\mu)}
W_{\mathcal{O},\mu}e^{i \sigma^3\phi_{\mathcal{O}}[\mathcal{O}(\bm{r}_\mu)]},
\end{equation}
In the case of U(1) PSG (with or without time reversal) and time reversal symmetric $\mathbb{Z}_2$ PSG in our chosen gauge, we can always write, with the help of the $\mathbb{Z}_2$ parameter $w_{\mathcal{O}}$, $W_{\mathcal{O},\mu} = (i\sigma^1)^{w_{\mathcal{O}}} e^{i\sigma^3 \phi^0_{\mathcal{O},\mu}}$, where $\phi^0_{\mathcal{O},\mu}$ is some phase that can be absorbed into the definition of $\phi_{\mathcal{O}}(\bm{r}_\mu)$, and we call $\bar\phi_{\mathcal{O}}(\bm{r}_\mu)=\phi_{\mathcal{O}}(\bm{r}_\mu)+\phi^0_{\mathcal{O},\mu}$. We have
\begin{equation}
G_{\mathcal{O}}\circ\mathcal{O}\colon
\left(\begin{array}{c} f_{\bm{r}_\mu}\\ f^\dag_{\bm{r}_\mu}\end{array}\right)
\rightarrow
V_{\mathcal{O},w_{\mathcal{O}}}e^{i \tau^3\bar\phi_{\mathcal{O}}[\mathcal{O}(\bm{r}_\mu)]}\left(\begin{array}{c} f_{\mathcal{O}(\bm{r}_\mu)}\\ f^\dag_{\mathcal{O}(\bm{r}_\mu)}\end{array}\right),
 \end{equation}
with
\begin{equation}
V_{\mathcal{O},w_{\mathcal{O}}}=\left(\begin{array}{cc} U^\dag_{\mathcal{O}}&\\&U^T_{\mathcal{O}} \end{array}\right)(-i\tau^2\sigma^2)^{w_{\mathcal{O}}},
\end{equation}
where we defined Pauli matrices $\tau^{1,2,3}$ to act on the subspace of $f_{\bm{r}_\mu}$ and $f^\dag_{\bm{r}_\mu}$. Note that $f^\dag_{\bm{r}_\mu}$ is understood as $(f^\dag_{\bm{r}_\mu})^T=\left(f^\dag_{\bm{r}_\mu\uparrow},f^\dag_{\bm{r}_\mu\downarrow}\right)^T$.

Then, Fourier transform gives
\begin{equation}\label{goook}
\begin{aligned}
\left(\begin{array}{c} f_{\mu,\bm{k}}\\ f^\dag_{\mu,-\bm{k}}\end{array}\right)
&=\frac{1}{\sqrt{N}}\sum_{\bm{r}} e^{-i \bm{k}\cdot \bm{r}_\mu}
\left(\begin{array}{c} f_{\bm{r}_\mu}\\ f^\dag_{\bm{r}_\mu}\end{array}\right)\\
&\xrightarrow{G_{\mathcal{O}}\circ \mathcal{O}}
\frac{V_{\mathcal{O},w_{\mathcal{O}}}}{\sqrt{N}}
\sum_{\bm{r}} e^{-i \bm{k}\cdot \bm{r}_\mu}
e^{i \tau^3 \bar\phi_{\mathcal{O}}[\mathcal{O}(\bm{r}_\mu)]}\left(\begin{array}{c} f_{\mathcal{O}(\bm{r}_\mu)}\\ f^\dag_{\mathcal{O}(\bm{r}_\mu)}\end{array}\right)\\
&= \frac{V_{\mathcal{O},w_{\mathcal{O}}}}{\sqrt{N}}
\sum_{\bm{r}} e^{-i [\bm{k}\cdot \mathcal{O}^{-1}(\bm{r}_{\mathcal{O}(\mu)})
-\tau^3 \bar\phi_{\mathcal{O}}(\bm{r}_{\mathcal{O}(\mu)})]}\left(\begin{array}{c} f_{\bm{r}_{\mathcal{O}(\mu)}}\\ f^\dag_{\bm{r}_{\mathcal{O}(\mu)}}\end{array}\right),
\end{aligned}
\end{equation}

Using the general structure
\begin{equation}\label{symmcs}
\begin{aligned}
\bar\phi_{\overline{C}_6}(\bm{r}_\mu) &= \phi_{\overline{C}_6T_3}r_1+\rho_\mu,\\
\bar\phi_S(\bm{r}_\mu) &= -\phi_{ST_1}r_1+3\phi_{ST_1}r_2
+\theta_\mu,
\end{aligned}
\end{equation}
we have
\begin{equation}
\begin{aligned}
&\left[\bm{k}\cdot \mathcal{O}^{-1}(\bm{r}_{\mathcal{O}(\mu)})
-\tau^3\bar\phi_{\mathcal{O}}(\bm{r}_{\mathcal{O}(\mu)})\right]\Big|_{\mathcal{O}=\overline{C}_6}\\
&=\left(\overline{C}_6(\bm{k})-\tau^3\phi_{\overline{C}_6T_3}\hat{\bm{b}}_1\right)\cdot \bm{r}- \bm{k}\cdot {\bm{e}}_\mu -\tau^3 \rho_{\overline{C}_6},
\end{aligned}
\end{equation}
and
\begin{equation}
\begin{aligned}
&\left[\bm{k}\cdot \mathcal{O}^{-1}(\bm{r}_{\mathcal{O}(\mu)})
-\tau^3\bar\phi_{\mathcal{O}}(\bm{r}_{\mathcal{O}(\mu)})\right]\Big|_{\mathcal{O}=S}\\
&=\left(S(\bm{k})-\tau^3(-\phi_{ST_1}\hat{\bm{b}}_1+3 \phi_{ST_1}\hat{\bm{b}}_2\right)\cdot \bm{r}-\bm{k}\cdot {\bm{e}}_\mu -\tau^3 \theta_{\overline{C}_6},
\end{aligned}
\end{equation}
we have
\begin{equation}
\begin{aligned}
G_{\overline{C}_6}&\circ \overline{C}_6\colon \left(\begin{array}{c} f_{\mu,\bm{k}}\\ f^\dag_{\mu,-\bm{k}}\end{array}\right)\rightarrow\\
&V_{\overline{C}_6,w_{\overline{C}_6}}e^{i(\bm{k}\cdot \bm{e}_\mu+\rho_{\overline{C}_6(\mu)})}\left(\begin{array}{c} f_{\overline{C}_6(\mu),\overline{C}_6(\bm{k})-\phi_{\overline{C}_6T_3}\hat{\bm{b}}_1}\\ f^\dag_{\overline{C}_6(\mu),\overline{C}_6(-\bm{k})-\phi_{\overline{C}_6T_3}\hat{\bm{b}}_1}\end{array}\right),
\end{aligned}
\end{equation}
and
\begin{equation}
\begin{aligned}
G_S &\circ S\colon \left(\begin{array}{c} f_{\mu,\bm{k}}\\ f^\dag_{\mu,-\bm{k}}\end{array}\right)\rightarrow\\
&V_{S,w_S}e^{i(\bm{k}\cdot \bm{e}_\mu+\theta_{S(\mu)})}
\left(\begin{array}{c} f_{S(\mu),S(\bm{k})+\phi_{ST_1}(\hat{\bm{b}}_1-3\hat{\bm{b}}_2)}\\ f^\dag_{S(\mu),S(-\bm{k})+\phi_{ST_1}(\hat{\bm{b}}_1-3\hat{\bm{b}}_2)}\end{array}\right),
\end{aligned}
\end{equation}

\section{\label{app:G}Derivation of the mean-field Hamiltonians}

The parameters for the $i$th-nearest-neighbor bonds for $i\leq 8$ are given in Table \ref{8NNbonds}. The PSG results in the following constraints for these parameters:
\begin{table}
\centering
\caption{Parameters for the representative $i$th-nearest-neighbor bonds for $i\leq 8$. The point group elements under which the representative bonds are invariant are given in the last column; see Eqs.~\eqref{S4group} for the explanation of the notations.}\label{8NNbonds}
\begin{ruledtabular}
\begin{tabular}{l|l|c|c}
Bond & \multirow{2}{*}{\qquad \;\;Bond} &\multirow{2}{*}{parameters} &invariant\\
type  & & &under\\
\hline
\multirow{8}{*}{Onsite}      &\multirow{8}{*}{$\bm{0}_0\rightarrow(0,0,0)_0$}  & \multirow{8}{*}{$\alpha$}  &$(1)$, $(12)$, $(23)$, \\
&&& $(123)$, $(132)$,\\
&&& $(+-)$, \\
&&& $(12)(+-)$, \\
&&& $(13)(+-)$, \\
&&& $(23)(+-)$, \\
&&& $(123)(+-)$,\\
&&& $(132)(+-)$\\
\hline
\multirow{3}{*}{NN}          &\multirow{3}{*}{$\bm{0}_0\rightarrow(0,0,0)_1$}  & \multirow{3}{*}{$a$, $b$, $c$, $d$}  &$(1)$, $(14)(23)$, \\
&&& $(14)(+-)$, \\
&&& $(23)(+-)$\\
\hline
NNN         &$\bm{0}_1\rightarrow(0,-1,0)_2$ & $A$, $B$, $C$, $D$  &$(1)$, $(12)$\\
\hline
\multirow{6}{*}{3nd NN} &\multirow{3}{*}{$\bm{0}_0\rightarrow(1,0,0)_0$} & \multirow{3}{*}{$A_3$, $B_3$, $C_3$, $D_3$}  &$(1)$, $(23)$, \\
&&& $(+-)$,\\
&&&  $(23)(+-)$\\
\cline{2-4}
            &\multirow{3}{*}{$\bm{0}_0\rightarrow(1,-1,0)_0$} & \multirow{3}{*}{$A'_3$, $B'_3$, $C'_3$, $D'_3$}  &$(1)$, $(12)$, \\
            &&&$(+-)$,\\
            &&& $(12)(+-)$\\
\hline
4th NN      &$\bm{0}_0\rightarrow(1,-1,0)_3$ & $A_4$, $B_4$, $C_4$, $D_4$  &$(1)$, $(12)(34)$\\
\hline
5th NN      &$\bm{0}_1\rightarrow(2,0,-1)_0$ & $A_5$, $B_5$, $C_5$, $D_5$  &$(1)$\\
\hline
\multirow{2}{*}{6th NN}      &\multirow{2}{*}{$\bm{0}_0\rightarrow(-1,-1,1)_0$}& \multirow{2}{*}{$A_6$, $B_6$, $C_6$, $D_6$}  &$(1)$, $(12)$, \\
&&& $(+-)$, $(12)(+-)$\\
\hline
\multirow{4}{*}{7th NN} &\multirow{3}{*}{$\bm{0}_0\rightarrow(-2,0,0)_1$} & \multirow{3}{*}{$A_7$, $B_7$, $C_7$, $D_7$}  &$(1)$, $(14)(23)$, \\
&&& $(14)(+-)$,\\
&&& $(23)(+-)$\\
\cline{2-4}
            &$\bm{0}_0\rightarrow(-2,1,1)_1$ & $A'_7$, $B'_7$, $C'_7$, $D'_7$  &$(1)$, $(23)(+-)$\\
\hline
8th NN      &$\bm{0}_0\rightarrow(-2,0,1)_2$ & $A_8$, $B_8$, $C_8$, $D_8$ &$(1)$, $(24)$
\end{tabular}
\end{ruledtabular}
\end{table}
\begin{subequations}\label{S4group}
\begin{align}
(1)&=C_3\circ C_3\circ C_3, \nonumber\\
(12)&=\Sigma\circ C_3\circ \Sigma\circ C^{-1}_3\circ \Sigma\circ C_3,\nonumber\\
(13)&=\Sigma\circ C_3\circ \Sigma\circ C^{-1}_3\circ \Sigma, \nonumber\\
(14)&=\Sigma\circ C_3\circ \Sigma\circ C^{-1}_3\circ \Sigma\circ C^{-1}_3 \circ \Sigma\circ C_3\circ \Sigma\circ C_3, \nonumber\\
(23)&=\Sigma\circ C_3\circ \Sigma\circ C^{-1}_3\circ \Sigma \circ C^{-1}_3,\nonumber\\
(24)&=C_3\circ \Sigma\circ C^{-1}_3\circ \Sigma\circ C^{-1}_3\circ \Sigma, \nonumber\\
(34)&=\Sigma,\nonumber\\
(123)&=C_3, \nonumber\\
(132)&=C^{-1}_3, \nonumber\\
(124)&=\Sigma\circ C_3\circ \Sigma,\nonumber\\
(142)&=\Sigma\circ C^{-1}_3\circ \Sigma,\nonumber\\
(134)&=\Sigma\circ C_3\circ \Sigma\circ C^{-1}_3, \nonumber\\
(143)&=\Sigma\circ C_3\circ \Sigma\circ C^{-1}_3\circ \Sigma\circ C_3\circ \Sigma\circ C^{-1}_3,\nonumber\\
(234)&=C^{-1}_3\circ \Sigma\circ C_3\circ \Sigma\circ C^{-1}_3\circ \Sigma\circ C_3\circ \Sigma, \nonumber\\
(243)&
=C^{-1}_3\circ \Sigma\circ C_3\circ \Sigma,\nonumber\\
(1243)&=\Sigma\circ C_3,\nonumber\\
(14)(23)&=\Sigma\circ C_3\circ \Sigma\circ C_3,\nonumber\\
(1342)&=\Sigma\circ C_3\circ \Sigma\circ C_3\circ \Sigma\circ C_3,\nonumber\\
(1234)&=C_3\circ \Sigma,\nonumber\\
(13)(24)&=C_3\circ \Sigma\circ C_3\circ \Sigma,\nonumber\\
(1432)&=C_3\circ \Sigma\circ C_3\circ \Sigma\circ C_3\circ \Sigma, \nonumber\\
(1324)&=C_3\circ \Sigma\circ C_3,\nonumber\\
(12)(34)&=C_3\circ \Sigma\circ C^{-1}_3\circ \Sigma\circ C_3, \nonumber\\
(1423)&=C_3\circ \Sigma\circ C^{-1}_3\circ \Sigma\circ C^{-1}_3\circ \Sigma\circ C_3.\tag{\ref{S4group}}
\end{align}
\end{subequations}

For U(1) PSG without time reversal, we have the following four cases:
\begin{itemize}
\item $(w_{\overline{C}_6},w_S)=(0,0)$:
\begin{subequations}\label{u100}
\begin{align}
\alpha=&\alpha,\nonumber\\
(a,b,c,d)=&e^{i\chi_{\overline{C}_6S}}e^{-i \chi_{ST_1}}(-a^*,b^*,-c^*,-d^*)\nonumber\\
         =&e^{-i\chi_{ST_1}}(-a^*,-b^*,d^*,c^*),\nonumber\\
(A,B,C,D)=&(-A^*,-C^*,-B^*,-D^*),\nonumber\\
(A_3,B_3,C_3,D_3)=&e^{-i\chi_{ST_1}}(-A_3^*,-B_3^*,-D_3^*,-C_3^*)\nonumber\\
                 =&e^{-i\chi_{ST_1}}(-A_3^*,B_3^*,C_3^*,D_3^*),\nonumber\\
(A'_3,B'_3,C'_3,D'_3)=&e^{i\chi_1}(A'_3,-C'_3,-B'_3,-D'_3)\nonumber\\
                     =&e^{i(\chi_1+\chi_{ST_1})}(-A'^*_3,B'^*_3,C'^*_3,D'^*_3),\nonumber\\
(A_4,B_4,C_4,D_4)=&e^{i\chi_{\overline{C}_6S}}e^{-i\chi_{ST_1}}(-A^*_4,-B^*_4,-C^*_4,-D^*_4),\nonumber\\
(A_5,B_5,C_5,D_5)=&\text{arbitrary},\nonumber\\
(A_6,B_6,C_6,D_6)=&e^{-i\chi_{ST_1}}(-A_6^*,-C_6^*,-B_6^*,-D_6^*)\nonumber\\
                 =&e^{i(\chi_1-\chi_{ST_1})}(-A_6^*,B_6^*,C_6^*,D_6^*),\nonumber\\
(A_7,B_7,C_7,D_7)=&e^{i\chi_{\overline{C}_6S}}e^{i\chi_{ST_1}}(-A^*_7,B^*_7,-C^*_7,-D^*_7)\nonumber\\
                 =& e^{i\chi_{ST_1}}(-A_7^*,-B_7^*,D_7^*,C_7^*),\nonumber\\
(A'_7,B'_7,C'_7,B'_7)=&e^{i(\chi_{\overline{C}_6S}+\chi_1)}(A'_7,-B'_7,-D'_7,-C'_7),\nonumber\\
(A_8,B_8,C_8,D_8)=&e^{-i\chi_{ST_1}}(-A_8^*,D_8^*,-C_8^*,B_8^*); \tag{\ref{u100}}
\end{align}
\end{subequations}
\item $(w_{\overline{C}_6},w_S)=(0,1)$:
\begin{subequations}\label{u101}
\begin{align}
\alpha=&\alpha^*,\nonumber\\
(a,b,c,d)=&e^{i\chi_{\overline{C}_6S}}(-a^*,b^*,-c^*,-d^*)\nonumber\\
         =&(-a,-b,d,c),\nonumber\\
(A,B,C,D)=&(-A,-C,-B,-D),\nonumber\\
(A_3,B_3,C_3,D_3)=&(-A_3,-B_3,-D_3,-C_3)\nonumber\\
                 =&e^{2i\chi_1}(-A_3^*,B_3^*,C_3^*,D_3^*),\nonumber\\
(A'_3,B'_3,C'_3,D'_3)=&e^{-i \chi_1}(A'^*_3,-C'^*_3,-B'^*_3,-D'^*_3)\nonumber\\
                     =&e^{-i\chi_1}(-A'^*_3,B'^*_3,C'^*_3,D'^*_3),\nonumber\\
(A_4,B_4,C_4,D_4)=&e^{2i\chi_1}e^{i\chi_{\overline{C}_6S}}(-A_4^*,-B_4^*,-C_4^*,D_4^*),\nonumber\\
(A_5,B_5,C_5,D_5)=&\text{arbitrary},\nonumber\\
(A_6,B_6,C_6,D_6)=&(-A_6,-C_6,-B_6,-D_6)\nonumber\\
                 =&e^{-i \chi_1}(-A_6^*,B_6^*,C_6^*,D_6^*),\nonumber\\
(A_7,B_7,C_7,D_7)=&e^{i\chi_{\overline{C}_6S}}(-A^*_7,B^*_7,-C^*_7,-D^*_8)\nonumber\\
                 =&(-A_7,-B_7,D_7,C_7),\nonumber\\
(A'_7,B'_7,C'_7,D'_7)=&e^{i(\chi_{\overline{C}_6S}+\chi_1)}(A^{\prime*}_7,-B^{\prime*}_7,-D^{\prime*}_7,-C^{\prime*}_7),\nonumber\\
(A_8,B_8,C_8,D_8)=&(-A_8,D_8,-C_8,B_8);\tag{\ref{u101}}
\end{align}
\end{subequations}
\item $(w_{\overline{C}_6},w_S)=(1,0)$:
\begin{subequations}\label{u110}
\begin{align}
\alpha=&\alpha^*,\nonumber\\
(a,b,c,d)=&e^{i\chi_{\overline{C}_6S}}(-a^*,b^*,-c^*,-d^*)\nonumber\\
         =&(-a,-b,d,c),\nonumber\\
(A,B,C,D)=&(-A^*,-C^*,-B^*,-D^*),\nonumber\\
(A_3,B_3,C_3,D_3)=&(-A_3^*,-B_3^*,-D_3^*,-C_3^*)\nonumber\\
                 =&(-A_3,B_3,C_3,D_3),\nonumber\\
(A'_3,B'_3,C'_3,D'_3)=&e^{i\chi_1}(A'_3,-C'_3,-B'_3,-D'_3)\nonumber\\
                     =&e^{i\chi_1}(-A'_3,B'_3,C'_3,D'_3),\nonumber\\
(A_4,B_4,C_4,D_4)=&e^{i\chi_{\overline{C}_6S}}(-A_4^*,-B_4^*,-C_4^*,D_4^*),\nonumber\\
(A_5,B_5,C_5,D_5)=&\text{arbitrary},\nonumber\\
(A_6,B_6,C_6,D_6)=&(-A_6^*,-C_6^*,-B_6^*,-D_6^*)\nonumber\\
                 =&e^{i\chi_1}(-A_6,B_6,C_6,D_6),\nonumber\\
(A_7,B_7,C_7,D_7)=&e^{i\chi_{\overline{C}_6S}}(-A_7^*,B_7^*,-C_7^*,-D_7^*)\nonumber\\
                 =&(-A_7,-B_7,D_7,C_7),\nonumber\\
(A'_7,B'_7,C'_7,D'_7)=&e^{i(\chi_{\overline{C}_6S}+\chi_1)}(A'^*_7,-B'^*_7,-D'^*_7,-C'^*_7),\nonumber\\
(A_8,B_8,C_8,D_8)=&(-A_8^*,D_8^*,-C_8^*,B_8^*);\tag{\ref{u110}}
\end{align}
\end{subequations}
\item $(w_{\overline{C}_6},w_S)=(1,1)$:
\begin{subequations}\label{u111}
\begin{align}
\alpha=&\alpha^*,\nonumber\\
(a,b,c,d)=&e^{i\chi_{\overline{C}_6S}}(-a^*,b^*,-c^*,-d^*)\nonumber\\
         =&(-a^*,-b^*,d^*,c^*),\nonumber\\
(A,B,C,D)=&(-A,-C,-B,-D),\nonumber\\
(A_3,B_3,C_3,D_3)=&(-A_3,-B_3,-D_3,-C_3)\nonumber\\
                 =&(-A_3,B_3,C_3,D_3),\nonumber\\
(A'_3,B'_3,C'_3,D'_3)=&e^{i\chi_1}(A'^*_3,-C'^*_3,-B'^*_3,-D'^*_3)\nonumber\\
                     =&e^{i\chi_1}(-A'_3,B'_3,C'_3,D'_3),\nonumber\\
(A_4,B_4,C_4,D_4)=&e^{i\chi_{\overline{C}_6S}}(-A_4^*,-B_4^*,-C_4^*,D_4^*),\nonumber\\
(A_5,B_5,C_5,D_5)=&\text{arbitrary},\nonumber\\
(A_6,B_6,C_6,D_6)=&(-A_6,-C_6,-B_6,-D_6)\nonumber\\
                 =&e^{i\chi_1}(-A_6,B_6,C_6,D_6),\nonumber\\
(A_7,B_7,C_7,D_7)=&e^{i\chi_{\overline{C}_6S}}(-A_7^*,B_7^*,-C_7^*,-D_7^*)\nonumber\\
                     =&(-A_7^*,-B_7^*,D_7^*,C_7^*),\nonumber\\
(A'_7,B'_7,C'_7,D'_7)=&e^{i(\chi_{\overline{C}_6S}+\chi_1)}(A_7,-B_7,-D_7,-C_7),\nonumber\\
(A_8,B_8,C_8,D_8)=&(-A_8,D_8,-C_8,B_8).\tag{\ref{u111}}
\end{align}
\end{subequations}
\end{itemize}

For $\mathbb{Z}_2$ without time reversal, we have the following six cases:
\begin{itemize}
\item $(\chi_{ST_1},\chi_{S\overline{C}_6},\chi_{\overline{C}_6}) = (\chi_1,0,0)$:
\begin{subequations}\label{z2case1}
\begin{align}
&(\alpha_h,\alpha_p)=\text{arbitrary},\nonumber\\
&(a_h,b_h,c_h,d_h,a_p,b_p,c_p,d_p)\nonumber\\&=\eta_{\overline{C}_6S}(-a_h^*,b_h^*,-c_h^*,-d_h^*,a_p,-b_p,c_p,d_p)\nonumber\\&=(-a_h^*,-b_h^*,d_h^*,c_h^*,a_p,b_p,-d_p,-c_p),\nonumber\\
&(A_h,B_h,C_h,D_h,A_p,B_p,C_p,D_p)\nonumber\\&=(-A_h^*,-C_h^*,-B_h^*,-D_h^*,A_p,C_p,B_p,D_p),\nonumber\\
&(A_{3h},B_{3h},C_{3h},D_{3h},A_{3p},B_{3p},C_{3p},D_{3p})\nonumber\\&=(-A_{3h}^*,-B_{3h}^*,-D_{3h}^*,-C_{3h}^*,A_{3p},B_{3p},D_{3p},C_{3p})\nonumber\\&=(-A_{3h}^*,B_{3h}^*,C_{3h}^*,D_{3h}^*,A_{3p},-B_{3p},-C_{3p},-D_{3p}),\nonumber\\
&(A’_{3h},B’_{3h},C’_{3h},D’_{3h},A’_{3p},B’_{3p},C’_{3p},D’_{3p})\nonumber\\&=\eta_1(A’_{3h},-C’_{3h},-B’_{3h},-D’_{3h},A’_{3p},-C’_{3p},-B’_{3p},-D’_{3p})\nonumber\\&=\eta_1(-A’^*_{3h},B’^*_{3h},C’^*_{3h},D’^*_{3h},A’_{3p},-B’_{3p},-C’_{3p},-D’_{3p}),\nonumber\\
&(A_{4h},B_{4h},C_{4h},D_{4h},A_{4p},B_{4p},C_{4p},D_{4p})\nonumber\\&=\eta_{\overline{C}_6S}(-A_{4h}^*,-B_{4h}^*,-C_{4h}^*,D_{4h}^*,A_{4p},B_{4p},C_{4p},-D_{4p}),\nonumber\\
&(A_{5h},B_{5h},C_{5h},D_{5h},A_{5p},B_{5p},C_{5p},D_{5p})=\text{arbitrary},\nonumber\\
&(A_{6h},B_{6h},C_{6h},D_{6h},A_{6p},B_{6p},C_{6p},D_{6p})\nonumber\\&=(-A_{6h}^*,-C_{6h}^*,-B_{6h}^*,-D_{6h}^*,A_{6p},C_{6p},B_{6p},D_{6p})\nonumber\\&=\eta_1(-A_{6h}^*,B_{6h}^*,C_{6h}^*,D_{6h}^*,A_{6p},-B_{6p},-C_{6p},-D_{6p}),\nonumber\\
&(A_{7h},B_{7h},C_{7h},D_{7h},A_{7p},B_{7p},C_{7p},D_{7p})\nonumber\\&=\eta_{\overline{C}_6S}(-A_{7h}^*,B_{7h}^*,-C_{7h}^*,-D_{7h}^*,A_{7p},-B_{7p},C_{7p},D_{7p})\nonumber\\&=(-A_{7h}^*,-B_{7h}^*,D_{7h}^*,C_{7h}^*,A_{7p},B_{7p},-D_{7p},-C_{7p}),\nonumber\\
&(A’_{7h},B’_{7h},C’_{7h},D’_{7h},A’_{7p},B’_{7p},C’_{7p},D’_{7p})\nonumber\\&=\eta_1\eta_{\overline{C}_6S}(A’_{7h},-B’_{7h},-D’_{7h},-C’_{7h},A’_{7p},-B’_{7p},-D’_{7p},-C’_{7p}),\nonumber\\
&(A_{8h},B_{8h},C_{8h},D_{8h},A_{8p},B_{8p},C_{8p},D_{8p})\nonumber\\&=(-A_{8h}^*,D_{8h}^*,-C_{8h}^*,B_{8h}^*,A_{8p},-D_{8p},C_{8p},-B_{8p});\tag{\ref{z2case1}}
\end{align}
\end{subequations}
\item $(\chi_{ST_1},\chi_{S\overline{C}_6},\chi_{\overline{C}_6}) = (\chi_1,\pi,\pi)$:
\begin{subequations}\label{z2case2}
\begin{align}
&(\alpha_h,\alpha_p)=(\alpha_h^*,\alpha_p^*),\nonumber\\
&(a_h,b_h,c_h,d_h,a_p,b_p,c_p,d_p)\nonumber\\&=\eta_{\overline{C}_6S}(a_h,-b_h,c_h,d_h,-a_p^*,b_p^*,-c_p^*,-d_p^*)\nonumber\\&=(a_h^*,b_h^*,-d_h^*,-c_h^*,-a_p,-b_p,d_p,c_p),\nonumber\\
&(A_h,B_h,C_h,D_h,A_p,B_p,C_p,D_p)\nonumber\\&=(A_h,C_h,B_h,D_h,-A_p^*,-C_p^*,-B_p^*,-D_p^*),\nonumber\\
&(A_{3h},B_{3h},C_{3h},D_{3h},A_{3p},B_{3p},C_{3p},D_{3p})\nonumber\\&=(-A_{3h}^*,-B_{3h}^*,-D_{3h}^*,-C_{3h}^*,A_{3p},B_{3p},D_{3p},C_{3p})\nonumber\\&=(-A_{3h},B_{3h},C_{3h},D_{3h},A_{3p}^*,-B_{3p}^*,-C_{3p}^*,-D_{3p}^*),\nonumber\\
&(A’_{3h},B’_{3h},C’_{3h},D’_{3h},A’_{3p},B’_{3p},C’_{3p},D’_{3p})\nonumber\\&=\eta_1(A’_{3h},-C’_{3h},-B’_{3h},-D’_{3h},A’_{3p},-C’_{3p},-B’_{3p},-D’_{3p})\nonumber\\&=\eta_1(-A’_{3h},B’_{3h},C’_{3h},D’_{3h},A’^*_{3p},-B’^*_{3p},-C’^*_{3p},-D’^*_{3p}),\nonumber\\
&(A_{4h},B_{4h},C_{4h},D_{4h},A_{4p},B_{4p},C_{4p},D_{4p})\nonumber\\&=\eta_{\overline{C}_6S}(-A_{4h}^*,-B_{4h}^*,-C_{4h}^*,D_{4h}^*,A_{4p},B_{4p},C_{4p},-D_{4p}),\nonumber\\
&(A_{5h},B_{5h},C_{5h},D_{5h},A_{5p},B_{5p},C_{5p},D_{5p})=\text{arbitrary},\nonumber\\
&(A_{6h},B_{6h},C_{6h},D_{6h},A_{6p},B_{6p},C_{6p},D_{6p})\nonumber\\&=(-A_{6h}^*,-C_{6h}^*,-B_{6h}^*,-D_{6h}^*,A_{6p},C_{6p},B_{6p},D_{6p})\nonumber\\&=\eta_1(-A_{6h},B_{6h},C_{6h},D_{6h},A_{6p}^*,-B_{6p}^*,-C_{6p}^*,-D_{6p}^*),\nonumber\\
&(A_{7h},B_{7h},C_{7h},D_{7h},A_{7p},B_{7p},C_{7p},D_{7p})\nonumber\\&=\eta_{\overline{C}_6S}(A_{7h},-B_{7h},C_{7h},D_{7h},-A_{7p}^*,B_{7p}^*,-C_{7p}^*,-D_{7p}^*)\nonumber\\&=(A_{7h}^*,B_{7h}^*,-D_{7h}^*,-C_{7h}^*,-A_{7p},-B_{7p},D_{7p},C_{7p}),\nonumber\\
&(A’_{7h},B’_{7h},C’_{7h},D’_{7h},A’_{7p},B’_{7p},C’_{7p},D’_{7p})\nonumber\\&=\eta_1\eta_{\overline{C}_6S}(A’^*_{7h},-B’^*_{7h},-D’^*_{7h},-C’^*_{7h},A’^*_{7p},-B’^*_{7p},-D’^*_{7p},-C’^*_{7p}),\nonumber\\
&(A_{8h},B_{8h},C_{8h},D_{8h},A_{8p},B_{8p},C_{8p},D_{8p})\nonumber\\
&=(-A_{8h}^*,D_{8h}^*,-C_{8h}^*,B_{8h}^*,A_{8p},-D_{8p},C_{8p},-B_{8p});\tag{\ref{z2case2}}
\end{align}
\end{subequations}
\item  $(\chi_{ST_1},\chi_{S\overline{C}_6},\chi_{\overline{C}_6}) = (\chi_1+\pi,\pi,0)$:
\begin{subequations}\label{z2case3}
\begin{align}
&(\alpha_h,\alpha_p)=(\alpha_h^*,e^{i\frac{10\pi}{3}  {j} } \alpha_p^*)=(\alpha_h^*,e^{i\frac{2\pi}{3}  {j} } \alpha_p^*)=(\alpha_h^*,\alpha_p^*)\nonumber\\&=(\alpha_h,e^{-i\frac{8\pi}{3} {j} } \alpha_p)=(\alpha_h,e^{-i\frac{16\pi}{3}  {j}} \alpha_p)=(\alpha_h^*,e^{i\frac{16\pi}{3}  {j} } \alpha_p^*)\nonumber\\&=(\alpha_h^*,e^{i\frac{8\pi}{3}  {j} } \alpha_p^*)=(\alpha_h,e^{-i\frac{14\pi}{3}  {j} } \alpha_p)=(\alpha_h,e^{-i\frac{22\pi}{3} {j}} \alpha_p),\nonumber\\
&(a_h,b_h,c_h,d_h,a_p,b_p,c_p,d_p)\nonumber\\&=((-1)^{j} a_h,-(-1)^{{j}} b_h,(-1)^{j} c_h,(-1)^{j} d_h,\nonumber\\
&\quad\qquad -a_p^* e^{i \frac{5\pi}{3} {j}},b_p^* e^{i \frac{5\pi}{3}{j}},-c_p^* e^{i \frac{5\pi}{3}{j}},-d_p^* e^{i \frac{5\pi}{3}{j}})\nonumber\\&=\eta_{\overline{C}_6S}(- a_h^* e^{-i\frac{7\pi}{3}{j}},- b_h^* e^{-i\frac{7\pi}{3} {j}}, d_h^* e^{-\frac{7\pi}{3} {j}}, c_h^* e^{-i\frac{7\pi}{3} {j}},\nonumber\\
&\quad\qquad (-1)^{{j}} a_p,(-1)^{{j}} b_p,-(-1)^{{j}} d_p,-(-1)^{{j}} c_p),\nonumber\\
&(A_h,B_h,C_h,D_h,A_p,B_p,C_p,D_p)\nonumber\\&=(-A_h,-C_h,-B_h,-D_h,\nonumber\\
&\quad\qquad A_p^* e^{i\frac{10\pi}{3}{j}},C_p^* e^{i\frac{10\pi}{3}{j}},B_p^* e^{i\frac{10\pi}{3}{j}},D_p^* e^{i\frac{10\pi}{3}{j}}),\nonumber\\
&(A_{3h},B_{3h},C_{3h},D_{3h},A_{3p},B_{3p},C_{3p},D_{3p})\nonumber\\&=(-A_{3h},-B_{3h},-D_{3h},-C_{3h},A_{3p}^*,B_{3p}^*,D_{3p}^*,C_{3p}^*)\nonumber\\&=(-A_{3h}^*,B_{3h}^*,C_{3h}^*,D_{3h}^*,A_{3p},-B_{3p},-C_{3p},-D_{3p}),\nonumber\\
&(A’_{3h},B’_{3h},C’_{3h},D’_{3h},A’_{3p},B’_{3p},C’_{3p},D’_{3p})\nonumber\\&=\eta_1( A’^*_{3h},-C’^*_{3h},-B’^*_{3h},-D’^*_{3h},\nonumber\\
&\quad\qquad A’^*_{3p} e^{i\frac{10\pi}{3}{j}} ,-C’^*_{3p} e^{i\frac{10\pi}{3}{j}} ,-B’^*_{3p} e^{i\frac{10\pi}{3}{j}} ,-D’^*_{3p} e^{i\frac{10\pi}{3}{j}})\nonumber\\&=\eta_1(-A’^*_{3h},B’^*_{3h},C’^*_{3h},D’^*_{3h},A’_{3p},-B’_{3p},-C’_{3p},-D’_{3p}),\nonumber\\
&(A_{4h},B_{4h},C_{4h},D_{4h},A_{4p},B_{4p},C_{4p},D_{4p})\nonumber\\&=((-1)^{{j}} A_{4h},(-1)^{{j}} B_{4h},(-1)^{{j}} C_{4h},-(-1)^{{j}} D_{4h},\nonumber\\
&\quad\qquad -A_{4p}^* e^{i \frac{\pi}{3}{j}},-B_{4p}^* e^{i \frac{\pi}{3}{j}}, -C_{4p}^* e^{i \frac{\pi}{3}{j}},D_{4p}^* e^{i \frac{\pi}{3}{j}}),\nonumber\\
&(A_{5h},B_{5h},C_{5h},D_{5h},A_{5p},B_{5p},C_{5p},D_{5p})=\text{arbitrary},\nonumber\\
&(A_{6h},B_{6h},C_{6h},D_{6h},A_{6p},B_{6p},C_{6p},D_{6p})\nonumber\\&=(-A_{6h},-C_{6h},-B_{6h},-D_{6h},\nonumber\\
&\quad\qquad A_{6p}^* e^{i \frac{10\pi}{3}{j}},C_{6p}^* e^{i \frac{10\pi}{3}{j}},B_{6p}^* e^{i \frac{10\pi}{3}{j}},D_{6p}^* e^{i \frac{10\pi}{3}{j}})\nonumber\\&=\eta_1(-A_{6h}^*,B_{6h}^*,C_{6h}^*,D_{6h}^*,A_{6p},-B_{6p},-C_{6p},-D_{6p}),\nonumber\\
&(A_{7h},B_{7h},C_{7h},D_{7h},A_{7p},B_{7p},C_{7p},D_{7p})\nonumber\\&=((-1)^{j} A_{7h},-(-1)^{{j}} B_{7h},(-1)^{j} C_{7h},(-1)^{j} D_{7h},\nonumber\\
&\qquad\quad -A_{7p}^* e^{i\frac{5\pi}{3} j},B_{7p}^* e^{i\frac{5\pi}{3} j},-C_{7p}^* e^{i\frac{5\pi}{3} j},-D_{7p}^* e^{i\frac{5\pi}{3} j})\nonumber\\&=\eta_{\overline{C}_6S}(- A_{7h}^* e^{-i\frac{7 \pi}{3}{j}},- B_{7h}^* e^{-i\frac{7 \pi}{3}{j}},D_{7h}^* e^{-i\frac{7 \pi}{3}{j}},C_{7h}^* e^{-i\frac{7 \pi}{3}{j}},\nonumber\\
&\qquad\quad(-1)^{j} A_{7p},(-1)^{j} B_{7p},-(-1)^{j}) D_{7p},-(-1)^{j}) C_{7p}),\nonumber\\
&(A’_{7h},B’_{7h},C’_{7h},D’_{7h},A’_{7p},B’_{7p},C’_{7p},D’_{7p})\nonumber\\&=\eta_1\eta_{\overline{C}_6S}(-A’^*_{7h} e^{-i\frac{4\pi}{3}{j}},B’^*_{7h} e^{-i\frac{4\pi}{3}{j}},D’^*_{7h} e^{-i\frac{4\pi}{3}{j}},C’^*_{7h} e^{-i\frac{4\pi}{3}{j}},\nonumber\\
&\qquad\quad -A’^*_{7p} e^{i\frac{20\pi}{3}{j}},B’^*_{7p} e^{i\frac{20\pi}{3}{j}},D’^*_{7p} e^{i\frac{20\pi}{3}{j}},C’^*_{7p} e^{i\frac{20\pi}{3}{j}}),\nonumber\\
&(A_{8h},B_{8h},C_{8h},D_{8h},A_{8p},B_{8p},C_{8p},D_{8p})\nonumber\\&=(A_{8h}^* e^{i\frac{4\pi}{3} j},-D_{8h}^* e^{i\frac{4\pi}{3} j},C_{8h}^* e^{i\frac{4\pi}{3} j},-B_{8h}^* e^{i\frac{4\pi}{3} j},\nonumber\\
&\quad\qquad -A_{8p},D_{8p},-C_{8p},B_{8p});\tag{\ref{z2case3}}
\end{align}
\end{subequations}
\item $(\chi_{ST_1},\chi_{S\overline{C}_6},\chi_{\overline{C}_6}) = (\chi_1+\pi,0,\pi)$:
\begin{subequations}\label{z2case4}
\begin{align}
&(\alpha_h,\alpha_p)=(\alpha_h^*,\alpha_p^*),\nonumber\\
&(a_h,b_h,c_h,d_h,a_p,b_p,c_p,d_p)\nonumber\\&=\eta_{\overline{C}_6S}(-a_h^*,b_h^*,-c_h^*,-d_h^*,a_p,-b_p,c_p,d_p)\nonumber\\&=(-a_h^*,-b_h^*,d_h^*,c_h^*,a_p,b_p,-d_p,-c_p),\nonumber\\
&(A_h,B_h,C_h,D_h,A_p,B_p,C_p,D_p)\nonumber\\&=(A_h^*,C_h^*,B_h^*,D_h^*,-A_p,-C_p,-B_p,-D_p),\nonumber\\
&(A_{3h},B_{3h},C_{3h},D_{3h},A_{3p},B_{3p},C_{3p},D_{3p})\nonumber\\&=(-A_{3h},-B_{3h},-D_{3h},-C_{3h},A_{3p}^*,B_{3p}^*,D_{3p}^*,C_{3p}^*)\nonumber\\&=(-A_{3h},B_{3h},C_{3h},D_{3h},A_{3p}^*,-B_{3p}^*,-C_{3p}^*,-D_{3p}^*),\nonumber\\
&(A’_{3h},B’_{3h},C’_{3h},D’_{3h},A’_{3p},B’_{3p},C’_{3p},D’_{3p})\nonumber\\&=\eta_1(A’^*_{3h},-C’^*_{3h},-B’^*_{3h},-D’^*_{3h},A’^*_{3p},-C’^*_{3p},-B’^*_{3p},-D’^*_{3p})\nonumber\\&=\eta_1(-A’_{3h},B’_{3h},C’_{3h},D’_{3h},A’^*_{3p},-B’^*_{3p},-C’^*_{3p},-D’^*_{3p}),\nonumber\\
&(A_{4h},B_{4h},C_{4h},D_{4h},A_{4p},B_{4p},C_{4p},D_{4p})\nonumber\\&=\eta_{\overline{C}_6S}(A_{4h},B_{4h},C_{4h},-D_{4h},-A_{4p}^*,-B_{4p}^*,-C_{4p}^*,D_{4p}^*),\nonumber\\
&(A_{5h},B_{5h},C_{5h},D_{5h},A_{5p},B_{5p},C_{5p},D_{5p})=\text{arbitrary},\nonumber\\
&(A_{6h},B_{6h},C_{6h},D_{6h},A_{6p},B_{6p},C_{6p},D_{6p})\nonumber\\&=(-A_{6h},-C_{6h},-B_{6h},-D_{6h},A_{6p}^*,C_{6p}^*,B_{6p}^*,D_{6p}^*)\nonumber\\&=\eta_1(-A_{6h},B_{6h},C_{6h},D_{6h},A_{6p}^*,-B_{6p}^*,-C_{6p}^*,-D_{6p}^*),\nonumber\\
&(A_{7h},B_{7h},C_{7h},D_{7h},A_{7p},B_{7p},C_{7p},D_{7p})\nonumber\\&=\eta_{\overline{C}_6S}(-A_{7h}^*,B_{7h}^*,-C_{7h}^*,-D_{7h}^*,A_{7p},-B_{7p},C_{7p},D_{7p})\nonumber\\&=(-A_{7h}^*,-B_{7h}^*,D_{7h}^*,C_{7h}^*,A_{7p},B_{7p},-D_{7p},-C_{7p}),\nonumber\\
&(A’_{7h},B’_{7h},C’_{7h},D’_{7h},A’_{7p},B’_{7p},C’_{7p},D’_{7p})\nonumber\\&=\eta_1\eta_{\overline{C}_6S}(A’_{7h},-B’_{7h},-D’_{7h},-C’_{7h},A’_{7p},-B’_{7p},-D’_{7p},-C’_{7p}),\nonumber\\
&(A_{8h},B_{8h},C_{8h},D_{8h},A_{8p},B_{8p},C_{8p},D_{8p})\nonumber\\&=(A_{8h}^*,-D_{8h}^*,C_{8h}^*,-B_{8h}^*,-A_{8p},D_{8p},-C_{8p},B_{8p});\tag{\ref{z2case4}}
\end{align}
\end{subequations}
\item $(\chi_{ST_1},\chi_{S\overline{C}_6},\chi_{\overline{C}_6},j) = (\chi_1+\pi,\pi,\pi,0)$:
\begin{subequations}\label{z2case5}
\begin{align}
&(\alpha_h,\alpha_p)=(\alpha_h^*,e^{i\frac{2\pi}{3}} \alpha_p^*)=(\alpha_h^*,-e^{i\frac{\pi}{3}} \alpha_p^*)\nonumber\\&=(\alpha_h^*,\alpha_p^*)
=(\alpha_h, e^{i\frac{2\pi}{3}} \alpha_p)=(\alpha_h,-e^{i\frac{\pi}{3}} \alpha_p)\nonumber\\&=(\alpha_h,-\alpha_p)=(\alpha_h^*,-e^{i\frac{2\pi}{3}} \alpha_p^*)=(\alpha_h^*,e^{i\frac{\pi}{3}} \alpha_p^*)\nonumber\\
&=(\alpha_h^*,-\alpha_p^*)=(\alpha_h,-e^{i\frac{2\pi}{3}}\alpha_p)=(\alpha_h,e^{i\frac{\pi}{3}} \alpha_p),\nonumber\\
&(a_h,b_h,c_h,d_h,a_p,b_p,c_p,d_p)\nonumber\\&=\eta_{\overline{C}_6S}(-a_h,b_h,-c_h,-d_h,\nonumber\\
&\quad\qquad-e^{i \frac{4\pi}{3}} a_p^*,e^{i \frac{4\pi}{3}} b_p^*,-e^{i \frac{4\pi}{3}} c_p^*,-e^{i \frac{4\pi}{3}} d_p^*)\nonumber\\&=(-e^{i\frac{\pi}{3}} a_h^*,-e^{i\frac{\pi}{3}} b_h^*,e^{i\frac{\pi}{3}} d_h^*,e^{i\frac{\pi}{3}} c_h^*,-a_p,-b_p,d_p,c_p),\nonumber\\
&(A_h,B_h,C_h,D_h,A_p,B_p,C_p,D_p)\nonumber\\&=-(A_h,C_h,B_h,D_h,e^{i\frac{2\pi}{3}} A_p^*,e^{i\frac{2\pi}{3}} C_p^*,e^{i\frac{2\pi}{3}} B_p^*,e^{i\frac{2\pi}{3}} D_p^*),\nonumber\\
&(A_{3h},B_{3h},C_{3h},D_{3h},A_{3p},B_{3p},C_{3p},D_{3p})\nonumber\\&=(-A_{3h},-B_{3h},-D_{3h},-C_{3h},A_{3p}^*,B_{3p}^*,D_{3p}^*,C_{3p}^*)\nonumber\\&=(-A_{3h}^*,B_{3h}^*,C_{3h}^*,D_{3h}^*,-A_{3p},B_{3p},C_{3p},D_{3p}),\nonumber\\
&(A’_{3h},B’_{3h},C’_{3h},D’_{3h},A’_{3p},B’_{3p},C’_{3p},D’_{3p})\nonumber\\&=\eta_1( A’^*_{3h},-C’^*_{3h},-B’^*_{3h},-D’^*_{3h},\nonumber\\
&\quad\qquad e^{i\frac{2\pi}{3}} A’^*_{3p},-e^{i\frac{2\pi}{3}}C’^*_{3p},-e^{i\frac{2\pi}{3}}B’^*_{3p},-e^{i\frac{2\pi}{3}}D’^*_{3p})\nonumber\\&=\eta_1(-A’^*_{3h},B’^*_{3h},C’^*_{3h},D’^*_{3h},-A’_{3p},B’_{3p},C’_{3p},D’_{3p}),\nonumber\\
&(A_{4h},B_{4h},C_{4h},D_{4h},A_{4p},B_{4p},C_{4p},D_{4p})\nonumber\\&=\eta_{\overline{C}_6S}(-A_{4h},-B_{4h},-C_{4h},D_{4h},\nonumber\\
&\quad\qquad e^{i\frac{2\pi}{3}} A_{4p}^*,e^{i\frac{2\pi}{3}} B_{4p}^*,e^{i\frac{2\pi}{3}}C_{4p}^* ,-e^{i\frac{2\pi}{3}}D_{4p}^*),\nonumber\\
&(A_{5h},B_{5h},C_{5h},D_{5h},A_{5p},B_{5p},C_{5p},D_{5p})=\text{arbitrary},\nonumber\\&(A_{6h},B_{6h},C_{6h},D_{6h},A_{6p},B_{6p},C_{6p},D_{6p})\nonumber\\&= (-A_{6h},-C_{6h},-B_{6h},-D_{6h},e^{i\frac{2\pi}{3}} A_{6p}^*,e^{i\frac{2\pi}{3}} C_{6p}^*,e^{i\frac{2\pi}{3}} B_{6p}^*,e^{i\frac{2\pi}{3}} D_{6p}^*)\nonumber\\&=\eta_1(-A_{6h}^*,B_{6h}^*,C_{6h}^*,D_{6h}^*,-A_{6p},B_{6p},C_{6p},D_{6p}),\nonumber\\
&(A_{7h},B_{7h},C_{7h},D_{7h},A_{7p},B_{7p},C_{7p},D_{7p})\nonumber\\&=\eta_{\overline{C}_6S}(-A_{7h}, B_{7h},-C_{7h},-D_{7h},\nonumber\\
&\quad\qquad e^{i\frac{\pi}{3}} A_{7p}^*,-e^{i\frac{\pi}{3}}B_{7p}^*,e^{i\frac{\pi}{3}} C_{7p}^*,e^{i\frac{\pi}{3}}D_{7p}^*)\nonumber\\&=(-e^{i\frac{\pi}{3}} A_{7h}^*,-e^{i\frac{\pi}{3}} B_{7h}^*,e^{i\frac{\pi}{3}} D_{7h}^*,e^{i\frac{\pi}{3}} C_{7h}^*,\nonumber\\
&\quad\qquad -A_{7p},-B_{7p},D_{7p},C_{7p}),\nonumber\\
&(A’_{7h},B’_{7h},C’_{7h},D’_{7h},A’_{7p},B’_{7p},C’_{7p},D’_{7p})\nonumber\\&=e^{i\frac{\pi}{3}}\eta_1\eta_{\overline{C}_6S}(A’^*_{7h},-B’^*_{7h},-D’^*_{7h},-C’^*_{7h},-A’^*_{7p},B’^*_{7p},D’^*_{7p},C’^*_{7p}),\nonumber\\
&(A_{8h},B_{8h},C_{8h},D_{8h},A_{8p},B_{8p},C_{8p},D_{8p})\nonumber\\&=(e^{i\frac{2\pi}{3}}A_{8h}^*,-e^{i\frac{2\pi}{3}} D^*_{8h},e^{i\frac{2\pi}{3}}C^*_{8h},-e^{i\frac{2\pi}{3}} B^*_{8h},\nonumber\\
&\quad\qquad -A_{8p},D_{8p},-C_{8p},B_{8p});\tag{\ref{z2case5}}
\end{align}
\end{subequations}
\item $(\chi_{ST_1},\chi_{S\overline{C}_6},\chi_{\overline{C}_6},j) = (\chi_1+\pi,\pi,\pi,1)$:
\begin{subequations}\label{z2case6}
\begin{align}
&(\alpha_h,\alpha_p)=(\alpha_h^*,\alpha_p^*)=(\alpha_h,-\alpha_p)=(\alpha_h^*,-\alpha_p^*),\nonumber\\
&(a_h,b_h,c_h,d_h,a_p,b_p,c_p,d_p)\nonumber\\&=\eta_{\overline{C}_6S}(-a_h, b_h,-c_h,-d_h,-a_p^*,b_p^*,-c_p^*,-d_p^*)\nonumber\\&=(a_h^*,b_h^*,-d_h^*,-c_h^*,-a_p,-b_p,d_p,c_p),\nonumber\\
&(A_h,B_h,C_h,D_h,A_p,B_p,C_p,D_p)\nonumber\\&=-(A_h,C_h,B_h,D_h,A_p^*,C_p^*,B_p^*,D_p^*),\nonumber\\
&(A_{3h},B_{3h},C_{3h},D_{3h},A_{3p},B_{3p},C_{3p},D_{3p})\nonumber\\&=(-A_{3h},-B_{3h},-D_{3h},-C_{3h},A_{3p}^*,B_{3p}^*,D_{3p}^*,C_{3p}^*)\nonumber\\&=(-A_{3h}^*,B_{3h}^*,C_{3h}^*,D_{3h}^*,-A_{3p},B_{3p},C_{3p},D_{3p}),\nonumber\\
&(A’_{3h},B’_{3h},C’_{3h},D’_{3h},A’_{3p},B’_{3p},C’_{3p},D’_{3p})\nonumber\\&=\eta_1( A’^*_{3h},-C’^*_{3h},-B’^*_{3h},-D’^*_{3h},A’^*_{3p},-C’^*_{3p},-B’^*_{3p},-D’^*_{3p}),\nonumber\\&=\eta_1(-A’^*_{3h},B’^*_{3h},C’^*_{3h},D’^*_{3h},-A’_{3p},B’_{3p},C’_{3p},D’_{3p}),\nonumber\\
&(A_{4h},B_{4h},C_{4h},D_{4h},A_{4p},B_{4p},C_{4p},D_{4p})\nonumber\\&=\eta_{\overline{C}_6S}(-A_{4h},-B_{4h},-C_{4h},D_{4h},A_{4p}^*,B_{4p}^*,C_{4p}^*,-D_{4p}^*),\nonumber\\
&(A_{5h},B_{5h},C_{5h},D_{5h},A_{5p},B_{5p},C_{5p},D_{5p})=\text{arbitrary},\nonumber\\
&(A_{6h},B_{6h},C_{6h},D_{6h},A_{6p},B_{6p},C_{6p},D_{6p})\nonumber\\&=(-A_{6h},-C_{6h},-B_{6h},-D_{6h},A_{6p}^*,C_{6p}^*,B_{6p}^*,D_{6p}^*)\nonumber\\&=\eta_1(-A_{6h}^*,B_{6h}^*,C_{6h}^*,D_{6h}^*,-A_{6p},B_{6p},C_{6p},D_{6p}),\nonumber\\
&(A_{7h},B_{7h},C_{7h},D_{7h},A_{7p},B_{7p},C_{7p},D_{7p})\nonumber\\&=\eta_{\overline{C}_6S}(-A_{7h},B_{7h},-C_{7h},-D_{7h},-A_{7p}^*,B_{7p}^*,-C_{7p}^*,-D_{7p}^*)\nonumber\\&=(A_{7h}^*,B_{7h}^*,-D_{7h}^*,-C_{7h}^*,-A_{7p},-B_{7p},D_{7p},C_{7p}),\nonumber\\
&(A’_{7h},B’_{7h},C’_{7h},D’_{7h},A’_{7p},B’_{7p},C’_{7p},D’_{7p}),\nonumber\\&=\eta_1\eta_{\overline{C}_6S}(-A’^*_{7h},B’^*_{7h},D’^*_{7h},C’^*_{7h},A’^*_{7p}, -B’^*_{7p},-D’^*_{7p},-C’^*_{7p}),\nonumber\\
&(A_{8h},B_{8h},C_{8h},D_{8h},A_{8p},B_{8p},C_{8p},D_{8p}),\nonumber\\&=(A_{8h}^*,-D_{8h}^*,C_{8h}^*,-B_{8h}^*,-A_{8p},D_{8p},-C_{8p},B_{8p}).\tag{\ref{z2case6}}
\end{align}
\end{subequations}
\end{itemize}

\section{\label{app:H}U(1) 0-flux MF  ans\"atze with $m_S=1$ has nodal star: a proof}

Let us first state a Lemma: define a $6\times 6$ matrix
\begin{equation}
A(x,y,z) \equiv \left(\begin{array}{ccc}x(\sigma_2-\sigma_3) &y(\sigma_1-\sigma_2)&z(\sigma_3-\sigma_1)\\
z(\sigma_1-\sigma_2) & x(\sigma_3-\sigma_1) &y(\sigma_2-\sigma_3)\\
y(\sigma_3-\sigma_1)&z(\sigma_2-\sigma_3)&x(\sigma_1-\sigma_2)\end{array}\right),
\end{equation}
(what we will use later is a special case with $(x,y,z) = (c,c',c'^*)$.)
then its inverse is
\begin{equation}
\left(A(x,y,z)\right)^{-1} = \frac{A(x^2-yz,z^2-xy,y^2-xz)}{2(x^3+y^3+z^3-3xyz)}.
\end{equation}
Furthermore,  define a $2\times 6$ matrix
\begin{equation}
\begin{aligned}
&B(\alpha,\beta,\gamma,\delta) \equiv \\
&\left(
\alpha\sigma^0+(\beta,\gamma,\delta)\cdot \bm{\sigma},\alpha \sigma^0+(\delta,\beta,\gamma)\cdot \bm{\sigma},\alpha \sigma^0+(\gamma,\delta,\beta)\cdot \bm{\sigma}\right)
\end{aligned}
\end{equation}
and a $6\times 2$ matrix
\begin{equation}
C(\alpha',\beta',\gamma',\delta') \equiv \left(\begin{array}{c}
\alpha'\sigma^0+(\beta',\gamma',\delta')\cdot \bm{\sigma}\\
\alpha' \sigma^0+(\delta',\beta',\gamma')\cdot \bm{\sigma}\\
\alpha' \sigma^0+(\gamma',\delta',\beta')\cdot \bm{\sigma}\end{array}\right)
\end{equation}
then for any complex $\zeta$ we have (be wary of the switching $\delta\leftrightarrow \gamma$ between $B$ and $C$ below)
\begin{equation}
B(\alpha,\beta,\gamma,\delta)\left(A(x,y,z)\right)^{-1}C(\zeta \alpha,\zeta \beta,\zeta \delta,\zeta \gamma) = 0.
\end{equation}
The proof of this Lemma is elementary.

Now we use the lemma to prove that the existence of nodal star. The U(1) 0-flux mean-field  ans\"atze correspond to an $8\times 8$ matrix $\mathcal{H}_{\text{U(1)}}(\bm{k})$ in the momentum space, with basis the parton operators $\left(f_{0\bm{k}\uparrow},f_{0\bm{k}\downarrow},
f_{1\bm{k}\uparrow},f_{1\bm{k}\downarrow},
f_{2\bm{k}\uparrow},f_{2\bm{k}\downarrow},
f_{3\bm{k}\uparrow},f_{3\bm{k}\downarrow}\right)^T$. We abbreviate $\mathcal{H}_{\text{U(1)}}(\bm{k})$ by $\mathcal{H}(\bm{k})$ for the rest of this appendix. For U(1) 0-flux states with the PSG number $w_S=1$, we have the three-fold rotation symmetry $C_3$ along $(1,1,1)$ axis, and the screw symmetry $S$:
\begin{equation}
\begin{aligned}
W^\dag_{C_3}(\bm{k}) \mathcal{H}(\bm{k}) W_{C_3}(\bm{k}) &= \mathcal{H}(k_z,k_x,k_y),\\
W^\dag_S(\bm{k}) \mathcal{H}(\bm{k}) W_S(\bm{k}) &= -\mathcal{H}^T(-k_y,-k_x,k_z),
\end{aligned}
\end{equation}
notice the second line is \emph{specific} to $w_S=1$. Now define the operation $R\equiv S\circ C_3\circ C_3\circ S\circ C_3\circ S$. Then we notice that $R$ and $C_3$ both map the momentum $(k,k,k)$ back to itself:
\begin{equation}
\begin{aligned}
W^\dag_{C_3}(k,k,k)\mathcal{H}(k,k,k)W_{C_3}(k,k,k) &= \mathcal{H}(k,k,k),\\
W^\dag_R(k,k,k) \mathcal{H}(k,k,k)W_R(k,k,k) &= -\mathcal{H}^T(k,k,k).
\end{aligned}
\end{equation}

Now, assume the most general form of an $8\times 8$ hermitian matrix $\mathcal{H}_{\text{U(1)}}(k,k,k) = \left[ h_{\mu\nu}\right]$, where $h_{\mu\nu}$ are $2\times 2$ blocks with $\mu,\nu=0,1,2,3$. Using the special form of $W_{C_3}(k,k,k)$ and $W_R(k,k,k)$, we can show step by step the following:

\begin{itemize}
\item $h_{00} = 0$, $h_{11} = c(\sigma^2-\sigma^3)$, $h_{22} = c(\sigma^3-\sigma^1)$, $h_{33} = c(\sigma^1-\sigma^2)$, where $c$ is some real parameter;
\item $h_{12} = c'(\sigma^1-\sigma^2)$, $h_{13} = c'^*(\sigma^3-\sigma^1)$, $h_{23} = c'(\sigma^2-\sigma^3)$, where $c'$ is some complex parameter;
\item $h_{01} = a'' \sigma^0+(b'',c'',d'')\cdot \bm{\sigma}$, $h_{02} = a'' \sigma^0 +(d'',b'',c'') \cdot \bm{\sigma}$, and $h_{03} = a''\sigma^0+(c'',d'',b'')\cdot \bm{\sigma}$, where $(a'',b'',c'',d'')$ are complex parameters satisfying $ a'' e^{ik} = -a''^*$, $e^{ik}(-b'',-d'',-c'') = (b''^*,c''^*,d''^*)$.
\end{itemize}

Therefore the Hamiltonian can be written in the form
\begin{equation}
\mathcal{H}(k,k,k) = \left(\begin{array}{cc} 0 & B \\ C & A \end{array}\right),
\end{equation}
where $A$ is a $6\times 6$ block containing $h_{ij}$ blocks with $i,j=1,2,3$, $B$ is a $2\times 6$ block containing $h_{01},h_{02}$ and $h_{03}$, and $C = B^\dag$. Using the above Lemma, we can show that $B A^{-1} C = 0_{2\times 2}$, therefore using the standard matrix decomposition, we have
\begin{equation}
\mathcal{H}(k,k,k)
 = \left(\begin{array}{cc} 1_{2\times 2} & BA^{-1}\\ 0& 1_{6\times 6} \end{array}\right) \left(\begin{array}{cc} 0 & 0 \\ 0 & A \end{array}\right)\left(\begin{array}{cc} 1_{2\times 2} & 0 \\ A^{-1}C & 1_{6\times 6} \end{array}\right),
 \end{equation}
 we see that the rank of $\mathcal{H}_{\text{U(1)}}(k,k,k)$ is smaller or equal to 6, i.e. $\mathcal{H}_{\text{U(1)}}(k,k,k)$ at least has two zero eigenvalues. The existence of zero eigenvalues of $\mathcal{H}_{\text{U(1)}}(-k,-k,k)$, $\mathcal{H}_{\text{U(1)}}(-k,k,-k)$ and $\mathcal{H}_{\text{U(1)}}(k,-k,-k)$ then follows.


\onecolumngrid

\section{Gauge invariance at one-loop level \label{app:H}}

It is know that for a generic Hamiltonian couple to a U(1) gauge field, gauge invariance requires that 1) the photon self energy $\Pi(q)$ vanishes when the photon external momentum $\bm{q}$ vanishes and that 2) Ward identity holds. These statements holds  perturbatively at each loop level. Here we explicitly proof these two statements for non-interacting fermions coupled to a U(1) gauge field at one-loop level. For a generic tight binding Hamiltonian
 \begin{equation}
H_0 = \sum_{\bm{r}_\mu,\bm{r}'_\nu} c^\dag_{\bm{r}_\mu} h_{\bm{r}_\mu,\bm{r}'_\nu} c_{\bm{r}'_\nu},
\end{equation}
the U(1) gauge coupling is introduced via the Peierl's substitution:
\begin{equation}
H[A] =  \sum_{\bm{r}_\mu,\bm{r}'_\nu} c^\dag_{\bm{r}_\mu} h_{\bm{r}_\mu,\bm{r}'_\nu} e^{i A_{\bm{r}_\mu,\bm{r}'_\nu}} c_{\bm{r}'_\nu} + \sum_{\bm{r}_\mu}A_{0,\bm{r}_\mu}n_{\bm{r}_\mu}
\end{equation}
the spatial fluctuation of the gauge field is small at short distance, suggesting that we can expand the exponential for the gauge field. To quadratic order of $A$ we obtain
\begin{equation}
H[A]
 = H_0 + \sum_{\bm{k},\bm{q}} A^i(-q) c^\dag_{\bm{k}+\bm{q}/2}\frac{\partial h(\bm{k})}{\partial k_i} c_{\bm{k} - \bm{q}/2} + \sum_{\bm{k},\bm{q}} iA^0(-q) c^\dag_{\bm{k}+\bm{q}/2} c_{\bm{k}- \bm{q}/2} + \sum_{\bm{k},\bm{q}} A^i(q)A^j(q') c^\dag_{\bm{k}+\bm{q}'}\frac{\partial^2 h(\bm{k})}{\partial k_j\partial k_i} c_{\bm{k} - \bm{q}} + O(A^3),
\end{equation}
up to this order we have the usual minimal coupling vertex $Ac^\dag c$ as well as a \emph{diamagnetic} vertex $A^2 c^\dag c$. These two vertices lead to the two diagrams at one-loop shown in Fig.~\ref{feynman}: the usual vacuum polarization bubble (left) and the ``tadpole'' diagram (right). Note that the diamagnetic term vanishes for a Dirac Hamiltonian since it is linear in momentum. In the following we show that the contribution of these two one-loop diagrams cancel each other at $q=0$, and furthermore the sum of the them at finite momentum and frequency satisfies the Ward identity.

The vacuum polarization bubble diagram in Fig.~\ref{feynman} originates from the $Ac^\dag c$ term. The vertex expression $\gamma_\mu(\bm{k}) = \delta_{\mu 0} + \delta_{\mu,i} \partial_{k_i}h$, i.e. the vertex is unity for the temporal component $\mu=0$ and is $\partial_{k_i} h$ for the spatial component. The ``tadpole'' diagram originates from the $A^2 c^\dag c$ term. The vertex expression is $\gamma_{\mu\nu}(\bm{k}) = \delta_{\mu,i} \delta_{\nu ,j}\partial_{k_i}\partial_{k_j} h$, i.e. the vertex only exists for $\mu,\nu$ both being spatial indices, with vertex expression $\partial_{k_i}\partial_{k_j} h$. The two diagrams have the following expression
\begin{equation}
\Pi^{(1)}_{1,\mu\nu} (q) = \int \frac{ d^4 k}{(2\pi)^4} \mathrm{Tr}[\gamma_\mu(\bm{k}) G_0(k+q/2) \gamma_\nu(\bm{k}) G_0(k-q/2)],\quad \Pi^{(1)}_{2,\mu\nu}(q) = \int \frac{d^4 k}{(2\pi)^4} \mathrm{Tr} [\gamma_{\mu\nu}(\bm{k})G_0(k-q)].
\end{equation}
We now show that $\Pi^{(1)}_{1,\mu\nu}(q=0) + \Pi^{(1)}_{2,\mu\nu}(q=0) =0 $.
First of all, when one of the $\mu,\nu$ is a temporal component, say $\nu=0$, then $\Pi^{(1)}_{2,\mu\nu}=0$ and we are only left with $\Pi^{(1)}_{1,\mu0}(\bm{q}=0)$: in the following we write $k_0=\omega$. We have
\begin{equation}
\begin{aligned}
\Pi^{(1)}_{1,\mu0}(q=0)
&=
 \int \frac{ d^4 k}{(2\pi)^4} \mathrm{Tr}[\gamma_\mu(\bm{k}) G^2_0(\bm{k})]
= \int \frac{d^4 k}{(2\pi)^4} \mathrm{Tr} \left[\gamma_\mu(\bm{k})\left(\frac{1}{\omega - h(\bm{k})}\right)^2\right]
= -\int \frac{d^4 k}{(2\pi)^4} \mathrm{Tr} \left[\gamma_\mu(\bm{k})\frac{\partial}{\partial \omega}\left(\frac{1}{\omega - h(\bm{k})}\right)\right]\\
&=-\int \frac{d^4 k}{(2\pi)^4}  \frac{\partial}{\partial \omega} \left\{\mathrm{Tr} \left[\gamma_\mu(\bm{k})\frac{1}{\omega - h(\bm{k})}\right]\right\} =
-\int \frac{d^3\bm{k}}{(2\pi)^4}  \mathrm{Tr} \left[\gamma_\mu(\bm{k})\frac{1}{\omega - h(\bm{k})}\right] \Bigg|^{\omega = +\infty} _{\omega = -\infty}=0,
\end{aligned}
\end{equation}
where we have used the fact that if $h$ is diagonalized as $h = U^\dag \Lambda U$, then $\frac{1}{(\omega - h)^2} = U^\dag \frac{1}{(\omega - \Lambda)^2} U = U^\dag\left( - \frac{\partial}{\partial \omega} \left( \frac{1}{\omega - \Lambda}\right)\right) U =- \frac{\partial}{\partial \omega} \left( \frac{1}{\omega - h}\right)$. We therefore see that $\Pi^{(1)}_{1,\mu0}(\bm{q}=0) = 0$ for any $\mu$. This means that $\Pi^{(1)}_{1,0\nu}(q=0) = 0$ for any $\nu$. Then, we look at spatial components (note we have suppressed the arguments $\bm{k}$ below):
\begin{equation}
\begin{aligned}
\Pi^{(1)}_{1,ij}(q=0) + \Pi^{(1)}_{2,ij}(q=0)
&=\int \frac{d^4 k}{(2\pi)^4}\mathrm{Tr}
\left[ \partial_{k_i} h  \frac{1}{\omega - h} \partial_{k_j} h \frac{1}{\omega - h}\right] + \int \frac{d^4 k}{(2\pi)^4}\mathrm{Tr}
\left[ \partial_{k_i} \partial_{k_j } h  \frac{1}{\omega - h} \right]\\
&=\int \frac{d^4 k}{(2\pi)^4}\mathrm{Tr}
\left[ \partial_{k_i} h  \frac{1}{\omega - h} \partial_{k_j} h \frac{1}{\omega - h}\right] -\int \frac{d^4 k}{(2\pi)^4}\mathrm{Tr}
\left[ \partial_{k_j } h  \partial_{k_i} \left(\frac{1}{\omega - h}\right) \right]\\
&=\int \frac{d^4 k}{(2\pi)^4}\mathrm{Tr}
\left[ \partial_{k_i} h  \frac{1}{\omega - h} \partial_{k_j} h \frac{1}{\omega - h}\right] -\int \frac{d^4 k}{(2\pi)^4}\mathrm{Tr}
\left[ \partial_{k_j } h\frac{1}{\omega - h}\partial_{k_i}h\frac{1}{\omega - h} \right]\\
&=0,
\end{aligned}
\end{equation}
where we have used the fact that $\partial(K^{-1}) = - K^{-1} (\partial K) K^{-1}$ for any (non-singular) matrix $K$. Therefore we have proved that the photon self-energy vanishes at one-loop level when photon external momentum is zero.

Next we show that Ward identity holds at one-loop level $q_\mu\left( \Pi^{(1)}_{1,\mu i}({q}) - \Pi^{(1)}_{2,\mu i}({q})\right)=0$. First, only the vacuum polarization diagram contributes to the $\mu 0$ component:
\begin{equation}
q_\mu \Pi^{(1)}_{1,\mu0}({q})
= \int \frac{d^4 k}{(2\pi)^4}\mathrm{Tr}\left[  \frac{-q_0 + q_i \partial_{k_i} h(\bm{k})}{(k_0+q_0/2 - h(\bm{k}+\bm{q}/2))(k_0-q_0/2 -h(\bm{k}-\bm{q}/2))}\right],
\end{equation}
note that
\begin{equation}
(k_0+q_0/2 - h(\bm{k}+\bm{q}/2))-(k_0-q_0/2 -h(\bm{k}-\bm{q}/2))
=q_0- q_i \partial_{k_i} h(\bm{k})+o(q),
\end{equation}
this means that
\begin{equation}
q_\mu \Pi^{(1)}_{1,\mu0}({q}) = \int \frac{d^4k}{(2\pi)^4}\mathrm{Tr} \left[\frac{1}{k_0+q_0/2 -h(\bm{k}+\bm{q}/2)} - \frac{1}{k_0-q_0/2-h(\bm{k}-\bm{q}/2)}\right],
\end{equation}
which gives zero since the two terms only differ by a shift. Similarly, we have
\begin{equation}
q_\mu \Pi^{(1)}_{1,\mu i} = \int \frac{d^4k}{(2\pi)^4}\mathrm{Tr}
\left[\left(-q_0 +  q_j \partial_{k_j} h(\bm{k})\right) \frac{1}{k_0+q_0/2-h(\bm{k}+\bm{q}/2)}\partial_{k_i} h(\bm{k}) \frac{1}{k_0-q_0/2-h(\bm{k}-\bm{q}/2)}\right],
\end{equation}
and similar to the $\mu0$ component case, we get
\begin{equation}
q_\mu \Pi^{(1)}_{1,\mu i} = \int \frac{d^4 k}{(2\pi)^4}\mathrm{Tr}\left[ \partial_{k_i} h(\bm{k})\left(\frac{1}{k_0+q_0/2 -h(\bm{k}+\bm{q}/2)} - \frac{1}{k_0-q_0/2-h(\bm{k}-\bm{q}/2)}\right)\right],
\end{equation}
on the other hand we have $ \frac{1}{2} q_i \partial_{k_i} \partial_{k_j} h(\bm{k}) = \partial_{k_j}h(\bm{k}+\bm{q}/2)-\partial_{k_j}h(\bm{k}) + o(q) = \partial_{k_j}h(\bm{k}) - \partial_{k_j}h(\bm{k} - \bm{q}/2)+o(q)$, and
\begin{equation}
\begin{aligned}
q_j \Pi^{(1)}_{2,ji} &=  \int \frac{d^4 k}{(2\pi)^4}\mathrm{Tr}
\left[\frac{ q_j \partial_{k_j} \partial_{k_i} h(\bm{k})}{k_0-q_0 - h(\bm{k}-\bm{q})}\right]
=
\int \frac{d^4 k}{(2\pi)^4}
\mathrm{Tr}\left[\frac{ \partial_{k_j}h(\bm{k}-\bm{q}/2)-\partial_{k_j}h(\bm{k}-\bm{q}) + \partial_{k_j}h(\bm{k}-\bm{q})-\partial_{k_j}h(\bm{k} - 3\bm{q}/2)}{k_0-q_0 - h(\bm{k}-\bm{q})}\right]\\
&=
\int \frac{d^4 k}{(2\pi)^4}
\mathrm{Tr}\left[\frac{ \partial_{k_j}h(\bm{k})-\partial_{k_j}h(\bm{k}-\bm{q}/2)}{k_0-q_0/2 - h(\bm{k}-\bm{q}/2)}\right]
+
\int \frac{d^4 k}{(2\pi)^4}
\mathrm{Tr}\left[\frac{ \partial_{k_j}h(\bm{k}+\bm{q}/2)-\partial_{k_j}h(\bm{k})}{k_0+q_0/2 - h(\bm{k}+\bm{q}/2)}\right],
\end{aligned}
\end{equation}
where we have shifted the integral variables. Therefore we see that (denote $\Pi^{(1)}_{2,0 \mu} = 0$)
\begin{equation}
q_\mu\left( \Pi^{(1)}_{1,\mu i}({q}) - \Pi^{(1)}_{2,\mu i}({q})\right)
=
\int \frac{d^4 k}{(2\pi)^4}
\mathrm{Tr}\left[ \frac{\partial _{k_i} h(\bm{k}+\bm{q}/2)}{k_0+q_0/2  - h(\bm{k}+\bm{q}/2)} - \frac{\partial _{k_i} h(\bm{k}-\bm{q}/2)}{k_0-q_0/2  - h(\bm{k}-\bm{q}/2)}\right],
\end{equation}
which gives zero on the Brillouin zone. Therefore Ward identity holds at one-loop level.

\section{Deriving the photon vacuum bubble: scaling analysis\label{app:I}}

We study the $00$ component of the vacuum polarization diagram: after completing the frequency integral, we have (again in imaginary time)
\begin{equation}
D(q) \equiv \Pi^{(1)}_{1,00}(q) = - \pi \mathrm{Re} \left[\int \frac{d^3\bm{k}}{(2\pi)^3} \frac{ 1- \bm{\hat d}_{\bm{k}+\bm{q}/2} \cdot \bm{\hat d}_{\bm{k}- \bm{q}/2}}{|\bm{d}_{\bm{k}+\bm{q}/2}| + | \bm{d}_{\bm{k}- \bm{q}/2}| + i q_0}\right],
\end{equation}
with $\bm{\hat d} = \bm{d}/|\bm{d}|$. To separate the contribution along the nodal line and that in the vicinity of the $\Gamma$ point, we use the identity $1 = \frac{x}{a+x} + \frac{a}{a+x}$, where we set $a = c \bm{q}^2$ and $x = |\bm{d}_{\bm{k}+\bm{q}/2}| + | \bm{d}_{\bm{k}- \bm{q}/2}| + i q_0$, so that
\begin{equation}
D(q) = D_1(q)+D_2(q),
\end{equation}
with
\begin{subequations}
\begin{align}
D_1(q) &=
- \pi \mathrm{Re} \left[\int \frac{d^3\bm{k}}{(2\pi)^3} \frac{ 1- \bm{\hat d}_{\bm{k}+\bm{q}/2} \cdot \bm{\hat d}_{\bm{k}- \bm{q}/2}}{|\bm{d}_{\bm{k}+\bm{q}/2}| + | \bm{d}_{\bm{k}- \bm{q}/2}| + i q_0 + c \bm{q}^2}\right],\\
D_2(q) &=
- \pi \mathrm{Re} \left[\int \frac{d^3\bm{k}}{(2\pi)^3} \frac{ c \bm{q}^2(1- \bm{\hat d}_{\bm{k}+\bm{q}/2} \cdot \bm{\hat d}_{\bm{k}- \bm{q}/2})}{(|\bm{d}_{\bm{k}+\bm{q}/2}| + | \bm{d}_{\bm{k}- \bm{q}/2}| + i q_0)(|\bm{d}_{\bm{k}+\bm{q}/2}| + | \bm{d}_{\bm{k}- \bm{q}/2}| + i q_0 + c \bm{q}^2) }\right],\label{D2or}
\end{align}
\end{subequations}
the choice of $a = c \bm{q}^2$ is made to agree with the scaling $\bm{d}_{\bm{k}} \propto \bm{q}^2$ near the $\Gamma$ point (see Eq.~\eqref{expdk}) and guarantees that $D_2$ extracts the contribution in vicinity of the $\Gamma$ point.
Having in mind that at small $\bm{q}$ the leading order $\bm{q}$ result is isotropic in $\bm{q}$. Therefore we choose a specific direction for $\bm{q}$: $\bm{q} = q_z \bm{\hat z}$.

We first look at $D_2(q)$: since $D_2(q)$ is supported in vicinity of the $\Gamma$ point, we expand $\bm{d}_{\bm{k}}$ as in Eq.~\eqref{expdk}. Rescaling $\bm{k} = \bm{x} q_z$ and $\bm{d}_{\bm{k}} = q_z^2 \bm{\epsilon}_{\bm{k}}$, we have
\begin{equation}
D_2(q) \sim - c\pi |q_z| \mathrm{Re} \left[\int \frac{d^3 \bm{x}}{(2\pi)^3} \frac{ 1-\bm{\hat \epsilon}_{\bm{x}+\bm{\hat z}/2}\cdot \bm{\hat \epsilon}_{\bm{x}-\bm{\hat z}/2}}{(|\bm{\epsilon}_{\bm{x}+\bm{\hat z}/2}| + | \bm{\epsilon}_{\bm{x}-\bm{\hat z}/2}| + i q_0/\bm{q}^2)(|\bm{\epsilon}_{\bm{x}+\bm{\hat z}/2}| + | \bm{\epsilon}_{\bm{x}-\bm{\hat z}/2}| + i q_0/|\bm{q}| + c ) }\right]+O (q^2),
\end{equation}
First look at the case of $q_0=0$. In this case, the integral in $D_2(q)$ is well behaved, which can be easily seen in the original expression \eqref{D2or}: the only singularity comes from $|\bm{d}_{\bm{k}+\bm{q}/2}|+|\bm{d}_{\bm{k}+\bm{q}/2}|=0$, or $|\bm{\epsilon}_{\bm{k}+\bm{q}/2}||\bm{\epsilon}_{\bm{k}-\bm{q}/2}|=0$, which gives isolated points $(x_1,x_2,x_3)=(\pm 1/2,\pm 1/2,0)$. Expand around these points: $x_1 = \pm 1/2 + \eta \xi_1$, $x_2 = \pm 1/2 + \eta \xi_2$, and $x_3 = \eta \xi_3$, we see that $|\bm{\epsilon}_{\bm{k}+\bm{q}/2}||\bm{\epsilon}_{\bm{k}-\bm{q}/2}|\sim \eta f(\xi_1,\xi_2,\xi_3)$ where the function $f$ is well behaved; therefore these singularities are integrable. We can also relax the integral for $\bm{x}$ to the infinite plane and still get finite results. Therefore the integral in $D_2(q)$ is well behaved, and $D_2(\bm{q},q_0=0)$ scales linearly with $|\bm{q}|$.

Next, we need to extract the scaling behavior at finite frequency; analytic continuation $iq_0\rightarrow \omega + i \delta$ is needed. First, the real part of $D_2(q)$ is of the form $D_2(q) = |\bm{q}| f(\omega/\bm{q}^2)$, where $f$ is a well-behaved function whose value is always finite (according to the $q_0=0$ analysis above). However it might be useful to see how the actual scaling looks like. What we care is when $w = \omega/\bm{q}^2\ll 1$ since this is the regime that $D_2(q)$ has both real and imaginary parts. In this regime, $1- \bm{\hat \epsilon}_{\bm{x}+\bm{\hat z}/2}\cdot \bm{\hat \epsilon}_{\bm{x}-\bm{\hat z}/2}$ is finite (numerically verified), therefore the scaling is determined simply by $ \int \frac{d^3 \bm{x}}{(2\pi)^3}  \frac{1}{(|\bm{\epsilon}_{\bm{x}+\bm{\hat z}/2}| + | \bm{\epsilon}_{\bm{x}-\bm{\hat z}/2}| \pm (w+i \delta))}$. The result is $ \sim \int \frac{r^2 dr}{r  \pm (w+i \delta)}\sim (\mp wr + r^2/2 + w^2 \ln (\pm w+r))|^R_0 + i \pi w^2 \mathrm{sgn}(w)$, we see that we have scaling $w^2 \ln |w| + i \pi w^2 \text{sgn}(w)$. Note that the real part $w^2 \ln |w|$ is in addition to other contributions in $f(w)$. Therefore in the limit $\omega\rightarrow 0$, we recover the scaling $D_2(q)\sim |q_z|$.

Next we deal with $D_1(q)$. The numerator can be simplified: notice that $\bm{d}^2_{\bm{k}+\bm{q}/2} + \bm{d}^2_{\bm{k}-\bm{q}/2} - 2
|\bm{d}_{\bm{k}+\bm{q}/2}||\bm{d}_{\bm{k}-\bm{q}/2}| = (|\bm{d}_{\bm{k}+\bm{q}/2}| - |\bm{d}_{\bm{k}-\bm{q}/2}|)^2 \sim (\sin q \sin k)^2$, and $(\bm{d}_{\bm{k}+\bm{q}/2}- \bm{d}_{\bm{k}-\bm{q}/2})^2= 8  \sin^2 k_z \sin^2\frac{q_z}{2} $, therefore we can substitute
\begin{equation}\label{qk22dd}
1- \bm{\hat d}_{\bm{k}+\bm{q}/2} \cdot \bm{\hat d}_{\bm{k}- \bm{q}/2}
= \frac{-(|\bm{d}_{\bm{k}+\bm{q}/2}| - |\bm{d}_{\bm{k}-\bm{q}/2}|)^2+(\bm{d}_{\bm{k}+\bm{q}/2}-\bm{d}_{\bm{k}-\bm{q}/2})^2}{2|\bm{d}_{\bm{k}+\bm{q}/2}||\bm{d}_{\bm{k}-\bm{q}/2}|} \sim \frac{ \sin^2 q \sin^2 k}{2|\bm{d}_{\bm{k}+\bm{q}/2}||\bm{d}_{\bm{k}-\bm{q}/2}|}.
\end{equation} 
since $D_1(q)$ receives contribution mainly along the line, we can make the rescaling $k_1 = k_3 + q x_1$, $k_2 = k_3 + q x_2$ and the approximation $|d_{\bm{k}\pm \bm{q}/2}| = |\bm{q}| | \sin k_3| f_\pm (x_1,x_2) +O(q^3)$, where $f_\pm(x_1,x_2) = \sqrt{ (1\mp 2(x_1+x_2) + 4 x_1^2 + 4 x_2^2 - 4 x_1 x_2)/2}$, we further have
\begin{equation}
1- \bm{\hat d}_{\bm{k}+\bm{q}/2} \cdot \bm{\hat d}_{\bm{k}- \bm{q}/2}
\sim \frac{1}{f_+(\bm{x})f_-(\bm{x})}.
\end{equation}
Then after analytic continuation to real frequency we have
\begin{equation}
D_1(q \bm{\hat z},\omega)
=- \frac{|\bm{q}|}{2\pi^2} \int d^2\bm{x} \frac{1}{f_+(\bm{x})f_-(\bm{x})}
\int^\pi_0 d k_3\left(\frac{1}{| \sin k_3| (f_++f_-)+\frac{\omega}{|\bm{q}|}+c|\bm{q}|+i \delta}
+\frac{1}{| \sin k_3| (f_++f_-)-\frac{\omega}{|\bm{q}|}+c|\bm{q}|-i \delta}\right)
\end{equation}
(following Leon) the first integral gives (here we only consider $\omega \ll q\ll 1$)
\begin{equation}
D_1^{(1)}=-\frac{|\bm{q}|}{\pi^2}
\ln\left(\frac{1}{|\frac{\omega}{|\bm{q}|} + c |\bm{q}||}\right) C_0
-i\frac{|\bm{q}|}{2\pi^2}\int d^2\bm{x} \frac{\Theta(0<-\frac{\omega}{|\bm{q}|}-c|\bm{q}|<f_++f_-)}{f_+(\bm{x})f_-(\bm{x}) \sqrt{(f_+(\bm{x})+f_-(\bm{x}))^2 -(\frac{\omega}{|\bm{q}|}+c|\bm{q}|)^2}},
\end{equation}
where we defined
\begin{equation}
C_0 = \int d^2 \bm{x} \frac{1}{f_+(\bm{x})f_-(\bm{x}) (f_+(\bm{x})+f_-(\bm{x}))}.
\end{equation}
the imaginary part  vanishes for $|\omega| \ll \bm{q}^2 $ since in this regime the Heaviside function has zero support. When $ |\bm{q}| \gg \omega \gg \bm{q}^2$, we can ignore the $c|\bm{q}|^2$ term, and we have
\begin{equation}
D_1(q\bm{\hat z},\omega) = -|\bm{q}| \ln\left(\frac{1}{|c^2\bm{q}^2 - \frac{\omega^2}{q^2}|}\right) C_0 -i|\bm{q}|\text{sgn}(\omega) \left[g\left(\frac{\omega}{|\bm{q}|}\right)  \Theta(|\bm{q}| \gg |\omega| > c\bm{q}^2) +\frac{\omega^2}{\bm{q}^4}\Theta(|\omega|\ll  \bm{q}^2)\right],
\end{equation}
where $g(x)$ is a function of $x=\omega/g$. We verify numerically that $g(\frac{\omega}{|\bm{q}|})= C_0 + \frac{1}{2}(\frac{\omega}{|\bm{q}|})^2$. As we will show in the next appendix, a pure nodal line approximation can recover the calculation here by introducing a cutoff $\theta_0$ for integrals along the nodal line at the $\Gamma$ point of the form $\theta_0 \sim \bm{q}$.

To summarize, the $00$ component $D(q)=\Pi^{00}(q)$ receives contribution of $D_2(q)\sim |\bm{q}|f(\omega/q^2)$ near the $\Gamma$ point and receives contribution $D_1 = - |\bm{q}|\ln\left(\frac{1}{| c^2 |\bm{q}|^2 - \omega^2/q^2|}\right) -i|\bm{q}|g(\omega/q) \Theta(|\bm{q}| \gg |\omega| > c\bm{q}^2)$ along the nodal lines. At small frequency $\omega \ll q^2$, $D(q)$ is real and is dominated by the nodal line ($ q \ln(1/q)$ vs $q$).

The $0i$ and $ij$ components can be analyzed in the same way. We have
\begin{subequations}
\begin{align}
\Pi^{(1)}_{1,0i}(q) &= -i\pi \mathrm{Re} \left[\int \frac{d^3\bm{k}}{(2\pi)^3} \frac{
\frac{| \bm{d}_{\bm{k}+\bm{q}/2}|-| \bm{d}_{\bm{k}-\bm{q}/2}|}
{| \bm{d}_{\bm{k}-\bm{q}/2}|+| \bm{d}_{\bm{k}+\bm{q}/2}|}
\frac{{C}_i(\bm{k},\bm{q})}{| \bm{d}_{\bm{k}-\bm{q}/2}|| \bm{d}_{\bm{k}+\bm{q}/2}|}
- \frac{D_i(\bm{k},\bm{q})}{| \bm{d}_{\bm{k}-\bm{q}/2}|| \bm{d}_{\bm{k}+\bm{q}/2}|}}
{|\bm{d}_{\bm{k}+\bm{q}/2}| + | \bm{d}_{\bm{k}- \bm{q}/2}| + i q_0} \right]q_0 \sin k_i,\label{exprpi0i}\\
\Pi^{(1)}_{1,ii}(q) &= - 2\pi \mathrm{Re} \left[\int \frac{d^3\bm{k}}{(2\pi)^3} \frac{ 1- \hat{B}_i(\bm{k},\bm{q})}{|\bm{d}_{\bm{k}+\bm{q}/2}| + | \bm{d}_{\bm{k}- \bm{q}/2}| + i q_0}\right] \sin^2k_i,\\
\Pi^{(1)}_{1,ij}(q) &= \pi \mathrm{Re} \left[\int \frac{d^3\bm{k}}{(2\pi)^3} \frac{ 1- 2 \hat{B}_i(\bm{k},\bm{q})-2\hat{B}_j(\bm{k},\bm{q})}{|\bm{d}_{\bm{k}+\bm{q}/2}| + | \bm{d}_{\bm{k}- \bm{q}/2}| + i q_0} \right]\sin k_i \sin k_j,
\end{align}
\end{subequations}
where (it is understood $i+3 \equiv i$)
\begin{subequations}
\begin{align}
\hat{B}_i &= \bm{\hat d}_{\bm{k}-\bm{q}/2} \cdot \bm{\hat d}_{\bm{k}+\bm{q}/2} - 3
\hat{d}^i_{\bm{k}-\bm{q}/2}\hat{d}^i_{\bm{k}+\bm{q}/2},\\
{C}_i &= \frac{1}{2}[({d}^{i+2}_{\bm{k}-\bm{q}/2}+{d}^{i+2}_{\bm{k}+\bm{q}/2})
- ({d}^{i+1}_{\bm{k}-\bm{q}/2}+{d}^{i+1}_{\bm{k}+\bm{q}/2})],\\
{D}_i &= \frac{1}{2}[({d}^{i+2}_{\bm{k}-\bm{q}/2}-{d}^{i+2}_{\bm{k}+\bm{q}/2})
- ({d}^{i+1}_{\bm{k}-\bm{q}/2}-{d}^{i+1}_{\bm{k}+\bm{q}/2})],
\end{align}
\end{subequations}

Due to the appearance of $\sin k_i$ coming from the vertex expressions, the $\Gamma$ point will contribute at a much higher order: $\Pi^{(1)}_{1,0i} \sim q^2$, and $\Pi^{(1)}_{1,ij} \sim q^3$.  Furthermore, the nodal line contribution becomes
\begin{equation}
\Pi^{(1)}_{1,0i}
=- i q_0\frac{1}{2\pi^2} \int d^2\bm{x} \frac{c_{0i}(\bm{x})}{f_+(\bm{x})f_-(\bm{x})}
\int^\pi_0 d k_3\left(\frac{ 1}{| \sin k_3| (f_++f_-)+\frac{\omega}{|\bm{q}|}+c|\bm{q}|+i \delta}
+\frac{1}{| \sin k_3| (f_++f_-)-\frac{\omega}{|\bm{q}|}+c|\bm{q}|-i \delta}\right)
\end{equation}
Note that the numerator of Eq.~\eqref{exprpi0i} gives
\begin{equation}\label{sdsdfsd}
\sim \frac{\sin q \sin k}{|\bm{d}_{\bm{k}+\bm{q}/2}| |\bm{d}_{\bm{k}-\bm{q}/2}|}
\end{equation}
 (c.f. Eq.~\eqref{qk22dd}), therefore it requires an extra $|\bm{q}|$ and an extra $\sin k_i$ to cancel the $|\bm{q}|^2 \sin^2k_3$ coming from $|\bm{d}_{\bm{k}+\bm{q}/2}| |\bm{d}_{\bm{k}-\bm{q}/2}|$ in Eq.~\eqref{sdsdfsd}. Then, for the $ij$ components
\begin{equation}
\Pi^{(1)}_{1,ij}
=- \frac{|\bm{q}|}{2\pi^2} \int d^2\bm{x} \frac{c_{ij}(\bm{x})}{f_+(\bm{x})f_-(\bm{x})}
\int^\pi_0 d k_3\left(\frac{ \sin^2 k_3}{| \sin k_3| (f_++f_-)+\frac{\omega}{|\bm{q}|}+c|\bm{q}|+i \delta}
+\frac{\sin^2 k_3}{| \sin k_3| (f_++f_-)-\frac{\omega}{|\bm{q}|}+c|\bm{q}|-i \delta}\right)
\end{equation}
which gives
\begin{equation}
\Pi^{(1)}_{1,0i}\sim  \omega n_{0i}(\bm{\hat q}) \ln(1/|w|^2)  + i \omega g(w),
\qquad
\Pi^{(1)}_{1,ij}\sim |\bm{q}|\left(n_{ij}(\bm{\hat q})(2- \pi w) + m_{ij}(\bm{\hat q})w^2 \ln(1/w)\right)   - i | \bm{q}| w^2 g(w),
\end{equation}
where $w = \omega/|\bm{q}| \ll 1$ and $|\bm{q}| \gg |\omega | \gg c \bm{q}^2$ (The discrete poles when $|\omega| \ll \bm{q}^2$ are not included here). We see that the nodal line contribution also dominates in these components. Note the above expressions hold only in the scaling sense; for example, the $|\bm{q}|$ factor in $\Pi^{(1)}_{1,ij}$ must have a component dependent form in order for Ward identity to hold. On the other hand, these expressions already have the right scaling for Ward identity to hold. The detailed form of these scaling functions can be found in the next section.

Therefore in photon thermodynamics we just have to concentrate on the nodal line and not the $\Gamma$ region. This validates the QED calculation in the next section.

Finally, we mention that the momentum dependent part of $\Pi_{2,\ij}$ starts to contribute at quadratic order in $q^2$: this is because due to the special form of $h(\bm{k})$ we have
\begin{equation}
\Pi^{(1)}_{2,ij} = \int \frac{d^4 k}{(2\pi)^4}\mathrm{Tr}\left[ \frac{\partial_{k_i}\partial_{k_j} h(\bm{k})}{k_0-q_0-h(\bm{k}-\bm{q})}\right] =  \int \frac{d^4 k}{(2\pi)^4}\mathrm{Tr}\left[ \frac{\partial_{k_i}\partial_{k_j}h(\bm{k}+\bm{q}))}{k_0-h(\bm{k})}\right]
=
\delta_{ij}\int \frac{d^4 k}{(2\pi)^4}\mathrm{Tr}\left[ \frac{\partial^2_{k_i}h(\bm{k}+\bm{q})}{k_0-h(\bm{k})}\right],
\end{equation}
since $h(-\bm{k}) = h(\bm{k})$, if we expand $\partial^2_{k_i}h(\bm{k}+\bm{q})$, the term linear in $\bm{q}$ is odd in $\bm{k}$ vanishes after the integral over $\bm{k}$, leaving the leading order contribution quadratic in $\bm{q}$.

\section{Deriving the photon vacuum bubble: nodal line approximation\label{app:J}}

We concluded in the last Appendix that the calculation of photon self-energy at one-loop level amounts to calculating the momentum dependent part of the vacuum polarization bubble. This is the diagram resulted from the minimal coupling term $A_i(-\bm{q})\partial_{k_i} \mathcal{H}(\bm{k}) \psi^\dag_{\bm{k}+\bm{q}/2} \psi_{\bm{k}-\bm{q}/2}$. The expressions for this diagram and the vertex have been given in Eqs.~\eqref{po} and \eqref{vert}. The bare Green's function for the spinons has the explicit form
\begin{equation}
G_0(k) = \frac{1}{ik_0 - \mathcal{H}} = \frac{i k_0 + \mathcal{H}}{-k_0^2 - E^2}
=
-\frac{ ik_0 + \cos k_1 (\sigma^2-\sigma^3) + \cos k_2 (\sigma^3 - \sigma^1) + \cos k_3 (\sigma^1 - \sigma^2)}{k_0^2 + (\cos k_1 -\cos k_2)^2+(\cos k_2 - \cos k_3)^2 + (\cos k_3 - \cos k_1)^2}.
\end{equation}

To evaluate the vacuum polarization bubble near the nodal star region, we first need to write the momentum in local coordinates. Denote the nodal line by $(\varsigma_1,\varsigma_2,\varsigma_3)$, where $\varsigma_{1,2,3} = \pm 1$ labels different nodal lines. For each nodal line, we denote
\begin{equation}
\bm{\varepsilon}_3 = \frac{1}{\sqrt{3}}(\varsigma_1 \hat{x} + \varsigma_2 \hat{y} + \varsigma_3 \hat{z}),\quad
\bm{\varepsilon}_1 = \frac{1}{\sqrt{2}}(\varsigma_1 \hat{x} - \varsigma_2 \hat{y}),\quad
\bm{\varepsilon}_2 = \frac{1}{\sqrt{6}}(\varsigma_1 \hat{x} + \varsigma_2 \hat{y} - 2 \varsigma_3 \hat{z}),
\end{equation}
any momentum will be expanded in these coordinates: denote $\bm{k} = k_1 \hat{x}+k_2 \hat{y}+k_3 \hat{z} = c \bm{\varepsilon}_3 +a (\bm{\varepsilon}_1 \cos\theta + \bm{\varepsilon}_2 \sin \theta) + b (- \bm{\varepsilon}_1 \sin \theta + \bm{\varepsilon}_2 \cos \theta)$, then we have
\begin{subequations}
\begin{eqnarray}
k_1 &=& \varsigma_1\left(\frac{c}{\sqrt{3}} +  \eta \left(\frac{a}{\sqrt{2}} + \frac{b}{\sqrt{6}}\right) \cos \theta + \eta\left( \frac{a}{\sqrt{6}}- \frac{b}{\sqrt{2}}\right) \sin \theta \right),\\
k_2 &=& \varsigma_2\left( \frac{c}{\sqrt{3}} + \eta\left(- \frac{a}{\sqrt{2}} + \frac{b}{\sqrt{6}}\right)\cos \theta + \eta \left( \frac{a}{\sqrt{6}}+ \frac{b}{\sqrt{2}}\right) \sin \theta\right),\\
k_3 &=& \varsigma_3 \left( \frac{c}{\sqrt{3}} - \eta\frac{2b}{\sqrt{6}} \cos \theta - \eta \frac{2 a}{\sqrt{6}} \sin \theta \right).
\end{eqnarray}
\end{subequations}
First, expand the Hamiltonian \eqref{heffh} to first order of $\eta$ and then set $\eta=1$, we obtain
\begin{equation}
\mathcal{H} = \sin \frac{c}{\sqrt{3}} \left( \frac{ (-a +\sqrt{3} b) \cos \theta + (\sqrt{3} a+b) \sin \theta}{\sqrt{2}} \sigma^1  - \frac{(a+\sqrt{3}b) \cos \theta+(\sqrt{3} a - b) \sin \theta}{\sqrt{2}} \sigma^2 + \sqrt{2} ( a \cos \theta - b \sin \theta) \sigma^3\right)
\end{equation}
with the energy $E^2(\bm{k}) = 3\sin^2 \frac{c}{\sqrt{3}} (a^2+b^2) = v^2(a^2+b^2)$ where we defined $v \equiv \sqrt{3} \sin\frac{c}{\sqrt{3}}$. From now on, for any $\bm{k}$ dependent function $f=f(\bm{k})$, we will introduce the notation $f_\pm \equiv f(\bm{k}\pm \bm{q}/2)$.

Using the Feynman parameterization we have
\begin{equation}
\Pi^{\mu\nu}(q)=\int \frac{d^4k}{(2\pi)^4} \int^1_0 du \frac{Z^{\mu\nu}(k,q)}{\left[u (k_{0+}^2 + v_+^2(a_+^2+b_+^2)) + (1-u)(k_{0-}^2+v_-^2(a_-^2+b_-^2))\right]^2},
\end{equation}
where we defined
\begin{equation}
Z^{\mu\nu} = \mathrm{Tr}[\Gamma^\mu(\bm{k}) (i(k_0+q_0/2)+H(\bm{k}+\bm{q}/2))\Gamma^\nu(\bm{k}) (i(k_0-q_0/2)+H(\bm{k}-\bm{q}/2))].
\end{equation}
In local coordinates, $\Pi^{\mu\nu}(q)$ then can be written as
\begin{equation}\label{pieq1}
\Pi^{\mu\nu}(q)=\frac{\varsigma_1\varsigma_2\varsigma_3}{(2\pi)^4}\int d\kappa_0  \int \frac{d^2 \bm{\kappa}_\perp}{g} \int d\kappa_\parallel \int^1_0 du \frac{Z^{\mu\nu}(k,q)}{\left(\kappa^2_0+\bm{\kappa}^2_\perp + \Delta \right)^2},
\end{equation}
where we defined
\begin{equation}\label{kappakk}
\kappa_0 = k_0 + (u-1/2)q_0,\quad \bm{\kappa}_\perp = \sqrt{g} \left(\bm{k}_\perp + \frac{1}{2} f \bm{q}_\perp\right),\quad \kappa_{\parallel} = \frac{c}{\sqrt{3}},
\end{equation}
with
\begin{equation}g = u v_+^2+(1-u)v_-^2,\quad h = u v_+^2-(1-u)v_-^2,\quad  f = \frac{h}{g} = \frac{u v_+^2-(1-u)v_-^2}{u v_+^2+(1-u)v_-^2},\quad \Delta = u(1-u)q^2_0 + \frac{1}{4}(1-f^2)g q^2_\perp.
\end{equation}
The denominator has spherical symmetry with respect to $(\kappa_0,\bm{\kappa}_\perp)$. If we set $v_+=v_-\equiv v$ then the isotropic (i.e. relativistic) limit is recovered: $\bm{\kappa}_\perp = \bm{k}_\perp +(u-1/2) \bm{q}_\perp$ which agrees with $\kappa_0 = k_0 + (u-1/2)q_0$; and $\frac{1}{4}g(1-f^2) q^2_\perp \rightarrow  u(1-u)q^2_\perp $ which agrees with $u(1-u)q_0^2$.

Now we simplify the numerator $Z^{\mu\nu}$. The vertex $\Gamma^\mu$ in principle needs expansion according to powers of $\eta$, however for our purpose it suffices to keep the zeroth order, i.e. $\Gamma^1(\bm{k}) = \sin k_1 (\sigma^2 - \sigma^3) \sim  \varsigma_1 \sin \frac{c}{\sqrt{3}}(\sigma^2 - \sigma^3)$, and similarly $\Gamma^2(\bm{k}) \sim \varsigma_2 \sin \frac{c}{\sqrt{3}} (\sigma^3-\sigma^1)$, and $\Gamma^3(\bm{k}) \sim \varsigma_3 \sin \frac{c}{\sqrt{3}}(\sigma^1-\sigma^2)$. For later convenience we further define $\varsigma_0\equiv 1$. This way in $Z^{\mu\nu}$ there will be prefactors of
$$\varsigma_\mu \varsigma_\nu\left( \sin \frac{c}{\sqrt{3}}\right)^{\delta_{\mu\neq 0}+\delta_{\nu \neq 0}} = \varsigma_\mu\varsigma_\nu \left(\frac{v}{\sqrt{3}}\right)^{\delta_{\mu\neq 0}+\delta_{\nu \neq 0}}.$$
We then apply Eq.~\eqref{kappakk} and $Z^{\mu\nu}$ will be written as polynomials of $\kappa$, up to quadratic order. Then only the constant terms in $\kappa$ and the squared terms in $\kappa$ needs to be kept since the other terms integrate to zero. The integral over $d^3\kappa = d\kappa_0 d^2\bm{\kappa}_\perp$ can then be evaluated. Using dimensional regularization
\begin{equation}
\int d^3 \kappa \frac{\{\kappa^2,1\}}{(\kappa^2+\Delta)^2} = 4 \pi \int d \kappa \frac{\{\kappa^4,\kappa^2\}}{(\kappa^2+\Delta)^2} = \pi^2\{ -3 \Delta^{1/2}, \frac{1}{\Delta^{1/2}}\}
\end{equation}
where the divergent part of the first term has been subtracted (this part is independent of the external momentum $\bm{q}$ and will cancel the divergence from the ``tadpole'' diagram). We then have
\begin{equation}\label{piiiip}
\Pi^{\mu\nu}(q) = \frac{\varsigma_1\varsigma_2\varsigma_3}{(2\pi)^4}\int d \kappa_\parallel  \varsigma_\mu\varsigma_\nu \left(\frac{v}{\sqrt{3}}\right)^{\delta_{\mu\neq 0}+\delta_{\nu \neq 0}} \int^1_0 du I^{\mu\nu}(\kappa_\parallel,q),
\end{equation}
where (note we have put back in $I^{\mu\nu}$ the extra $\frac{1}{g}$ in Eq.~\eqref{pieq1} resulted from the change of integral variables)
\begin{subequations}
\begin{align}
I^{00} &= 4\pi^2\left(-I_1 q_0^2+I_2 q_0^2 v_- v_+-\frac{1}{6} I_2 Q^2 v_-^2 v_+^2+ \frac{1}{2}I_3 Q^2 v_-^3 v_+^3\right),\\
I^{0i} &= -4\pi^2 q_0 Q_iv_-v_+ \frac{v_-+v_+}{2\sqrt{3}}I_2,\qquad\qquad\qquad\qquad\qquad\quad i \in\{1,2,3\},\\
I^{ii} &= 4\pi^2\left(2 I_1 q_0^2+\frac{1}{3} I_2 Q^2 v_-^2 v_+^2-\frac{1}{3}I_3 Q_{ii}^2 v_-^3 v_+^3\right),\qquad\qquad\quad i \in\{1,2,3\},\\
I^{ij} &= -4\pi^2\left(I_1 q_0^2+\frac{1}{6} I_2 Q^2 v_-^2 v_+^2+ \frac{1}{3}I_3 Q_{kk}^2 v_-^3 v_+^3\right),\;\;\quad \{i,j,k\} = \{1,2,3\},
\end{align}
\end{subequations}
where the definitions of $Q$, $Q_i$ and $Q_{ii}$ are in Eqs.~\eqref{qqqqqqq}, and we defined
\begin{equation}\label{I1to5}
I_1 = \frac{(1-u)u}{\sqrt{\Delta} g},\qquad I_2 = \frac{(1-u)u}{\sqrt{\Delta} g^2},\qquad
I_3 = \frac{(1-u)u}{\sqrt{\Delta} g^3}.
\end{equation}
We note that the integrals $\int^1_0 I_{1,2,3}du$ for general $v_+\neq v_-$ can be evaluated, and the result is written in terms of the Elliptic functions. However we are allowed to set $v_+=v_-=v$ in $I_{1,2,3}$ since the $\bm{q}$ dependence in $v_\pm$ does not affect the leading order of the photon self energy $\Pi_1$ which is linear in $\bm{q}$. Then we have
\begin{equation}
\int^1_0 I_1 du =\frac{\pi}{8v^2 \sqrt{q_0^2 + \frac{Q^2}{3}v^2}},\qquad
\int^1_0 I_2 du =\frac{\pi}{8v^4 \sqrt{q_0^2 + \frac{Q^2}{3}v^2}},\qquad
\int^1_0 I_3 du =\frac{\pi}{8v^6 \sqrt{q_0^2 + \frac{Q^2}{3}v^2}},
\end{equation}
This allows us to write
\begin{equation}
\int^1_0 I^{\mu\nu} du
=
\frac{\pi^3}{2\sqrt{q_0^2 + \frac{Q^2}{3}v^2}}
\left(\begin{array}{cccc}
\frac{1}{3} Q^2& -\frac{q_0 Q_1}{\sqrt{3}v}&-\frac{q_0 Q_2}{\sqrt{3}v}&-\frac{q_0 Q_3}{\sqrt{3}v}\\
-\frac{q_0 Q_1}{\sqrt{3}v}&2 \frac{q_0^2}{v^2}+\frac{1}{3} Q^2 -\frac{1}{3} Q_{11}^2&-\frac{q_0^2}{v^2}-\frac{1}{6} Q^2-\frac{1}{3} Q_{33}^2 &-\frac{q_0^2}{v^2}-\frac{1}{6} Q^2-\frac{1}{3} Q_{22}^2\\
-\frac{q_0 Q_2}{\sqrt{3}v}&-\frac{q_0^2}{v^2}-\frac{1}{6} Q^2-\frac{1}{3} Q_{33}^2&2 \frac{q_0^2}{v^2}+\frac{1}{3} Q^2-\frac{1}{3} Q_{22}^2&-\frac{q_0^2}{v^2}-\frac{1}{6} Q^2-\frac{1}{3} Q_{11}^2\\
-\frac{q_0 Q_3}{\sqrt{3}v}&-\frac{q_0^2}{v^2}-\frac{1}{6} Q^2-\frac{1}{3} Q_{22}^2&-\frac{q_0^2}{v^2}-\frac{1}{6} Q^2-\frac{1}{3} Q_{11}^2 &2 \frac{q_0^2}{v^2}+ \frac{1}{3}Q^2-\frac{1}{3} Q_{33}^2
\end{array}\right).
\end{equation}
The matrix on the right has two zero eigenvalues and two nonzero eigenvalues, $\frac{3\pi^3}{2v^2} \sqrt{q_0^2+ \frac{Q^2}{3} v^2}$ and $\frac{3\pi^3}{2v^2}\frac{q_0^2+\frac{1}{9}Q^2v^2}{\sqrt{q_0^2+\frac{Q^2}{3}v^2}}$. And we have
\begin{equation}\label{piiiipi}
\begin{aligned}
\Pi^{\mu\nu}(q) &= \frac{\varsigma_1\varsigma_2\varsigma_3}{(2\pi)^4}\int d \kappa_\parallel
\frac{\pi^3}{2\sqrt{q_0^2 + \frac{Q^2}{3}v^2}}\times \\
&\quad \,\left(\begin{array}{cccc}
\frac{1}{3} Q^2& -\varsigma_1\frac{q_0 Q_1}{3}&-\varsigma_2\frac{q_0 Q_2}{3}&-\varsigma_3\frac{q_0 Q_3}{3}\\
-\varsigma_1\frac{q_0 Q_1}{3}& \frac{2}{3} q_0^2+\frac{v^2}{9} (Q^2 -Q_{11}^2)&-\varsigma_1\varsigma_2\left[\frac{q_0^2}{3}+\frac{v^2}{18} (Q^2+2Q_{33}^2)\right] &-\varsigma_1\varsigma_3\left[\frac{q_0^2}{3}-\frac{v^2}{18} (Q^2+2Q_{22}^2)\right]\\
-\varsigma_2\frac{q_0 Q_2}{3}&-\varsigma_1\varsigma_2\left[\frac{q_0^2}{3}+\frac{v^2}{18} (Q^2+2Q_{33}^2)\right]&\frac{2}{3} q_0^2+\frac{v^2}{9} (Q^2 -Q_{22}^2)&-\varsigma_2\varsigma_3\left[\frac{q_0^2}{3}+\frac{v^2}{18} (Q^2+2Q_{11}^2)\right]\\
-\varsigma_3\frac{q_0 Q_3}{3}&-\varsigma_1\varsigma_3\left[\frac{q_0^2}{3}+\frac{v^2}{18} (Q^2+2Q_{22}^2)\right]&-\varsigma_1\varsigma_2\left[\frac{q_0^2}{3}+\frac{v^2}{18} (Q^2+2Q_{11}^2)\right]&\frac{2}{3} q_0^2+\frac{v^2}{9} (Q^2 -Q_{33}^2)
\end{array}\right).
\end{aligned}
\end{equation}
The matrix now has two zero eigenvalues and two nonzero eigenvalues $\frac{\pi^3(q_0^2+\frac{1}{3} Q^2)}{2 \sqrt{q_0^2+\frac{Q^2}{3} v^2}}$ and $\frac{\pi^3}{2} \sqrt{q_0^2+\frac{Q^2}{3} v^2}$.

The final integral is over $\kappa_\parallel$. To do this, cutoff must be imposed in the vicinity of $\Gamma$ and $\mathrm{L}$ points where the Dirac velocity vanishes:
\begin{subequations}
\begin{align}
\int^{\pi-\theta_0}_{\theta_0} d \kappa_\parallel \frac{1}{\sqrt{A+ \sin^2\kappa_\parallel}} &= \frac{F(\pi-\theta_0,-\frac{1}{A}) - F(\theta_0,-\frac{1}{A})}{\sqrt{A}},\\
\int^{\pi-\theta_0}_{\theta_0} d\kappa_\parallel \frac{\sin^2\kappa_\parallel}{\sqrt{A+ \sin^2\kappa_\parallel}} &=
\sqrt{A}\left(E(\pi-\theta_0, -\frac{1}{A}) - E(\theta_0, -\frac{1}{A})- F(\pi-\theta_0,-\frac{1}{A}) + F(\theta_0,-\frac{1}{A})\right),
\end{align}
\end{subequations}
with $A = \frac{q^2_0}{Q^2}$. Note that we can safely set $\theta_0=0$ in the second integral.

\twocolumngrid

\bibliography{pyrochlore_PSG} 

\end{document}